\pdfoutput=1
\documentclass[12pt,a4]{article}
\usepackage{relsize}
\usepackage{array}
\usepackage{a4wide}
\usepackage[hidelinks]{hyperref}
\usepackage{cite}
\usepackage{mathrsfs}
\usepackage{amssymb}
\usepackage{amsmath}
\usepackage{graphicx}
\usepackage{braket}
\usepackage{enumerate}
\usepackage[usenames,dvipsnames]{xcolor}
\usepackage{todonotes}

\newcommand{\cM}{\mathcal{M}}
\newcommand{\cN}{\mathcal{N}}

\newcommand{\1}{{\rm 1\hskip-0.25em I}}
\newcommand{\be}{\begin{equation}\label}
\newcommand{\ee}{\end{equation}}
\newcommand{\bea}{\begin{eqnarray}\label}
\newcommand{\eea}{\end{eqnarray}}

\newcommand{\lambdat}{\tilde{\lambda}}
\makeatletter
\newcommand*{\textoverline}[1]{$\overline{\hbox{#1}}\m@th$}
\makeatother
\newcommand{\MHVB}{\textoverline{MHV}$\;$}
\newcommand{\Res}[1]{\underset{#1}{\mbox{Res}}}
\newcommand{\ang}[1]{\langle #1\rangle}

\date{}
\begin{document}

\title{From 4d Ambitwistor Strings to On Shell Diagrams and Back}
\author{Joseph A. Farrow and Arthur E. Lipstein \vspace{7pt}\\ \normalsize \textit{
Department of Mathematical Sciences}\\\normalsize\textit{Durham University, Durham, DH1 3LE, United Kingdom}}
\maketitle
\begin{abstract}
We investigate the relation between 4d ambitwistor string theory and on-shell diagrams for planar $\mathcal{N}=4$ super-Yang-Mills and $\mathcal{N}=8$ supergravity, and deduce several new results about their scattering amplitudes at tree-level and 1-loop. In particular, we derive new Grassmannian integral formulae for tree-level amplitudes and obtain new worldsheet formulae for 1-loop amplitudes which are manifestly supersymmetric and supported on scattering equations refined by MHV degree.  

\end{abstract}

\pagebreak
\tableofcontents

\section{Introduction}

Traditional Feynman diagram techniques often obscure the underlying simplicity of on-shell scattering amplitudes.  In recent years several new approaches have been developed which compute amplitudes more efficiently and reveal new mathematical structure. In this paper, we will explore the relationship between two approaches known as 4d ambitwistor string theory and on-shell diagrams in $\mathcal{N}=4$ super-Yang-Mills (SYM) and $\mathcal{N}=8$ supergravity (SUGRA), which are believed to be the simplest quantum field theories in four dimensions. For example, the planar scattering amplitudes of $\mathcal{N}=4$ SYM enjoy Yangian symmetry, which is a hallmark of integrability \cite{Drummond:2009fd}, and the loop amplitudes of $\mathcal{N}=8$ SUGRA exhibit unexpected UV cancellations which suggest that the theory may be pertubatively finite \cite{Bern:2006kd}. Ultimately, we hope that the approaches we explore in this paper will lead to a deeper understanding of the remarkable properties of scattering amplitudes in $\mathcal{N}=4$ SYM and $\mathcal{N}=8$ SUGRA.  

Ambitwistor string theories were first proposed in \cite{Mason:2013sva}. These models are critical in ten dimensions and their spectra only contain field theory degrees of freedom and their tree-level correlation functions produce scattering amplitudes in the form discovered by Cachazo, He, and Yuan (CHY) \cite{Cachazo:2013hca}, notably they are expressed as worldsheet integrals which localize onto solutions of the scattering equations \cite{Fairlie:1972zz,Gross:1987kza}. One-loop amplitudes in 10d ambitwistor string theories were first proposed in \cite{Adamo:2013tsa} and were recently recast in terms of off-shell scattering equations on the Riemann sphere \cite{Geyer:2015bja,Geyer:2015jch}. It is also possible to define intrinsically four-dimensional ambitwistor string theories which can describe tree-level gauge and gravity amplitudes with any amount of supersymmetry and give rise to formulae that are manifestly supersymmetric and supported on scattering equations refined by MHV degree \cite{Geyer:2014fka}. These formulae are closely related to those arising from twistor string theory \cite{Nair:1988bq,Witten:2003nn,Berkovits:2004hg,Roiban:2004yf,Skinner:2013xp}. In particular, the 4d ambitwistor formulae can be obtained by integrating out moduli of the twistor string formulae \cite{Spradlin:2009qr}. 

On-shell diagrams were first proposed in \cite{ArkaniHamed:2012nw}. Unlike Feynman diagrams, on-shell diagrams do not contain virtual particles and are built out of 3-point vertices using BCFW recursion \cite{Britto:2004ap,Britto:2005fq,ArkaniHamed:2010kv}. They were first developed in the context of planar $\mathcal{N}=4$ SYM where they revealed an underlying Grassmannian structure \cite{ArkaniHamed:2009dn} which suggests a geometric interpretation of scattering amplitudes as the volume of an object known as the Amplituhedron \cite{Arkani-Hamed:2013jha}. More recently, on-shell diagrams were developed for tree-level amplitudes in $\mathcal{N}=8$ SUGRA, revealing new connections to planar $\mathcal{N}=4$ SYM \cite{Heslop:2016plj,Herrmann:2016qea}. For example, it is possible to compute $\mathcal{N}=8$ SUGRA amplitudes by decorating planar on-shell diagrams and summing over permutations of the external legs, giving rise to new Grassmannian integral formulae. Although it is possible to extend BCFW recursion to loop level in planar $\mathcal{N}=4$ SYM, it is not known how to generalize this beyond the planar limit or to other theories like $\mathcal{N}=8$ SUGRA. On the other hand, recent progress in this direction has been made using Q-cuts \cite{Baadsgaard:2015twa}, which are intrinsically $d>4$ dimensional and give rise to formulae closely related to those of 10d ambitwistor string theory.   

In this paper, we will investigate how to map worldsheet formulae of 4d ambitwistor string theory into Grassmannian integral formulae arising from on-shell diagrams, obtaining several new results at tree-level and 1-loop. In section \ref{ambiosdreview} we review 4d ambitwistor string theory and on-shell diagrams for $\mathcal{N}=4$ SYM and $\mathcal{N}=8$ SUGRA in greater detail. In section \ref{mhv}, we derive Grassmannian integral formulae for tree-level MHV amplitudes using on-shell diagrams and ambitwistor string theory, generalizing the $\mathcal{N}=8$ SUGRA results obtained in \cite{Heslop:2016plj} to any number of legs. In section \ref{nmhv}, we consider non-MHV amplitudes. In this case, one must specify a contour in the Grassmannian which will depend on the method one uses to compute the amplitudes. For the 6-point NMHV amplitude of $\mathcal{N}=4$ SYM, we show that the three contributing on-shell diagrams correspond to residues of single top form in the Grassmannian and can subsequently be encoded in a Grassmannian contour integral which can be mapped into a 4d ambitwistor string formula using a residue theorem (in agreement with previous results \cite{Spradlin:2009qr,Dolan:2009wf,Nandan:2009cc,ArkaniHamed:2009dg}). On the other hand, for $\mathcal{N}=8$ SUGRA we find that the three decorated planar on-shell diagrams from which the full amplitude can be derived do not correspond to residues of a single top form, so it is unclear how to relate the Grassmanninan contour integral obtained using on-shell diagrams to 4d ambitwistor string theory using residue theorems, although we suggest various other strategies for doing so. 

In section \ref{1loopsection} we use on-shell diagrams to obtain a new worldsheet formulae for the 1-loop four point amplitude of $\mathcal{N}=4$ SYM. Although the procedure can be extended tomore complicated amplitudes in $\mathcal{N}=4$ SYM using loop-level BCFW recursion, it is not yet clear how to do this for $\mathcal{N}=8$ SUGRA. Nevertheless, it is possible to describe the 1-loop 4-point amplitude of $\mathcal{N}=8$ SUGRA using a decorated on-shell diagram \cite{Heslop:2016plj}, from which we deduce a worldsheet formula as well. These formulae are manifestly supersymmetric and supported on 1-loop scattering equations refined by MHV degree.

We also include several appendices. In Appendix \ref{algorithm} we describe an algorithm for computing on-shell diagrams in $\mathcal{N}=8$ SUGRA. In Appendix \ref{bonusappendix} we explain how to incorporate the bonus relations for $\mathcal{N}=8$ SUGRA into on-shell diagrams and use this to solve the recursion relations in the planar MHV sector obtaining a simplified version of the BGK formula for tree-level MHV graviton scattering \cite{Berends:1988zp}. In Appendix \ref{identities} we derive useful identities relating spinor brackets to minors appearing in Grassmannian integral formulae. In Appendix \ref{1loopb}, we show how to map our worldsheet formula for the 1-loop 4-point amplitude of $\mathcal{N}=4$ SYM to the well-known expression in terms of a scalar box integral \cite{Green:1982sw}. Finally, in Appendix \ref{counting} we consider a generalization of the 1-loop scattering equations refined by MHV degree to any number of legs and analyze various properties of their solutions.       

\section{Review} \label{ambiosdreview}

\subsection{4d Ambitwistor Strings} \label{ambrev}

In this section, we will review the construction of 4d ambitwistor string theories for $\mathcal{N}=4$ SYM and $\mathcal{N}=8$ SUGRA \cite{Geyer:2014fka,Bandos:2014lja}. For $\mathcal{N}=4$ SYM, the worldsheet fields are
\[
Z^A=\left(\begin{array}{c}
\lambda_{\alpha}\\
\mu^{\dot{\alpha}}\\
\chi^{a}
\end{array}\right),\,\,\, W_A=\left(\begin{array}{c}
\tilde{\mu}^{\alpha}\\
\tilde{\lambda}_{\dot{\alpha}}\\
\tilde{\chi}_{a}
\end{array}\right)
\]
where $\lambda$ and $\tilde{\lambda}$ are commuting 2-component spinors and $\chi,\tilde{\chi}$ are fermions transforming in the fundamental representation of the R-symmetry group $SU(4)$. We use the following notation to denote spinor inner products: $\left\langle rs\right\rangle =r_{\alpha}s_{\beta}\epsilon^{\alpha\beta}$ and $\left[rs\right]=r^{\dot{\alpha}}s^{\dot{\beta}}\epsilon_{\dot{\alpha}\dot{\beta}}$, where $\epsilon$ is the Levi-Civita symbol. The Lagrangian for the worldsheet theory  is 
\begin{equation}
\mathcal{L}=W_{A}\bar{\partial}Z^{A}+uW_{A}Z^{A}\label{eq:ly}
\end{equation}
where $u$ is a $GL(1)$ gauge field. Note that this is the same action as in twistor string theory \cite{Witten:2003nn,Berkovits:2004hg}. The new feature of 4d ambitwistor strings are that the worldsheet fields have conformal weight $\left(\frac{1}{2},0\right)$, and vertex operators are defined for both positive and negative helicity particles.

In four dimensions, a null momentum can be written in bispinor form as 
\[
p_{i}^{\alpha\dot{\alpha}}=\lambda_{i}^{\alpha}\tilde{\lambda}_{i}^{\dot{\alpha}}.
\]
where $i$ is a particle label. Moreover one may also define the supermomentum as
\[
q_{i}^{\alpha a}=\lambda_{i}^{\alpha}\tilde{\eta}_{i}^{a}
\]
where $a$ is an R-symmetry index. The integrated vertex operators for $\mathcal{N}=4$ SYM with supermomentum parametrized by $\lambda_i,\tilde{\lambda}_i,\tilde{\eta}_i$ are then given by 
\[
\mathcal{V}_{i}=\int d\sigma_{i}\frac{dt_{i}}{t_{i}}\delta^{2}\left(\lambda_{i}-t_{i}\lambda(\sigma_{i})\right)e^{it_{i}\left(\left[\mu(\sigma_{i})\tilde{\lambda}_{i}\right]+\chi(\sigma_{i})\cdot\tilde{\eta}_{i}\right)}J(\sigma_{i})
\]
\[
\tilde{\mathcal{V}}_{i}=\int d\sigma_{i}\frac{dt_{i}}{t_{i}}\delta^{2|4}\left(\tilde{\lambda}_{i}-t_{i}\tilde{\lambda}(\sigma_{i})|\tilde{\eta}_{i}-t_{i}\chi(\sigma_{i})\right)e^{it_{i}\left\langle \tilde{\mu}(\sigma_{i})\lambda_{i}\right\rangle }J(\sigma_{i})
\]
where $J$ is a Kac-Moody current and $\tilde{\mathcal{V}}$ is obtained by complex conjugating $\mathcal{V}$ and Fourier transforming back to $\tilde{\eta}$ space. Note that the worldsheet coordinates can be thought of as homogenous coordinates on $CP^1$ with components $\sigma_{i}^{\alpha}=t_{i}^{-1}\left(1,\sigma_{i}\right)$, in terms of which we define the inner product
$(ij)=\sigma_{i}^{\alpha}\sigma_{j}^{\beta}\epsilon_{\alpha\beta}$.

The BRST cohomology also contains vertex operators corresponding to conformal supergravity states, but they can be neglected at tree-level. Schematically, a tree-level N$^{k-2}$MHV amplitude in $\mathcal{N}=4$ SYM can then be computed from a genus zero correlator with $k$ $\tilde{\mathcal{V}}$ vertex operators and $(n-k)$ $\mathcal{V}$ vertex operators to obtain
\begin{equation}\label{n=4ambi}
\mathcal{\mathcal{A}}_{n,k}^{(0)}=\int\frac{1}{GL(2)}\prod_{i=1}^{n}\frac{d^{2}\sigma_{i}}{\left(i\,i{+}1\right)}\prod_{l}\delta^{2|4}\left(\tilde{\lambda}_{l}-\sum_{r}\frac{\tilde{\lambda}_{r}}{\left(lr\right)}\right)\prod_{r}\delta^{2}\left(\lambda_{r}-\sum_{l}\frac{\lambda_{l}}{(rl)}\right)
\end{equation}
where $l\in\left\{ 1,...,k\right\} $ and $r\in\{k+1,...,n\}$. Note that $\delta^{2|4}$ contain fermionic delta functions and can therefore be written more precisely as
\[
\delta^{2|4}\left(\left.\tilde{\lambda}_{l}-\sum_{r}\frac{\tilde{\lambda}_{r}}{\left(lr\right)}\right|\tilde{\eta}_{l}-\sum_{r}\frac{\tilde{\eta}_{r}}{\left(lr\right)}\right)
\]
but we will use the notation in \eqref{n=4ambi} for brevity. The arguments of the delta functions are known as the 4d tree-level scattering equations refined by MHV degree. Note that the cyclic structure in $1/(i\, i+1)$ arises from contractions of the current algebra and encodes the formula for the gluon MHV amplitudes discovered by Parke and Taylor \cite{Parke:1986gb}. The $GL(2)$ symmetry can be used to fix four worldsheet coordinates following the usual Fadeev-Popov procedure. For example, it is conventional to fix $\sigma_i^{\alpha}=(1,0)$ and $\sigma_j^{\alpha}=(0,1)$ for some $i,j$.

For $\mathcal{N}=8$ SUGRA, the worldsheet theory has $Z,W$ fields with eight fermionic components, as well as the following additional fields:
\[
\rho^A=\left(\begin{array}{c}
\rho_{\alpha}\\
\rho^{\dot{\alpha}}\\
\omega^{a}
\end{array}\right),\,\,\,\tilde{\rho}_A=\left(\begin{array}{c}
\tilde{\rho}^{\alpha}\\
\tilde{\rho}_{\dot{\alpha}}\\
\tilde{\omega}_{a}
\end{array}\right)
\]
which have the opposite statistics of $(Z,W)$. The Lagrangian is \cite{Skinner:2013xp}
\begin{equation}
\mathcal{L}=W_{A}\bar{\partial}Z^{A}+\tilde{\rho}_{A}\bar{\partial}\rho^{A}+u^{B}K_{B}\label{eq:lg}
\end{equation}
where there are four bosonic and four fermionic currents given by  
\begin{equation}
K_{B}=\left\{ W_{A}Z^{A},\tilde{\rho}_{A}\rho^{A},\rho^{\alpha}\rho_{\alpha},\tilde{\rho}^{\dot{\alpha}}\tilde{\rho}_{\dot{\alpha}},\rho^{A}W_{A},Z^{A}\tilde{\rho}_{A},\lambda^{\alpha}\rho_{\alpha},\tilde{\lambda}^{\dot{\alpha}}\tilde{\rho}_{\dot{\alpha}}\right\}.
\label{currents}
\end{equation}
The integrated vertex operators are
\[
\mathcal{V}_{i}=\int d\sigma_{i}\left(\left[W,\frac{\partial h}{\partial Z}\right]+\left[\tilde{\rho},\frac{\partial}{\partial Z}\right]\rho\cdot\frac{\partial h}{\partial Z}\right)(\sigma_{i})
\]
and 
\[
\tilde{\mathcal{V}}_{i}=\int d\sigma_{i}\left(\left\langle Z,\frac{\partial\tilde{h}}{\partial W}\right\rangle +\left\langle \rho,\frac{\partial}{\partial W}\right\rangle \tilde{\rho}\cdot\frac{\partial\tilde{h}}{\partial W}\right)(\sigma_{i})
\]
where we define $\left\langle Z_{i}Z_{j}\right\rangle =\left\langle \lambda_{i}\lambda_{j}\right\rangle $ and $\left[W_{i}W_{j}\right]=\left[\tilde{\lambda}_{i}\tilde{\lambda}_{j}\right]$, and
\[
h(\sigma_{i})=\int\frac{dt_{i}}{t_{i}^{3}}\delta^{2}\left(\lambda_{i}-t_{i}\lambda(\sigma_{i})\right)e^{it_{i}\left(\left[\mu(\sigma_{i})\tilde{\lambda}_{i}\right]+\chi(\sigma_{i})\cdot\tilde{\eta}_{i}\right)}
\] 
\[
\tilde{h}(\sigma_{i})=\int\frac{dt_{i}}{t_{i}^{3}}\delta^{2|8}\left(\tilde{\lambda}_{i}-t_{i}\tilde{\lambda}(\sigma_{i})|\tilde{\eta}_{i}-t_{i}\chi(\sigma_{i})\right)e^{it\left\langle \tilde{\mu}(\sigma_{i})\lambda_{i}\right\rangle }.
\]

The BRST cohomology also contains unintegrated vertex operators constructed from ghosts associated with the fermionic currents in \eqref{currents}, but we will not discuss them for simplicity. In the end, a tree-level N$^{k-2}$MHV amplitude in $\mathcal{N}=8$ SUGRA can be computed from a genus zero correlator to obtain
\begin{equation}
\mathcal{M}_{n,k}^{(0)}=\int\frac{\prod_{i=1}^{n}d^{2}\sigma_{i}}{GL(2)}\det{}' H \, \det{}' \tilde{H} \, \prod_{l}\delta^{2|8}\left(\tilde{\lambda}_{l}-\sum_{r}\frac{\tilde{\lambda}_{r}}{\left(lr\right)}\right)\prod_{r}\delta^{2}\left(\lambda_{r}-\sum_{l}\frac{\lambda_{l}}{(rl)}\right)
\label{n=8ambi}
\end{equation}
where  $\det'$ indicates to remove one row and column and evaluate the determinant of the following matrices, which we refer to as Hodges matrices:
\[
H_{ll}=-\sum_{l'\neq l}\frac{\left\langle ll'\right\rangle }{\left(ll'\right)},\,\,\, H_{ll'}=\frac{\left\langle ll'\right\rangle }{\left(ll'\right)},\,\,\, l\neq l',
\]
with $l,l'\in\{1,...,k\}$, and
\[
\tilde{H}_{rr}=-\sum_{r'\neq r}\frac{\left[rr'\right]}{\left(rr'\right)}\,\,\,\tilde{H}_{rr'}=\frac{\left[rr'\right]}{\left(rr'\right)},\,\,\, r\neq r'
\]
with $r,r'\in\{k+1,...,n\}$. The determinants in this formula arise from  contractions of the $\rho,\tilde{\rho}$ fields and encode the formula for graviton MHV amplitudes discovered by Hodges \cite{Hodges:2012ym}.  

For later sections, it will be useful to describe how little group transformations are realized in the above worldsheet formulae. For an $n$-point N$^{k-2}$MHV amplitude, a general little group transformation can be written as follows:
\[
\left(\lambda_{i},\tilde{\lambda}_{i},\tilde{\eta}_{i}\right)\rightarrow\left(\alpha_{i}^{-1}\lambda_{i},\alpha_{i}\tilde{\lambda}_{i},\alpha_{i}\tilde{\eta}_{i}\right),
\]
where $i \in \{1,...,n\}$ and $\alpha_i \in GL(1)$. It is then easy to show that the formulae in \eqref{n=4ambi} and \eqref{n=8ambi} transform covariantly if the worldsheet coordinates transform as follows:
\[
\sigma_{l}\rightarrow\alpha_{l}^{-1}\sigma_{l},\,\,\,\sigma_{r}\rightarrow\alpha_{r}\sigma_{r},
\]
where $l\in\{1,...,k\}$ and $r\in\{k+1,...,n\}$. In particular, under this transformation the superamplitudes are rescaled by an overall factor of 
\[
\Pi_{i=1}^{n}\alpha_{i}^{2 s},
\]
where $s=1,2$ for $\mathcal{N}=4$ SYM and $\mathcal{N}=8$ SUGRA, respectively. Note that if we set $\left(\lambda_{l},\tilde{\lambda}_{l},\tilde{\eta}_{l}\right)=\left(-\lambda_{r},\tilde{\lambda}_{r},\tilde{\eta}_{r}\right)$ for some $l$ and $r$, then the discussion above implies that the inner product $(lr)$ will be invariant under little group transformations since $\sigma_l$ and $\sigma_r$ transform with opposite weight.

\subsection{On-Shell Diagrams}\label{osdrev}

\begin{figure}
\centering
\begin{tabular}{m{3cm} m{.5cm} m{5cm}}
       \includegraphics[scale=0.5]{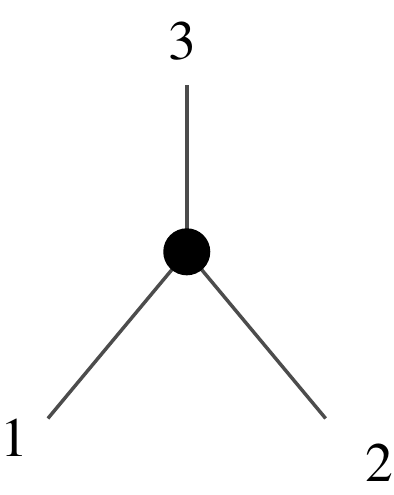} &=& $\dfrac{\delta^{4}(P)\delta^{2\cN}\left(\lambda_{1}\tilde{\eta}_{1}+\lambda_{2}\tilde{\eta}_{2}+\lambda_{3}\tilde{\eta}_{3}\right)}{\big(\ang{12}\ang{23}\ang{31}\big)^{\cN/4}}$\\
       \includegraphics[scale=0.5]{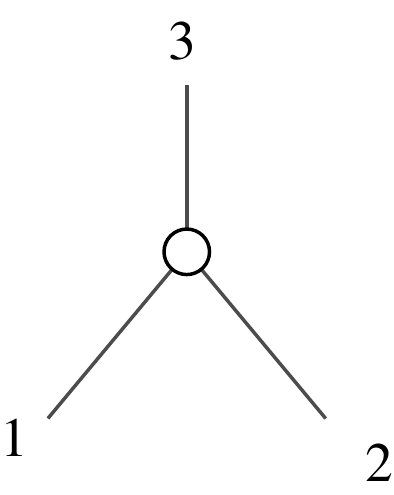} &=&  $\dfrac{\delta^{4}(P)\delta^{\cN}\left([12]\tilde{\eta}_{3}+[23]\tilde{\eta}_{1}+[31]\tilde{\eta}_{2}\right)}{\big([12][23][31]\big)^{\cN/4}}$ \\
       \rule{2.5cm}{.5pt}&=& $\int \dfrac{d^2\lambda d^2\tilde{\lambda}}{GL(2)}d^{\cN}\eta $
\end{tabular}
    \caption{Vertices and edges for on shell diagrams in $\mathcal{N}=4$
SYM and $\mathcal{N}=8$
SUGRA} 
    \label{fig:3ptosddefs}
\end{figure}

On-shell diagrams are graphs constructed from 3-point black
and white vertices which correspond to 3-point MHV and \MHVB superamplitudes
respectively, as shown in the upper part of Figure~\ref{fig:3ptosddefs}. Unlike ordinary Feynman diagrams, the
internal lines of on-shell diagrams do not contain virtual particles
and correspond to integrating over on-shell degrees of freedom,
as depicted in the lower part of Figure~\ref{fig:3ptosddefs}.

\begin{figure}
\centering
\begin{tabular}{m{4cm} 	m{.5cm}m{1cm} m{5cm} m{1cm} m{5cm}}
 \includegraphics[scale=0.85]{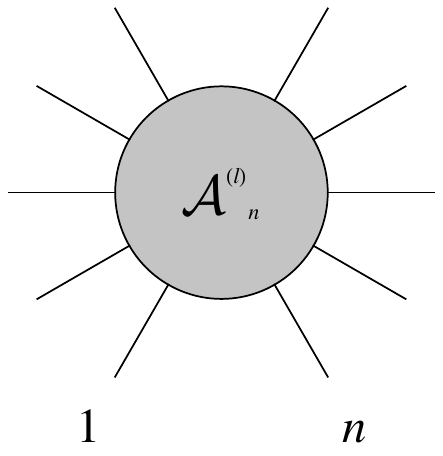} & = & $\mathlarger{\sum}_{\textrm{L,R}}$ &\includegraphics[scale=0.7]{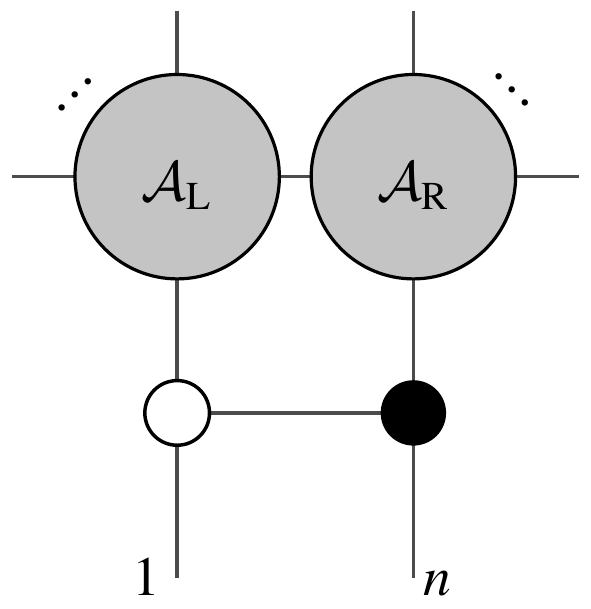} & + & \includegraphics[scale=0.45]{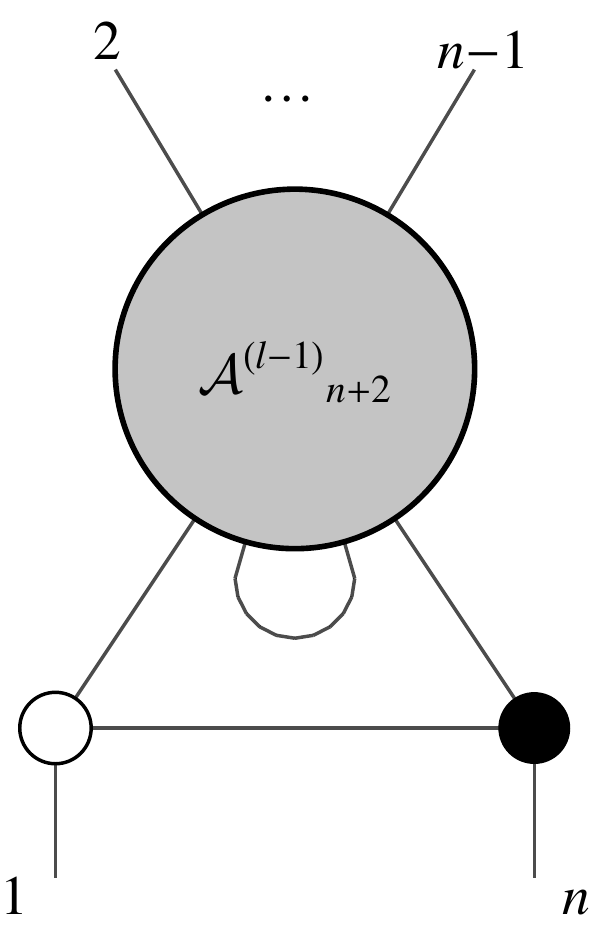}
 \end{tabular}
    \caption{Loop-level BCFW recursion for planar $\mathcal{N}=4$
SYM} 
    \label{n4recursion}
\end{figure}

\begin{figure}
\centering
       \includegraphics[scale=1]{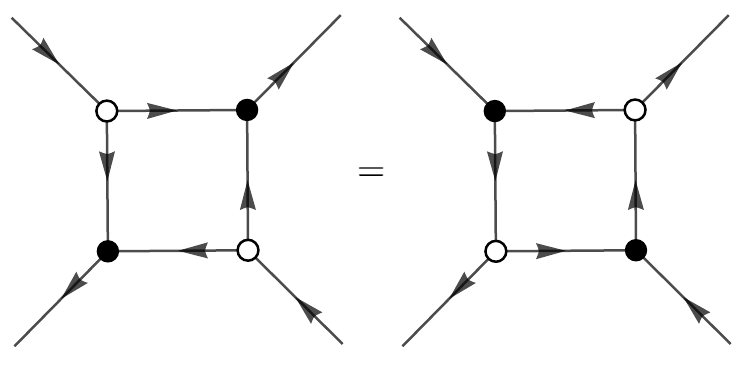}
    \caption{Square move equivalence relation
SYM} 
    \label{squaremove}
\end{figure}

\begin{figure}
\centering
       \includegraphics[scale=1]{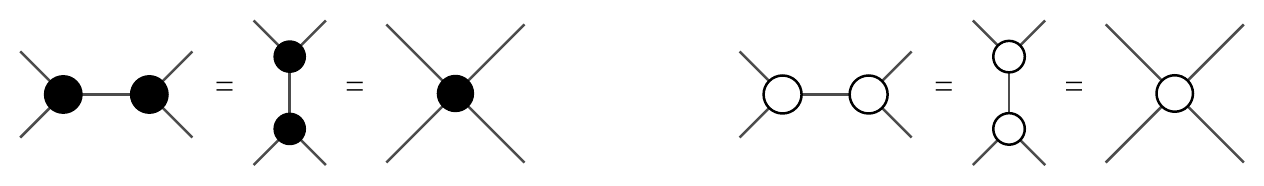}
    \caption{Merger equivalence relations for $\mathcal{N}=4$
SYM} 
    \label{merger}
\end{figure}

The planar scattering amplitudes in $\mathcal{N}=4$
SYM can be constructed from on-shell diagrams using the recursion
relation in Figure \ref{n4recursion} \cite{ArkaniHamed:2012nw}. If one neglects the second term on the right-hand side,
this encodes BCFW recursion for tree-level amplitudes. In particular,
the structure attaching legs $1$ and $n$ to the lower-point on-shell
diagrams implements the standard BCFW shift and is known as a BCFW
bridge. In planar $\mathcal{N}=4$ SYM, it is possible to extend the
recursion relation to loop-level, which is taken into account by the
second term on the right-hand side in Figure \ref{n4recursion}, which involves connecting two adjacent
legs of a lower-loop diagram (referred to as a forward limit) and
attaching a BCFW bridge. The on-shell diagrams of $\mathcal{N}=4$
SYM also enjoy various equivalence relations such as the square move
and mergers depicted in Figures \ref{squaremove} and \ref{merger}, which can often be used to simplify
calculations.

\begin{figure}
\centering
       \includegraphics[scale=.9]{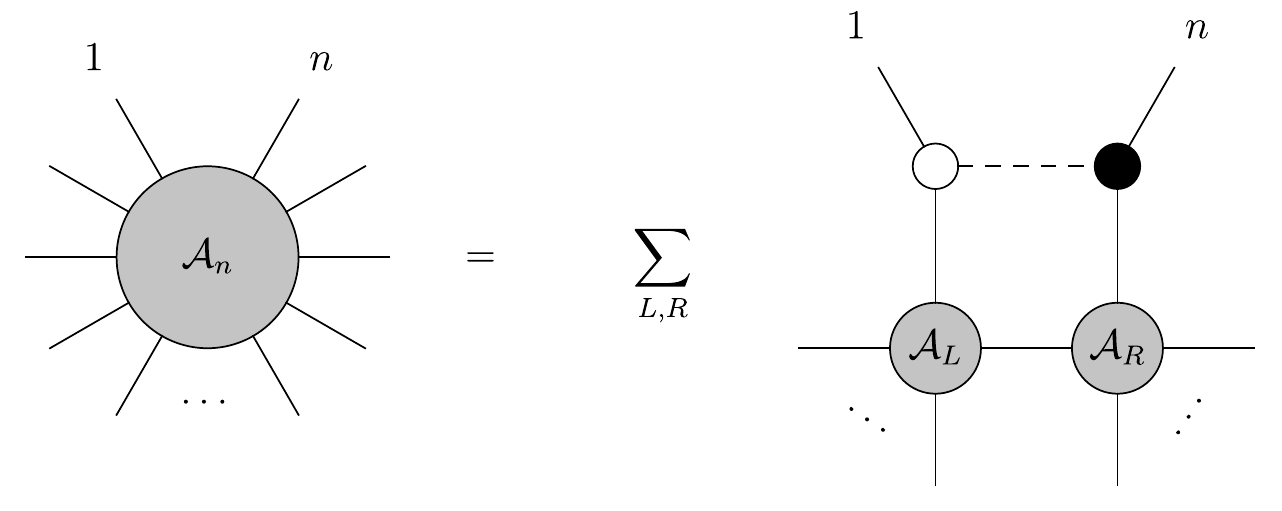}
    \caption{Tree-level BCFW recursion relations in $\mathcal{N}=8$
SUGRA} 
    \label{recursion8}
\end{figure}

\begin{figure}
\centering
       \includegraphics[scale=1]{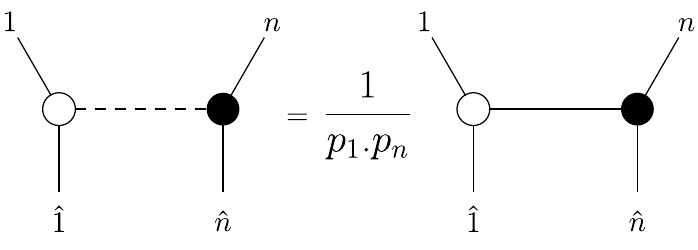}
    \caption{Definition of BCFW bridge decoration in $\mathcal{N}=8$
SUGRA} 
    \label{bridge}
\end{figure}

\begin{figure}
\centering
       \includegraphics[scale=1]{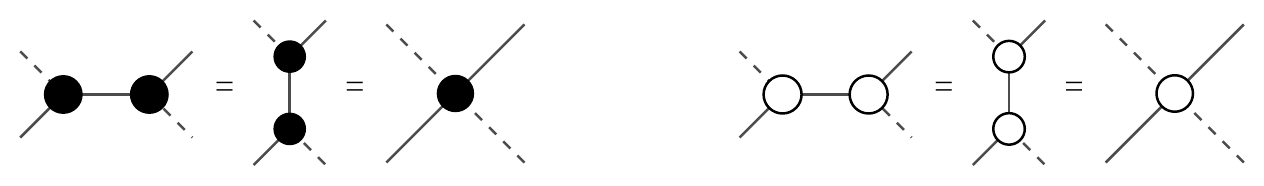}
    \caption{Merger equivalence relations in $\mathcal{N}=8$
SUGRA} 
    \label{merge2}
\end{figure}

In $\mathcal{N}=8$ SUGRA, it is possible to define a tree-level recursion
relation in terms of on-shell diagrams, as depicted in Figure \ref{recursion8} \cite{Heslop:2016plj}. In
this case, the BCFW bridge is decorated by a kinematic factor as shown
in Figure \ref{bridge}, and one sums over all partitions of the external legs
in the two subamplitudes holding legs $(1,n)$ fixed. In general, this will yield non-planar on-shell
diagrams, but it is possible to restrict the recursion to a planar
sector by attaching the fixed legs of each subdiagram to the bridge
or the other subdiagram at each step in the recursion. The full amplitude
can then be obtained by summing over permutations
of the unshifted external legs, implying nontrivial identities for non-planar
on-shell diagrams. Furthermore, the on-shell diagrams of $\mathcal{N}=8$
SUGRA enjoy equivalence relations similar to those of $\mathcal{N}=4$ SYM,
in particular the square move in Figure \ref{squaremove} and decorated mergers in Figure \ref{merge2}.

A remarkable feature of on-shell diagrams is that they naturally give
rise to formulae for N$^{k-2}$MHV amplitudes in the form of integrals
over $k$-planes in $n$ dimensions, also known as the Grassmannian
$Gr(k,n)$. These integrals can be represented as integral over a
$k\times n$ matrix $C$ modulo a left action of $GL(k)$ and are
supported on delta functions of the form
\begin{equation}
\delta^{k\times\left(2|\mathcal{N}\right)}\left(C\cdot\tilde{\lambda}|C\cdot\tilde{\eta}\right)\delta^{2\times(n-k)}\left(C^{\perp}\cdot\lambda\right)
\label{eq:deltafunctions}
\end{equation}
where $C^{\perp}$ is an $n\times(n-k)$ matrix satisfying $C^{\perp}\cdot C=0$
and
\[
\left(i_{1}...i_{n-k}\right)^{\perp}=\epsilon_{i_{1}...i_{n}}\left(i_{n-k+1}...i_{n}\right)
\]
where the left and right hand sides denote the minors of $C^{\perp}$
and $C$, respectively. The dot products appearing in the delta functions are with respect to particle number, so for example $C \cdot \tilde{\lambda}$ is written more precisely as $\sum_{j=1}^{n}C_{Ij}\tilde{\lambda}_{j}^{\alpha}$, where $I\in\left\{ 1,...,k\right\} $. 

It is often convenient to use the $GL(k)$ symmetry
to fix $C$ in such a way that $k$ columns form a $k\times k$ unit
matrix and one integrates over the remaining $k\times(n-k)$ elements.
This form is referred to as the link representation and is closely
related to the expressions arising from 4d ambitwistor string theory.
Indeed, in the link representation the delta functions in \eqref{eq:deltafunctions}
take the same form as the 4d scattering equations refined
by MHV degree.

There is a simple algorithm for deriving Grassmannian integral formulae
directly from on-shell diagrams, which we shall now describe schematically (more details can
be found in Appendix \ref{algorithm}). First one assigns variables $\alpha$ and arrows to the edges of the diagram such that there are two arrows entering and one arrow leaving every black node, and two arrows leaving and one arrow entering every white node. Then one sets an edge variable associated with each vertex to unity, leaving $2n-4$ edge variables. To construct the Grassmannian integral in $\mathcal{N}=4$ SYM, one then takes the product of $\rm{d} \alpha /\alpha$ for each edge variable and multiplies this by the delta functions in \eqref{eq:deltafunctions}, where the $C$ and $C^{\perp}$ matrices are computed by summing over paths through the on-shell diagram and taking the product of the edge variables encountered along each path, as described in more detail in Appendix \ref{algorithm}. The resulting formula can then be thought of as a gauge fixed Grassmannian integral formula (where the gauge symmetry corresponds to $GL(k)$). Lifting this to a covariant formula will often give the following expression or one of its residues: 
\begin{equation}
d^{k \times n} \Omega_\mathcal{N} := \frac{d^{k \times n} C}{\mbox{Vol(GL}(k))} 
\frac{\delta^{k\times (2|\mathcal{N})}(C \cdot\lambdat|C \cdot \eta)\delta^{(n-k)\times2}(\lambda\cdot C^{\perp})}
{\prod_{i=1}^{n}(i \: ... \: i{+}k{-}1)}.
\label{meas}
\end{equation}
Note a similar factor also appears in Grassmannian integral formulae for $\mathcal{N}=8$ SUGRA amplitudes, so we keep the supersymmetry parameter $\mathcal{N}$ unfixed. 

For $\mathcal{N}=8$ SUGRA, the algorithm for deriving Grassmannian integral formulae from on-shell diagrams is similar to that of $\mathcal{N}=4$ SYM, except that one includes a factor of $\rm{d} \alpha /\alpha^2$ for each edge variable leaving a white vertex or entering a black vertex and $\rm{d} \alpha /\alpha^3$ for each edge variable entering a white vertex or leaving a black vertex. Furthermore, one must include decorations for the BCFW bridges as depicted in Figure \ref{bridge} and spinor brackets for the vertices. In particular, for each black vertex include a factor of $\left\langle ij\right\rangle$ where $i$ and $j$ are the two edges with ingoing arrows, and for each white vertex include a factor of $\left[ij\right]$ where  $i$ and $j$ are the two edges with outgoing arrows. The  spinors in these brackets can be written in terms of the external spinors and edge variables by summing over paths in the on-shell diagram in a similar way to how one computes the $C$-matrix. In the final step, one includes the delta functions in \eqref{eq:deltafunctions} and lifts the integrand to a covariant expression. More details and various shortcuts for computing on-shell diagrams in $\mathcal{N}=8$ SUGRA are described in Appendix \ref{algorithm}. In Appendix \ref{bonusappendix} we also explain how to incorporate the bonus relations into the on-shell diagram recursion for MHV amplitudes.

\section{Tree-level MHV} \label{mhv}

In this section we will derive Grassmannian integral formulae for tree-level MHV amplitudes in $\mathcal{N}=4$ SYM and $\mathcal{N}=8$ SUGRA using on-shell diagrams and 4d ambitwistor string theory. Note that the 4d ambitwistor string formulae can already be thought of as integrals over the Grassmannian $Gr(2,n)$ if we arrange the worldsheet coordinates $\sigma_i^\alpha$ into a $2\times n$ matrix. For N$^{k-2}$MHV amplitudes, we must map this $Gr(2,n)$ into $Gr(k,n)$ via link variables in order to compare with the expressions we obtain from on-shell diagrams, so we will first describe this mapping for MHV amplitudes. We will generalize to non-MHV and 1-loop amplitudes in subsequent sections.   

\subsection{$\mathcal{N}=4$}
We will first derive the Grassmannian integral formula for MHV amplitudes in $\mathcal{N}=4$ SYM by mapping the 4d ambitwistor string formula in \eqref{n=4ambi} into link variables. This can be accomplished by inserting $1$ in the form
\begin{equation}
1=\int\prod_{l,r}dc_{lr}\delta\left(c_{lr}-\frac{1}{(lr)}\right)
\label{intc}
\end{equation} 
to obtain
\[
\mathcal{A}_{n,2}^{(0)}=\int\frac{1}{GL(2)}\prod_{i=1}^{n}\frac{d^{2}\sigma_{i}}{\left(i\,i{+}1\right)}\prod_{l,r}dc_{lr}\delta\left(c_{lr}-\frac{1}{(lr)}\right)\prod_{l}\delta^{2|4}\left(\tilde{\lambda}_{l}-c_{lr}\tilde{\lambda}_{r}\right)\prod_{r}\delta^{2}\left(\lambda_{r}+c_{lr}\lambda_{l}\right)
\]
where $l\in\left\{ 1,2\right\} $ and $r\in\{3,...,n\}$. If we use the $GL(2)$ symmetry to
to fix $\sigma_{1}=(1,0)$ and $\sigma_{2}=(0,1)$, then $(12)=1$, $(1r)=\sigma_{r}^{2}$,
$(2r)=-\sigma_{r}^{1}$, and the delta functions in the link variables can be written as 
\begin{equation}
\prod_{l,r}\delta\left(c_{lr}-\frac{1}{(lr)}\right)=\prod_{r}\frac{1}{c_{1r}^{2}c_{2r}^{2}}\delta\left(\sigma_{r}^{2}-1/c_{1r}\right)\delta\left(\sigma_{r}^{1}+1/c_{2r}\right).
\label{deltac}
\end{equation}
Furthermore, on the support of these delta functions we find that
\[
(i\,i{+}1)=\frac{c_{1i}c_{2i+1}-c_{1i+1}c_{2i}}{c_{1i}c_{2i}c_{1i+1}c_{2i+1}}
\]
for $3\leq i\leq n-1$. Hence, if we integrate the worldsheet coordinates against the delta functions in \eqref{deltac} we are left with the following integral over link variables:

\[
\mathcal{A}_{n,2}^{(0)}=\int\frac{d^{2\times(n-2)}C}{(12)...(n1)}\delta^{(2|4)\times2}\left(C\cdot\tilde{\lambda}\right)\delta^{2\times(n-2)}\left(\lambda\cdot C^{\perp}\right)
\]
where we have arranged the link variables into a $2\times n$ matrix
$C$

\begin{equation}
C=\left(\begin{array}{ccccc}
1 & 0 & c_{13} & ... & c_{1n}\\
0 & 1 & c_{23} & ... & c_{2n}
\end{array}\right)
\label{Cch}
\end{equation}
and $(ij)$ now refers to a minor of $C$ involving columns $i$ and $j$ rather than an inner product of worldsheet coordinates. If we think of $C$ as an element of $Gr(2,n)$, the formula above corresponds to a particular choice of coordinates on this space. The formula for MHV amplitudes can then be written covariantly as follows 
\begin{equation}
\mathcal{A}_{n,2}^{(0)}=\int\frac{d^{2\times n}C}{GL(2)}\frac{1}{(12)...(n1)}\delta^{2\times(2|4)}\left(C\cdot\tilde{\lambda}\right)\delta^{(n-2)\times 2}\left(\lambda\cdot C^{\perp}\right).\label{eq:n=00003D4mhvgrass}
\end{equation}
where the $GL(2)$ allows one to fix four elements of the $C$-matrix, as we did in \eqref{Cch}.

\begin{figure}
\centering
       \includegraphics[scale=0.8]{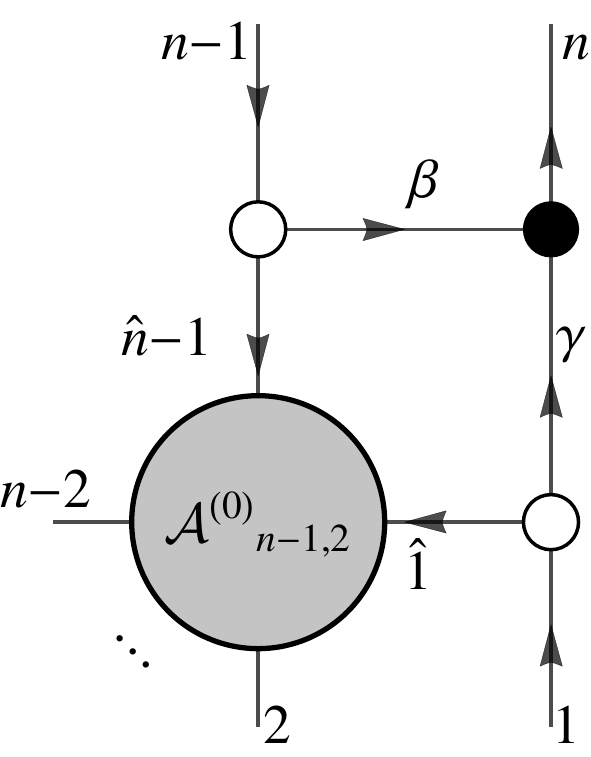}
    \caption{On-shell diagram for tree-level $n$-point MHV amplitude in $\mathcal{N}=4$ SYM in terms of $(n-1)$-point MHV amplitude} 
    \label{n4m}
\end{figure}

It not difficult to derive this expression directly from on-shell diagrams. Indeed for MHV amplitudes, there is only one on-shell diagram to consider, depicted in Figure~\ref{n4m}. At $n$ points, it is given by

\[
\mathcal{A}_{n,2}^{(0)}=\int\frac{d\beta}{\beta}\frac{d\gamma}{\gamma}\int\frac{d^{2\times (n{-}1)}C}{GL(2)}\delta^{2\times(2|4)}\left(C\cdot \tilde{\lambda}\right)\delta^{(n-2)\times 2}\left(\lambda\cdot C^{\perp}\right)\mathcal{I}\left(\hat{1},2,...,\hat{n}{-}1\right)
\]
where $\mathcal{I}$ is the integrand of the $(n-1)$-point sub-amplitude, without the delta functions. The
$C$ matrix can be computed in terms of edge variables following the
algorithm in section~\ref{algorithm}, and is given by 
\[
C=\left(\begin{array}{cccc}
1 & ... & 0 & \gamma\\
0 & ... & 1 & \beta
\end{array}\right),
\]
where the rows correspond to legs $(1,n-1)$ and the ellipsis encodes the edge variables of the subdiagram.
Noting that $(1\,n{-}1)=1$, $(1n)=\beta$, and $(n\,n{-}1)=\gamma$, we see that the integral over edge variables can be covariantized as follows:
\[
\mathcal{A}_{n,2}^{(0)}=\int\frac{d^{2\times n}C}{GL(2)}\frac{(1\,n{-}1)}{(1n)(n\,n{-}1)}\delta^{2\times(2|4)}\left(C\cdot \tilde{\lambda}\right)\delta^{(n-2)\times 2}\left(\lambda\cdot C^{\perp}\right)\mathcal{I}\left(\hat{1},2,...,\hat{n}{-}1\right).
\]
Using the $GL(2)$ symmetry to set 
\begin{equation}
C=\left(\lambda_{1}...\lambda_{n}\right)
\label{C}
\end{equation}
we then obtain the following recursion relation for MHV amplitudes
\[
\mathcal{A}_{n,2}^{(0)}=\frac{\left\langle 1\,n{-}1\right\rangle }{\left\langle 1n\right\rangle \left\langle n\,n{-}1\right\rangle }\mathcal{A}_{n{-}1,2}^{(0)}
\]
which is easily solved to give
\[
\mathcal{A}_{n,2}^{(0)}=\frac{\delta^{4|8}(P)}{\prod_{i=1}^{n}\left\langle i\,i{+}1\right\rangle }
\]
where $n+1 \sim 1$. It is easy to see that \eqref{eq:n=00003D4mhvgrass} is the unique Grassmannian uplift of the above formula, which can be seen by using the $GL(2)$ symmetry to choose $C$ as in \eqref{C}.

\subsection{$\mathcal{N}=8$}
In this section, we will derive a new Grassmannian integral formula for the MHV amplitudes of $\mathcal{N}=8$ SUGRA, generalizing the results obtained using on on-shell diagrams in \cite{Heslop:2016plj} to any number of legs. As we did in the previous subsection, start by inserting~(\ref{intc}) into the 4d ambitwistor formula in \eqref{n=8ambi}
\begin{equation}
\mathcal{\mathcal{M}}_{n,2}^{(0)}=\int\frac{\prod_{i=1}^{n}d^{2}\sigma_{i}}{GL(2)}\prod_{l,r}dc_{lr}\delta\left(c_{lr}-\frac{1}{(lr)}\right)\det{}' H\det{}'\tilde{H}\prod_{l}\delta^{2|8}\left(\tilde{\lambda}_{l}-c_{lr}\tilde{\lambda}_{r}\right)\prod_{r}\delta^{2}\left(\lambda_{r}+c_{lr}\lambda_{l}\right)
\label{Mc}
\end{equation}
where $l\in\left\{ 1,2\right\} $ and $r\in\{3,...,n\}$. Using the $GL(2)$ symmetry
to fix $\sigma_{1}=(1,0)$ and $\sigma_{2}=(0,1)$, we can once again write the delta functions in the link variables as in \eqref{deltac} and on the support of these delta functions we obtain 
\[
(ij)=\frac{c_{1j}c_{2i}-c_{1i}c_{2j}}{c_{1i}c_{2i}c_{1j}c_{2j}}
\]
for $i,j\in \{3,...,n\}$. Furthermore, if we choose to remove rows/columns 1 from $H$ and $n$ from $\tilde{H}$, we see that  $\det' H$ reduces to $\left\langle 12\right\rangle$ and after
rescaling the $i$'th row and $j$'th column of $\tilde{H}$ by $c_{1i}c_{2i}$
and $c_{1j}c_{2j}$ respectively, this brings out a factor of $\prod_{r=3}^{n-1}c_{1r}^{2}c_{2r}^{2}$
from $\det'\tilde{H}$ and $\tilde{H}$ reduces to 
\[
\tilde{H}_{rr}=-\sum_{r'\neq r}\frac{\left[rr'\right]}{\left(rr'\right)}\frac{c_{1r'}c_{2r'}}{c_{1r}c_{2r}},\,\,\,\tilde{H}_{rr'}=\frac{\left[rr'\right]}{\left(rr'\right)},\,\,\, r\neq r',
\]
where $r,r'\in\{3,...,n-1\}$ and $(ij)$ now refers to the minor of columns $i$ and $j$ of
the $2\times n$ C-matrix in \eqref{Cch}.
Integrating out the worldsheet coordinates in \eqref{Mc} against the delta functions in \eqref{deltac} then leaves the
following integral over link variables:

\[
\mathcal{M}_{n,2}^{(0)}=\int\frac{d^{2\times(n-2)}C}{GL(2)}\frac{\left\langle 12\right\rangle }{(12)}\frac{\det\tilde{H}}{(12)^{2}(2n)^{2}(n1)^{2}}\delta^{2\times(2|8)}\left(C\cdot\tilde{\lambda}\right)\delta^{(n-2)\times2}\left(\lambda\cdot C^{\perp}\right).
\]
Note that on the support of the delta functions, we can replace $\frac{\left\langle 12\right\rangle }{(12)}$
with any $\frac{\left\langle pq\right\rangle }{(pq)}$. For a derivation of these identities relating spinor brackets to minors and a generalization to higher MHV degree, see Appendix \ref{identities}. Covariantizing the above formula, we finally obtain
\begin{equation}
\mathcal{M}_{n,2}^{(0)}=\int\frac{d^{2\times n}C}{GL(2)}\frac{\left\langle pq\right\rangle }{\left(pq\right)}\frac{\det\tilde{H}}{(ab)^{2}(bc)^{2}(ca)^{2}}\delta^{2\times(2|8)}\left(C\cdot \tilde{\lambda} \right)\delta^{(n-2)\times2}\left(\lambda\cdot C^{\perp}\right),
\label{n8mhvgras}
\end{equation}
where $a,b,c$ are any three distinct particles and
\[
\tilde{H}_{ii}=-\sum_{j=1,j\notin\{a,b,c\}}^{n}\frac{\left[ij\right]}{\left(ij\right)}\frac{(aj)(bj)}{(ai)(bi)},\,\,\,\tilde{H}_{ij}=\frac{\left[ij\right]}{\left(ij\right)},\,\,\, i\neq j
\]
where $i,j\in\{1,...,n\}-\{a,b,c\}$.

\begin{figure}
\centering
       \includegraphics[scale=0.8]{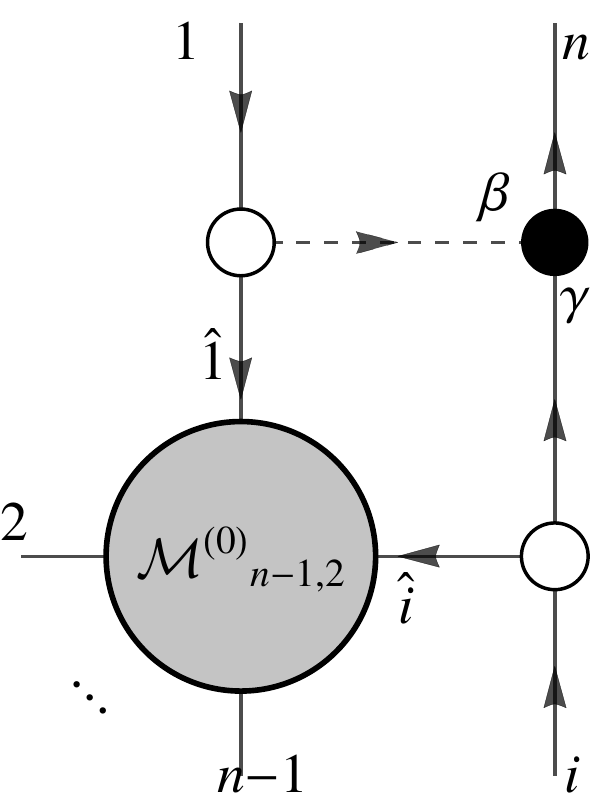}
    \caption{On-shell diagram contributing to tree-level $n$-point MHV tree amplitude in $\mathcal{N}=8$ SUGRA} 
    \label{n8f}
\end{figure}

This formula can also be obtained directly from on-shell diagrams as follows. In Appendix \ref{bonusappendix}, we explain how to incorporate the bonus relations of $\mathcal{N}=8$ SUGRA into on-shell diagram recursion for MHV amplitudes by modifying the bridge decoration. In particular, for the diagram in Figure~\ref{n8f}, the modified bridge decoration is given by

\[
B_{12n;i}=\frac{\left\langle i2\right\rangle }{\left\langle 1i\right\rangle \left\langle n2\right\rangle \left[1n\right]}
\]
Using this bridge decoration, the full amplitude is obtained by summing the diagram over $i\in\{3,...,n-1\}$. In terms of the edge variables, we then obtain
\begin{align*}
\mathcal{M}_{n,2}^{(0)}=\sum_{i=3}^{n-1}\int\frac{d\beta}{\beta^{2}}\frac{d\gamma}{\gamma^{2}}[\beta\hat{1}][\gamma\hat{i}]\left\langle \gamma\beta\right\rangle B_{12n;i}& \\
\times\int\frac{d^{2\times (n{-}1)}C}{GL(2)}&\delta^{2\times(2|8)}\left(C\cdot \tilde{\lambda}\right)\delta^{(n-2)\times 2}\left(\lambda\cdot C^{\perp}\right)\mathcal{I}_{n-1}\left(\hat{1},2,...,\hat{i},...n-1\right)
\end{align*}
where $\mathcal{I}$ is the integrand of the $(n-1)$-point amplitude, without the delta functions. Noting that 
\[
[\beta\hat{1}][\gamma\hat{i}]\left\langle \gamma\beta\right\rangle =\beta\gamma\left[n1\right]\left[ni\right]\left\langle i1\right\rangle 
\]
the equation above reduces to
\[
\mathcal{M}_{n,2}^{(0)}=\sum_{i=3}^{n-1}\frac{\left[ni\right]\left\langle i2\right\rangle }{\left\langle n2\right\rangle }\int\frac{d\beta}{\beta}\frac{d\gamma}{\gamma}\int\frac{d^{2\times (n{-}1)}C}{GL(2)}\delta^{2\times(2|8)}\left(C\cdot \tilde{\lambda}\right)\delta^{(n-2)\times 2}\left(\lambda\cdot C^{\perp}\right)\mathcal{I}_{n-1}\left(\hat{1},2,...,\hat{i},...n-1\right).
\]
For the diagram in Figure~\ref{n8f}, the $C$-matrix is given by
\[
C=\left(\begin{array}{ccccc}
1 & ... & 0 & ... & \beta\\
0 & ... & 1 & ... & \gamma
\end{array}\right)
\]
where the rows correspond to legs $(1,i)$ and the indicated columns correspond to legs $(1,i,n)$. For this $C$-matrix, we see that
$(ni)=\beta$, $(1n)=\gamma$, and $(1i)=1$, so the amplitude can be written covariantly as
\[
\mathcal{M}_{n,2}^{(0)}=\sum_{i=3}^{n-1}\frac{\left[ni\right]\left\langle i2\right\rangle }{\left\langle n2\right\rangle }\int\frac{d^{2\times n}C}{GL(2)}\frac{(1i)}{(in)(1n)}\delta^{2\times(2|8)}\left(C\cdot \tilde{\lambda}\right)\delta^{(n-2)\times 2}\left(\lambda\cdot C^{\perp}\right)\mathcal{I}_{n-1}\left(\hat{1},2,...,\hat{i},...n-1\right).
\]
If we use the $GL(2)$ symmetry to choose $C$ according to \eqref{C}, we obtain the following recursion relation for MHV amplitudes:
\[
\mathcal{M}_{n,2}^{(0)}=\sum_{i=3}^{n-1}\frac{\left[in\right]}{\left\langle in\right\rangle }\frac{\left\langle 1i\right\rangle \left\langle 2i\right\rangle }{\left\langle 1n\right\rangle \left\langle 2n\right\rangle }\mathcal{M}_{(n{-}1),2}^{(0)}\left(\hat{1},2,...,\hat{i},...n-1\right).
\]
This is precisely the recursion relation obtained by Hodges in \cite{Hodges:2011wm}. Moreover, he obtained the following solution in \cite{Hodges:2012ym}:
\[
\mathcal{M}_{n,2}^{(0)}=\frac{\delta^{4|16}(P){\rm det}\tilde{H}}{\left\langle 12\right\rangle ^{2}\left\langle 2n\right\rangle ^{2}\left\langle n1\right\rangle ^{2}}
\]
where
\[
\tilde{H}_{ii}=-\sum_{j=3}^{n-1}\frac{\left[ij\right]}{\left\langle ij\right\rangle }\frac{\left\langle 1j\right\rangle \left\langle 2j\right\rangle }{\left\langle 1i\right\rangle \left\langle 2i\right\rangle },\,\,\,\tilde{H}_{ij}=\frac{\left[ij\right]}{\left(ij\right)},\,\,\, i\neq j
\]
for $i,j\in\{3,...,n-1\}$. Once again, we find that equation \eqref{n8mhvgras} is the unique Grassmannian uplift, which can be seen by using the $GL(2)$ symmetry to choose $C$ as in \eqref{C} and choosing $\{a,b,c\}=\{1,2,n\}$.

In Appendix \ref{bonusappendix}, we use the bonus relations to solve the planar on-shell diagram recursion relations for MHV amplitudes and obtain the BGK formula \cite{Berends:1988zp} in a slightly simplified form. Our calculation shows that the BGK formula arises naturally from a planar object. The full MHV amplitude can then be obtained by summing this expression over permutations of $(n-3)$ legs, which we verify numerically.  Although the physical interpretation of this planar object is not clear, it would be interesting to see if it has a geometric interpretation as the volume of some object.    

\section{Tree-level NMHV} \label{nmhv}

In this section we will generalize our calculations to non-MHV amplitudes and find an additional subtlety. Whereas Grassmannian integrals for MHV amplitudes are completely localized by the bosonic delta functions in $C\cdot\lambda$ and $C^\perp\cdot\lambdat$, for non-MHV amplitudes there will be more integrals than delta functions so one must specify a contour in order to make the integrals well-defined. In particular, for an N$^{k-2}$MHV amplitude there will be $k(n-k)$ integrations and $2n-4$ bosonic delta functions (after subtracting four that impose momentum conservation), so the dimension of the contour will be $(k-2)(n-k-2)$. 

The precise form of the Grassmannian contour integral will depend on the method one uses to compute the amplitudes. For $\mathcal{N}=4$ SYM, the contour integral implied by BCFW will reduce to summing over residues of a single top form in the Grassmannian, each of which corresponds to an on-shell diagram, and can be related to the contour integral arising from ambitwistor string theory using global residue theorems. On the other hand, for $\mathcal{N}=8$ SUGRA we will show that the decorated planar on-shell diagrams (from which the full amplitude can be deduced by summing over permutations of external legs) do not correspond to residues of a single top form so the Grassmannian contour integral has a slightly more complicated form. It is also possible to derive such a formula using ambitwistor string theory although it is unclear how to map it into the contour integral arising from on-shell diagrams using global residue theorems.

To make the discussion as simple as possible we will focus on the example of the 6-point NMHV amplitude (which is the simplest example of a non-MHV amplitude since the contour in the Grassmannian is one-dimensional) and first review how to obtain its Grassmannian integral formula in $\mathcal{N}=4$ SYM, which was previously derived using various approaches in \cite{Spradlin:2009qr,Dolan:2009wf,Nandan:2009cc,Bullimore:2009cb,ArkaniHamed:2009dg}. We will then generalize the analysis to $\mathcal{N}=8$ SUGRA.     

\subsection{$\mathcal{N}=4$} \label{6pt4}

\begin{figure}[h]
\centering
\begin{tabular}{ccc}
\includegraphics[scale = 0.5]{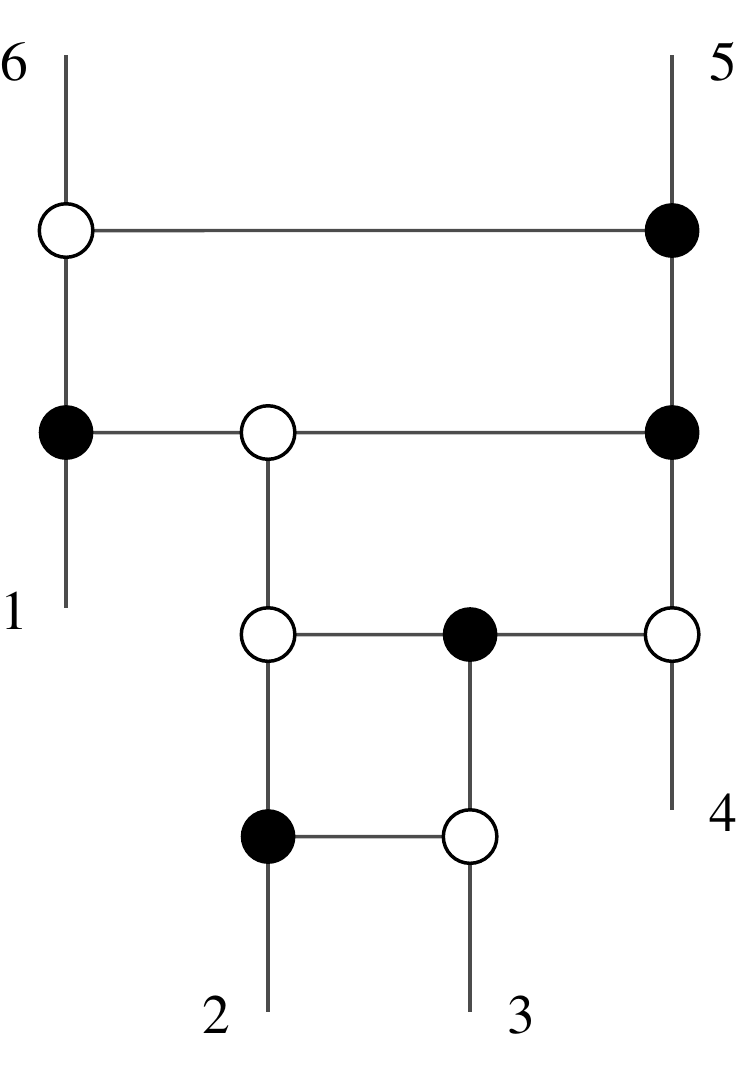}  &\hspace{1cm}
\includegraphics[scale = 0.5]{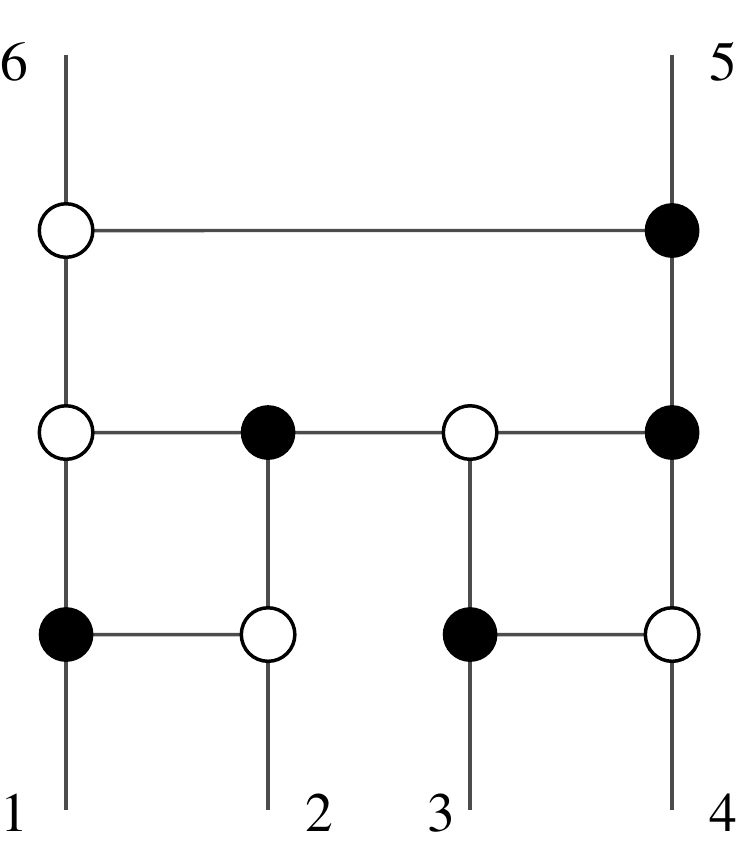}&\hspace{1cm}\includegraphics[scale = 0.5]{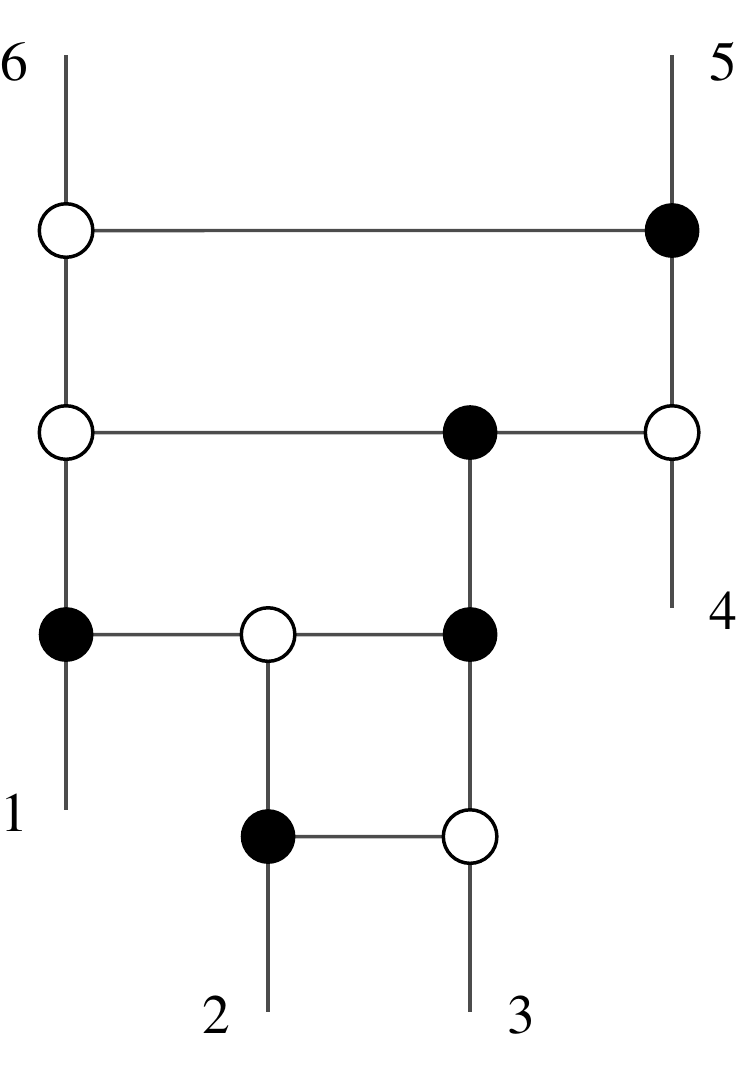}\\
$3{+}5$  & \hspace{1cm}$4{+}4$ & \hspace{1cm}$5{+}3$
\end{tabular}
\caption{On-shell diagrams contributing to the 6-point NMHV amplitude in $\mathcal{N} = 4$ SYM}
\label{fig:N=46ptnmhvbcfw}
\end{figure}

In this section we will derive the 6-point NMHV amplitude in $\mathcal{N}=4$ SYM in the form of a contour integral over the Grassmannian $Gr(3,6)$ using on-shell diagrams and then derive an alternative formula using ambitwistor string theory. We will then demonstrate how the two contour integrals can be mapped into each other using global residue theorems. 

Using the recursion relation defined in Figure \ref{n4recursion}, one finds that there are three on-shell diagrams contributing to the 6-point NMHV amplitude, which are shown in Figure~\ref{fig:N=46ptnmhvbcfw}. The first one corresponds to combining a three point \MHVB diagram with a five point MHV diagram which will be referred to as the $3{+}5$ channel diagram. Secondly we can paste together two four point diagrams, this channel will be called the $4{+}4$ channel. Finally we can paste together a five point \MHVB with a three point MHV diagram, and this will be referred to as the $5{+}3$ channel. 

\begin{figure}[h]
\centering
\includegraphics[scale = 0.65]{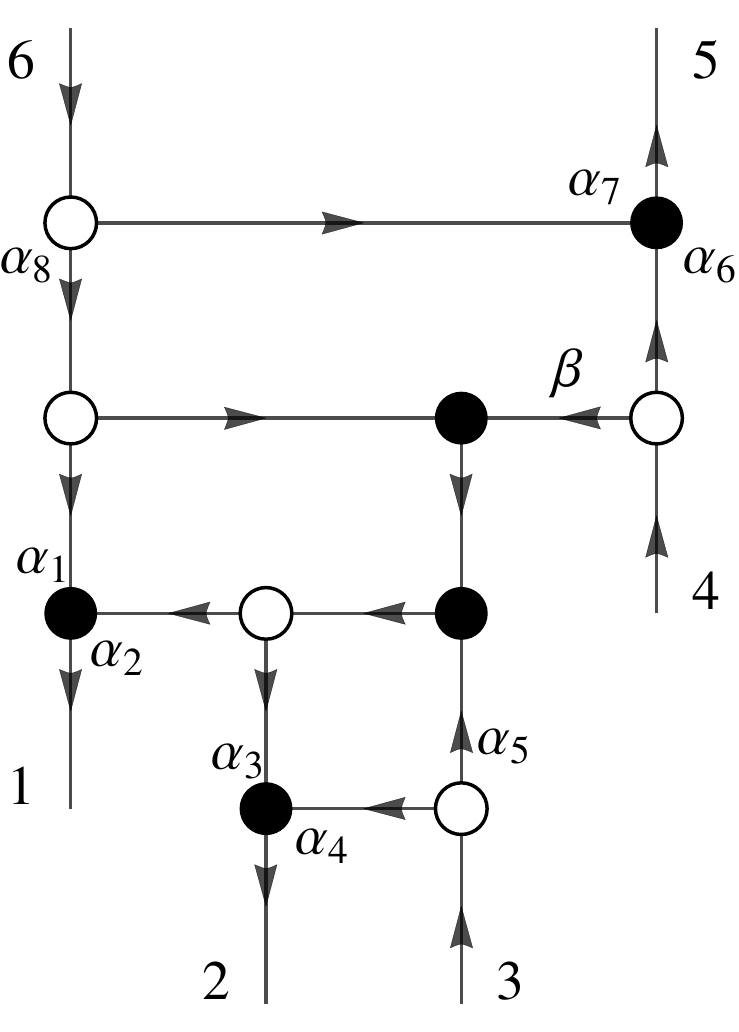} 
\caption{The $5{+}3$ channel BCFW diagram contributing to the 6 point NMHV in $\mathcal{N} = 4$ SYM. Edge variables are denoted as $\alpha_i$. The label $\beta$ is not an edge variable.}
\label{fig:N=46ptnmhv5+3labelled}
\end{figure}

On-shell diagrams in $\mathcal{N}=4$ SYM can be evaluated in terms of edge variables using the algorithm defined in section \ref{osdrev}. In particular, assigning arrows and variables to the edges of the $5{+}3$ diagram as shown in figure~\ref{fig:N=46ptnmhv5+3labelled}, one obtains the following formula for the $C$-matrix by summing over paths between external legs:
\begin{equation}
C_{5{+}3}=\left(
\begin{array}{cccccc}
 \alpha_2 \alpha_5 & \alpha_3 \alpha_5+\alpha_4 & 1 & 0 & 0 & 0 \\
 \alpha_2 & \alpha_3 & 0 & 1 & \alpha_6 & 0 \\
 \alpha_8 (\alpha_1+\alpha_2) & \alpha_3 \alpha_8 & 0 & 0 & \alpha_7 & 1 \\
\end{array}
\right),
\end{equation}
where the rows correspond to legs $3,4,6$ which have incoming arrows. This matrix has the minor $(456) = 0$, which will ultimately imply a contour in the Grassmannian when writing down a covariant formula for the $5{+}3$ diagram. In order to derive such a formula, first consider the following deformation of the $C$-matrix:
\begin{equation}
\tilde{C}_{5{+}3}=\left(
\begin{array}{cccccc}
 \alpha_2 \alpha_5 & \alpha_3 \alpha_5+\alpha_4 & 1 & 0 & \alpha & 0 \\
 \alpha_2 & \alpha_3 & 0 & 1 & \alpha_6 & 0 \\
 \alpha_8 (\alpha_1+\alpha_2) & \alpha_3 \alpha_8 & 0 & 0 & \alpha_7 & 1 \\
\end{array}
\right).
\label{c3p5}
\end{equation}
The deformed matrix now has $(456) = \alpha$ and depends on nine parameters so it can be used to define an integral over $Gr(3,6)$. Moreover, using the algorithm in \ref{osdrev} one finds that the $5{+}3$ diagram is given in terms of edge variables by
\begin{equation}
\mathcal{A}^{(0)}_{6,3\;(5{+}3)} = \Res{\alpha=0} \int \frac{d\alpha}{\alpha} \prod_{i = 1}^8 \frac{d\alpha_i}{\alpha_i} \delta^{3\times (2|4)}(\tilde{C} \cdot\lambdat
)\delta^{3\times2}(\lambda\cdot\tilde{C}^{\perp})
\label{35a}
\end{equation}
Noting that 
\[
d^9\tilde{C} = \alpha_2 \alpha_3\alpha_8 d\alpha \prod_{i = 1}^8 d\alpha_i
\]
and
\[
(123)(234)(345)(456)(561)(612) = \alpha \alpha_1 \alpha_2 \alpha_3^2 \alpha_4 (\alpha_2 \alpha_5\alpha_6 - \alpha \alpha_2) \alpha_7 \alpha_8^2
\]
one finds that~(\ref{35a}) can be uplifted to following covariant formula, with $d^{3\times6}\Omega_4$ defined in~(\ref{meas}):
\[
\mathcal{A}^{(0)}_{6,3\;(5{+}3)}=\Res{(456)=0}\int d^{3\times6}\Omega_4.
\]
In summary, we find that the $5{+}3$ diagram arises from a residue of the canonical volume form of $Gr(3,6)$. From this, we can immediately calculate the $3{+}5$ diagram by complex conjugating and permuting the external legs. Under this mapping, we send $[ij] \leftrightarrow \langle ij\rangle$, and $(ijk) \rightarrow \epsilon_{ijkabc}(abc)$ and apply the permutation $P = \left( \begin{smallmatrix} 1&2&3&4&5&6\\4&3&2&1&6&5 \end{smallmatrix} \right)$ to obtain
\begin{equation}
\mathcal{A}^{(0)}_{6,3\;(3{+}5)} = \Res{(234)=0} \int d^{3\times6}\Omega_4
\end{equation}

\begin{figure}[h]
\centering
\includegraphics[scale = 0.65]{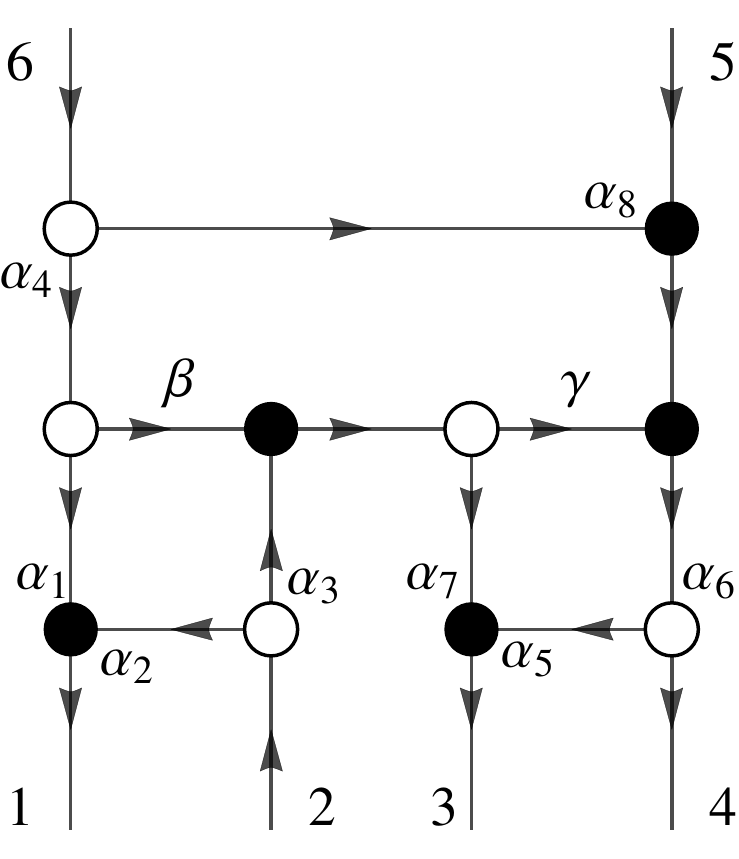} 
\caption{The $4{+}4$ channel BCFW diagram contributing to the 6 point NMHV amplitude in $\mathcal{N} = 4$ SYM. Edge variables are denoted as $\alpha_i$. The labels $\beta$ and $\gamma$ are not edge variables.}
\label{fig:N=46ptnmhv4+4labelled}
\end{figure}

Finally consider the $4{+}4$ channel diagram, which is oriented and labelled as in figure~\ref{fig:N=46ptnmhv4+4labelled}. 
In this case, the $(612)$ minor of the $C$-matrix vanishes, so we consider the following deformed matrix
\begin{equation}
\label{eqn:4+4Cmatrix}
\tilde{C} = \left(
\begin{array}{cccccc}
 \alpha_2 & 1 & \alpha_3 (\alpha_5
   \alpha_6+\alpha_7) & \alpha_3 \alpha_6 & 0 & 0 \\
 \alpha & 0 & \alpha_5 \alpha_6 & \alpha_6 & 1 & 0 \\
 \alpha_1 \alpha_4 & 0 & \alpha_4 (\alpha_5
   \alpha_6+\alpha_7)+\alpha_5 \alpha_6 \alpha_8 &
   \alpha_6 (\alpha_4+\alpha_8) & 0 & 1 \\
\end{array}
\right)
\end{equation}
which has been constructed to have the minor $(612) = \alpha$. In terms of edge variables, the diagram can be written 
\[
\mathcal{A}^{(0)}_{6,3\;(4{+}4)}= \Res{\alpha=0} \int \frac{d\alpha}{\alpha} \prod_{i = 1}^8 \frac{d\alpha_i}{\alpha_i} \delta^{3\times (2|4)}(\tilde{C} \cdot\lambdat
)\delta^{3\times2}(\lambda\cdot\tilde{C}^{\perp}).
\]
Noting that
\[
d^{3\times3} \tilde{C}_{4{+}4}= \alpha_3 \alpha_4\alpha_6^3\alpha_7 d\alpha \prod_{i = 1}^8 d\alpha_i
\]
and
\begin{align}
(123)&(234)(345)(456)(561)(612) = \nonumber\\ &\alpha \alpha_2 \alpha_3^2 \alpha_4 \alpha_6^3 \alpha_7^2 \alpha_8 (-\alpha_1 \alpha_4 \alpha_5 \alpha_6 + 
   \alpha (\alpha_4 (\alpha_5 \alpha_6 + \alpha_7) + \alpha_5 \alpha_6 \alpha_8))
\end{align}
we find that the $4+4$ diagram uplifts to the following covariant expression:
\begin{equation}
\mathcal{A}^{(0)}_{6,3\;(4{+}4)} = \Res{(612)=0} \int d^{3\times6}\Omega_4.
\label{44diag}
\end{equation}
Note that the $4{+}4$ must be self-conjugate under complex conjugation, and we have exactly that $(612)$ remains invariant under this transformation, paired with the permutation $P$ defined in the $5{+}3$ calculation. 

\begin{figure}
\centering
       \includegraphics[scale=0.6]{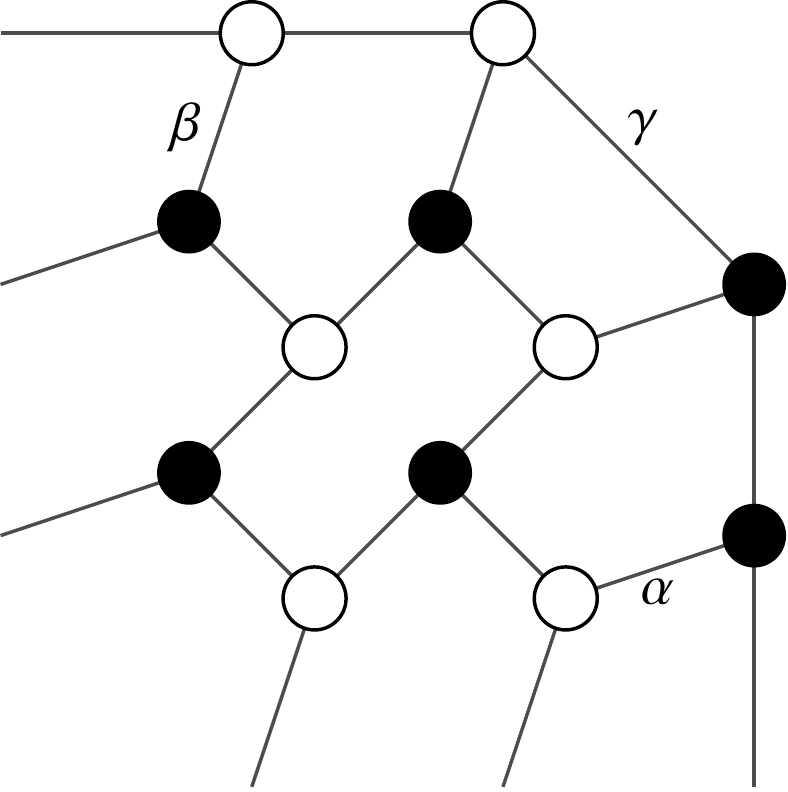}
    \caption{Postnikov diagram for the 6 point NMHV amplitude in $\mathcal{N}=4$ SYM} 
    \label{postnikov33}
\end{figure}

\begin{figure}
\centering
\begin{tabular}{m{6cm} m{2cm} m{6cm}}
\includegraphics[scale=0.5]{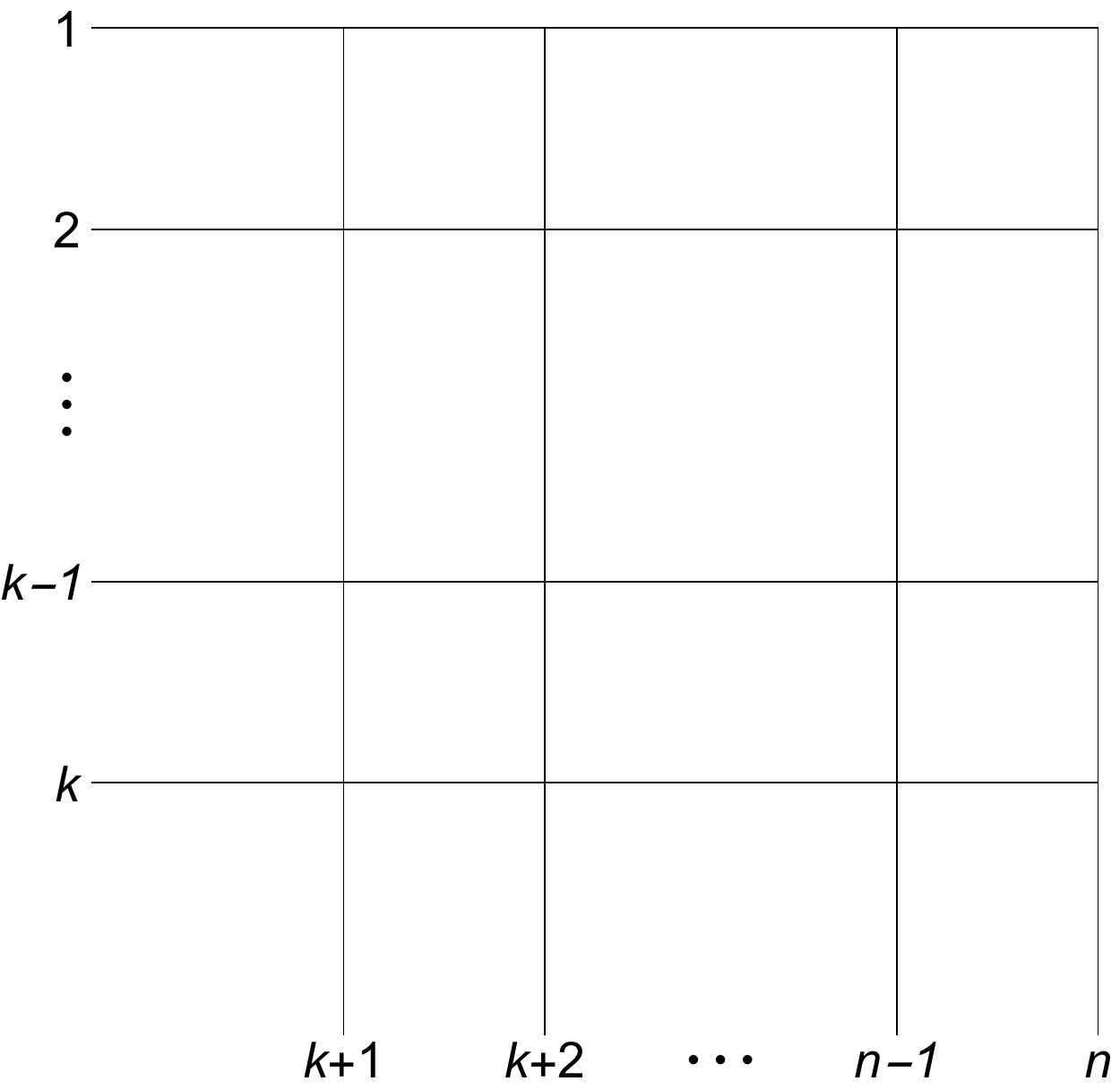}
&
&

\begin{tabular}{m{2cm} m{1cm} m{2cm}}
\includegraphics[scale=0.5]{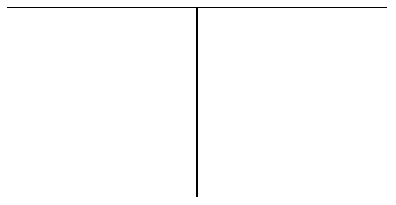} & $\rightarrow$ & \includegraphics[scale=0.5]{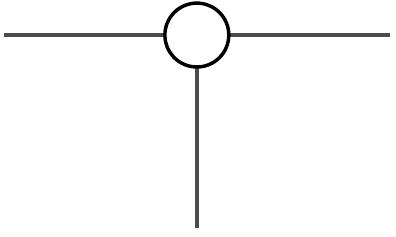} \\
\includegraphics[scale=0.5]{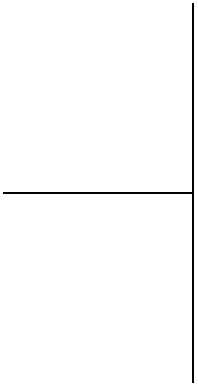} & $\rightarrow$ & \includegraphics[scale=0.5]{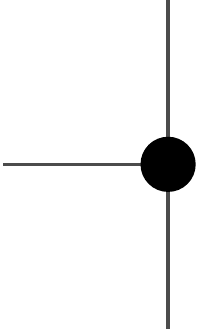}\\
\includegraphics[scale=0.5]{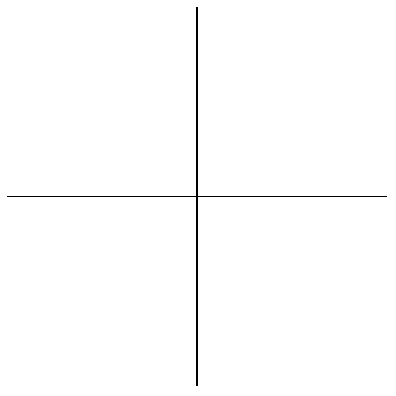} & $\rightarrow$ & \includegraphics[scale=0.5]{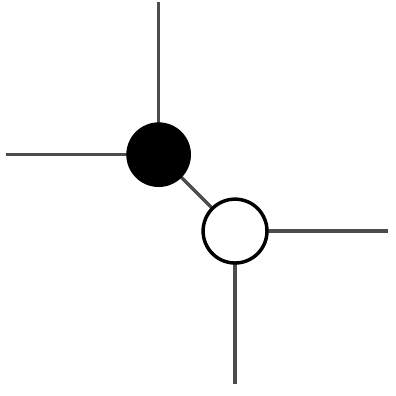}\\
\end{tabular}

\end{tabular}
    \caption{Postnikov diagram for $n$-point N$^{k-2}$MHV amplitude in $\mathcal{N}=4$ SYM} 
    \label{postnikov}
\end{figure}

Hence, we find that the full amplitude can be written as a sum of three residues of a single top form
\begin{equation}
\mathcal{A}^{(0)}_{6,3}= \left(\Res{(234)=0} + \Res{(456)=0}+\Res{(612)=0} \right)\int d^{3\times6}\Omega_4
\label{top}
\end{equation}
This can be written as a contour integral if one defines the contour to encircle the three poles in $(234)$, $(456)$, and $(612)$. The existence of such a formula relies on the fact that the three on-shell diagrams in Figure \ref{fig:N=46ptnmhvbcfw} can be embedded into a single diagram depicted in Figure \ref{postnikov33}, which we refer to as a Postnikov diagram \cite{Postnikov:2006kva}. In particular, the $3{+}5$, $5{+}3$ and $4{+}4$ diagrams in Figure~\ref{fig:N=46ptnmhvbcfw} correspond to residues with respect the edge variables $\alpha$, $\beta$ and $\gamma$ respectively, using the square moves and mergers. More generally, the Postnikov diagram for an $n$-point N$^{k-2}$MHV amplitude in $\mathcal{N}=4$ SYM can be constructed as in Figure \ref{postnikov} \cite{Ferro:2013dga}. In contrast, we will find that the decorated planar on-shell diagrams from which the non-MHV amplitudes of $\mathcal{N}=8$ supergravity can be derived cannot be embedded in a single decorated Postnikov diagram.

We will now derive a Grassmannian contour integral formula for the 6-point NMHV amplitude of $\mathcal{N}=4$ SYM using the 4d ambitwistor string formula
\[
\mathcal{A}_{6,3}^{(0)}=\int\frac{1}{GL(2)}\prod_{i=1}^{6}\frac{d^{2}\sigma_{i}}{(i\,i{+}1)}\prod_{l}\delta^{2|4}\left(\tilde{\lambda}_{l}-\sum_{r}\frac{\tilde{\lambda}_{r}}{(lr)}\right)\prod_{r}\delta^{2}\left(\lambda_{r}-\sum_{l}\frac{\lambda_{l}}{(rl)}\right)
\]
where $l\in\{1,3,5\}$ and $r\in\{2,4,6\}$. First insert $1$ in the form
of an integral over link variables 
\[
1=\int\prod_{l,r}dc_{lr}\delta\left(c_{lr}-\frac{1}{(lr)}\right)
\]
to obtain

\[
\mathcal{A}_{6,3}^{(0)}=\int\frac{1}{GL(2)}\prod_{i=1}^{6}\frac{d^{2}\sigma_{i}}{(i\,i{+}1)}\prod_{l,r}dc_{lr}\delta\left(c_{lr}-\frac{1}{(lr)}\right)\prod_{l}\delta^{2|4}\left(\tilde{\lambda}_{l}-c_{lr}\tilde{\lambda}_{r}\right)\prod_{r}\delta^{2}\left(\lambda_{r}+c_{lr}\lambda_{l}\right).
\]
Next use the $GL(2)$ symmetry to fix $\sigma_{1}=(1,0)$ and $\sigma_{3}=(0,1)$.
After doing so, the eight remaining worldsheet coordinates are fixed
by eight of the delta functions in the link variables. In particular, we can write
\[
\prod_{r}\delta\left(c_{1r}-\frac{1}{(1r)}\right)\delta\left(c_{3r}-\frac{1}{(3r)}\right)=\prod_{r}\frac{1}{c_{1r}^{2}c_{3r}^{2}}\delta\left(\sigma_{r}^{2}-\frac{1}{c_{1r}}\right)\delta\left(\sigma_{r}^{1}+\frac{1}{c_{3r}}\right)
\]
and
\[
\delta\left(c_{52}-\frac{1}{(52)}\right)\delta\left(c_{54}-\frac{1}{(54)}\right)=\frac{c_{12}c_{34}c_{32}c_{14}}{c_{52}^{2}c_{54}^{2}\left(c_{32}c_{14}-c_{12}c_{34}\right)}\delta^{2}\left(\sigma_{5}-\sigma_{5}^{*}\right)
\]
where
\begin{equation}
\sigma_{5}^{*}=\frac{1}{c_{52}c_{54}\left(c_{32}c_{14}-c_{12}c_{34}\right)}\left(\begin{array}{c}
c_{12}c_{14}\left(c_{32}c_{54}-c_{34}c_{52}\right)\\
c_{32}c_{34}\left(c_{12}c_{54}-c_{14}c_{52}\right)
\end{array}\right).
\label{sigma5}
\end{equation}
Note that there is one remaining delta function in the link variables which will not be integrated out and provides a constraint on the $c_{lr}$

\[
\delta\left(c_{56}-\frac{1}{(56)}\right)=\frac{c_{52}c_{54}c_{16}c_{36}\left(c_{32}c_{14}-c_{12}c_{34}\right)}{c_{56}}\delta(S)
\]
where

\[
S=c_{52}c_{36}\left(c_{54}c_{16}-c_{56}c_{14}\right)\left(c_{12}c_{34}-c_{14}c_{32}\right)-c_{32}c_{56}\left(c_{14}c_{36}-c_{16}c_{34}\right)\left(c_{52}c_{14}-c_{54}c_{12}\right).
\]
Putting everything together then gives
\begin{equation}
\mathcal{A}_{6,3}^{(0)}=\int d^{3\times3}C\frac{(135)\delta(S)}{(123)(345)(561)}\delta^{3(2|4)}\left(C\cdot\tilde{\lambda}\right)\delta^{2\times3}\left(\lambda\cdot C^{\perp}\right)\label{eq:lin}
\end{equation}
where

\[
C=\left(\begin{array}{cccccc}
1 & c_{12} & 0 & c_{14} & 0 & c_{16}\\
0 & c_{32} & 1 & c_{34} & 0 & c_{36}\\
0 & c_{52} & 0 & c_{54} & 1 & c_{56}
\end{array}\right)
\]
and

\begin{equation}
S=(123)(561)(346)(245)-(125)(136)(456)(234).\label{eq:S}
\end{equation}

Covariantizing \eqref{eq:lin} gives a contour integral in
the Grassmannian, where one takes $\delta(S)\rightarrow1/S$
and defines the contour to encircle this pole:

\begin{equation}
\mathcal{A}_{6,3}^{(0)}=\Res{S=0}\int\frac{d^{3\times6}C}{GL(3)}\frac{1}{S}\frac{(135)}{(123)(345)(561)}\delta^{3\times(2|4)}\left(C\cdot\tilde{\lambda}\right)\delta^{3\times2}\left(\lambda\cdot C^{\perp}\right).\label{eq:twistorgrass}
\end{equation}
We can now apply a global residue theorem to wrap the contour around the other poles of the integrand to obtain

\begin{equation}
\mathcal{A}_{6,3}^{(0)}=\left(\Res{(123)=0}+\Res{(345)=0}+\Res{(561)=0}\right)\int\frac{d^{3\times6}C}{GL(3)}\frac{1}{S}\frac{(135)}{(123)(345)(561)}\delta^{3\times(2|4)}\left(C\cdot\tilde{\lambda}\right)\delta^{3\times2}\left(\lambda\cdot C^{\perp}\right).\label{eq:bcf}
\end{equation}
Using Pl\"{u}cker identities, we can write $S$ in equation (\ref{eq:S})
as 

\[
S=(135)(234)(456)(612)-(246)(123)(345)(561).
\]
Noting that the second term in $S$ can be discarded on support of each of the
residues in \eqref{eq:bcf}, we see that \eqref{eq:bcf} is equivalent to \eqref{top}, which was deduced from on-shell diagrams.

In summary, we have obtained two Grassmannian contour integral formulae for
the 6-point NMHV amplitude in $\mathcal{N}=4$ SYM using on-shell diagrams and 4d ambitwistor
string theory, given by equations \eqref{top} and \eqref{eq:twistorgrass} respectively. Remarkably, these two contour integrals are
related by a global residue theorem.

\subsection{$\mathcal{N}=8$}

\begin{figure}[h]
\centering
\begin{tabular}{ccc}
\includegraphics[scale = 0.5]{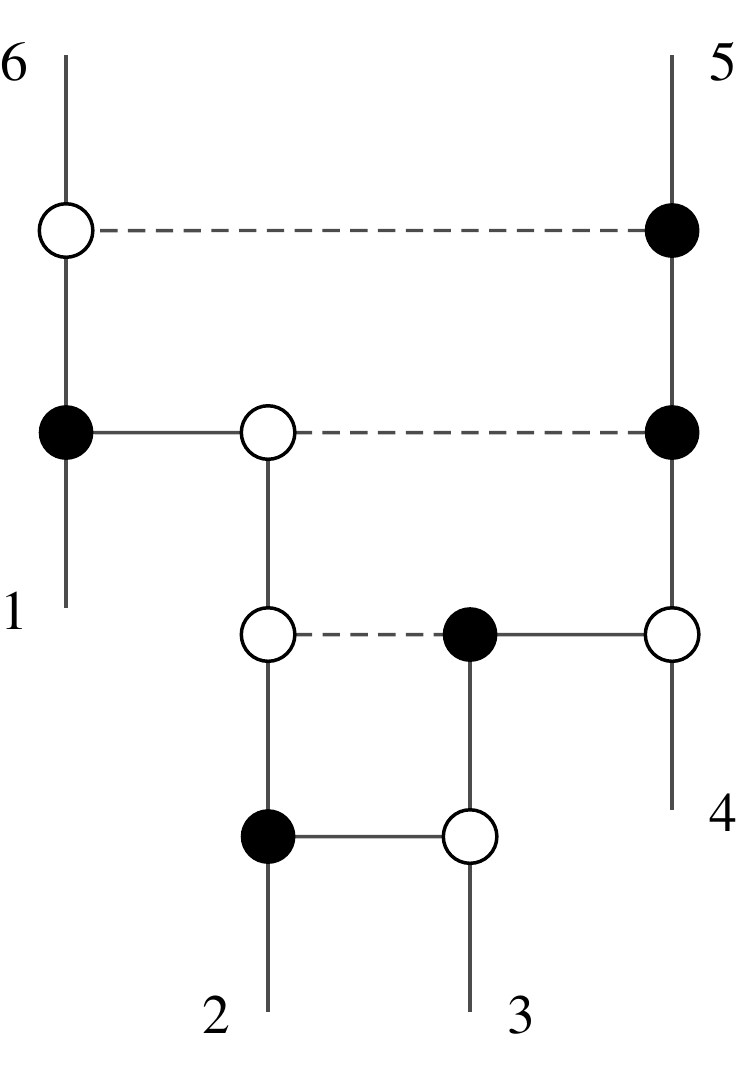}  &\hspace{1cm}
\includegraphics[scale = 0.5]{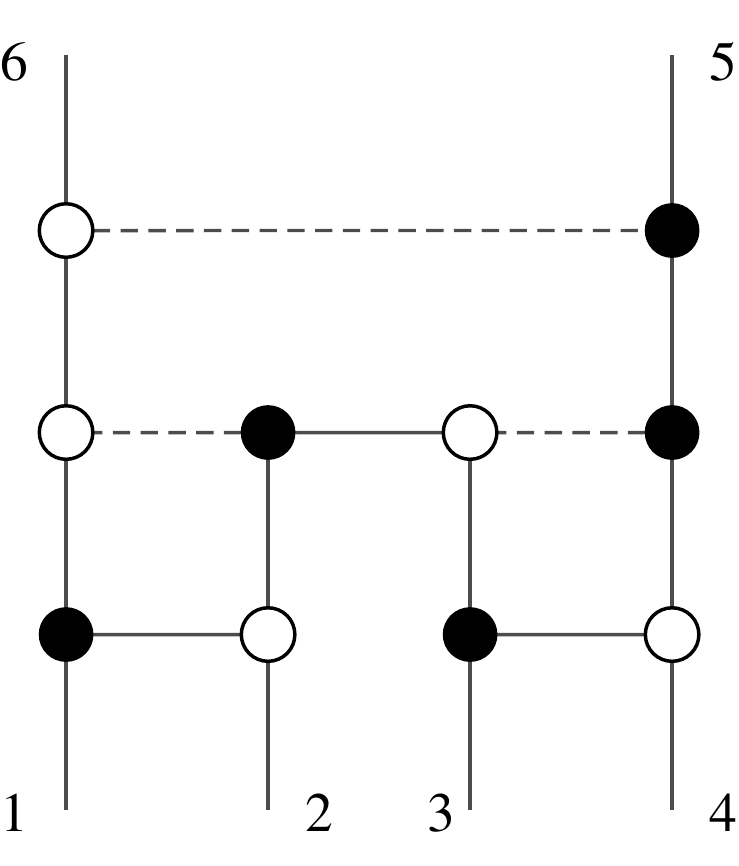}&\hspace{1cm}\includegraphics[scale = 0.5]{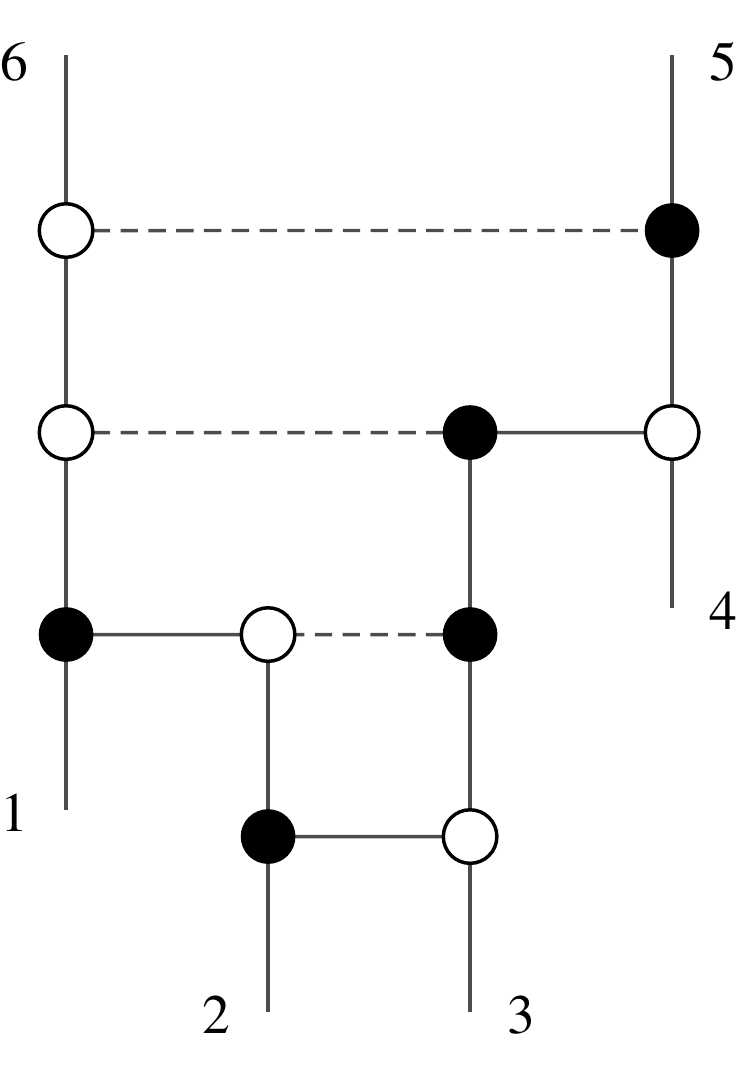}\\
$3{+}5$  & \hspace{1cm}$4{+}4$ & \hspace{1cm}$5{+}3$
\end{tabular}
\caption{Decorated planar on-shell diagrams contributing to the 6 point NMHV amplitude in $\mathcal{N} = 8$ SUGRA. The full amplitude can be obtained by summing over permutations of legs $1$ to $4$.}
\label{fig:N=86ptnmhvbcfw}
\end{figure}

Now that we have understood the details of the 6-point NMHV amplitude of $\mathcal{N} = 4$ SYM, we extend the calculation to  $\mathcal{N} = 8$ SUGRA. Since the on-shell diagram recursion can be restricted to a planar sector, the calculation can be reduced to computing three planar diagrams which are essentially decorated versions of the ones appearing in $\mathcal{N}=4$ SYM, as shown in Figure~\ref{fig:N=86ptnmhvbcfw}. The full 6-point NMHV amplitude can then be obtained by summing over permutations of legs $1$ to $4$. We will use the same orientation and labelling as the $\mathcal{N} = 4$ SYM diagrams, so the $C$ matrices will remain the same. 
 
Let us first compute the $3+5$ diagram in Figure \ref{fig:N=86ptnmhvbcfw}. Using the orientation and labelling from figure~\ref{fig:N=46ptnmhv5+3labelled} and following the algorithm in Appendix \ref{algorithm}, we obtain 
\begin{equation}
\mathcal{A}^{(0)}_{6,3\;(3{+}5)} = \Res{(456)=0} \int d^{3\times6}\Omega_8 \frac{\alpha_7}{\alpha_6\alpha_8} \frac{\langle\alpha_1\alpha_2\rangle \langle\alpha_3\alpha_4\rangle [\alpha_4\alpha_5] [\beta\alpha_6] }{\prod_{i = 1}^8\alpha_i}
\label{a1}
\end{equation}
where $d^{3\times6}\Omega_8$ is defined in~(\ref{meas}). We can then relate the internal spinors to external spinors by summing over paths connecting them using them as described in the algorithm, finding 
\begin{align*}
 \langle\alpha_1\alpha_2\rangle &= \frac{\alpha_8}{\alpha_2}\langle 16\rangle \\
  \langle\alpha_3\alpha_4\rangle &= \frac{1}{\alpha_3}\langle 23\rangle \\
  [\alpha_4\alpha_5] &= \frac{\alpha_4}{\alpha_5} [23] \\
    [\beta\alpha_6] &= \alpha_6 [45].
\end{align*}
Substituting these relations into \eqref{a1} and simplifying gives
\begin{equation}
\mathcal{A}^{(0)}_{6,3\;(3{+}5)} = \Res{(456)=0} \int d^{3\times6}\Omega_8 \frac{\langle 16\rangle [45] \langle 23\rangle[32]}{\alpha_1 \alpha_2^2\alpha_3^2\alpha_5^2\alpha_6\alpha_8}
\end{equation}
which can be uplifted to the following covariant expression:
\begin{equation}
\mathcal{A}^{(0)}_{6,3\;(3{+}5)} = \Res{(456)=0} \int d^{3\times6}\Omega_8 \frac{\langle 16\rangle [45] \langle 23\rangle[32]}{(123)(561)(146)(236)}
\label{a35}
\end{equation}
where the minors are computed using \eqref{c3p5}. 

The $5{+}3$ channel in Figure \ref{fig:N=86ptnmhvbcfw} can be calculated directly as the complex conjugate of the $3{+}5$ channel. Under the mapping, we send $[ij] \leftrightarrow \langle ij\rangle$, and $(ijk) \rightarrow \epsilon_{ijkabc}(abc)$. To keep the cyclic definition of the legs consistent , we then apply the permutation $P = \left( \begin{smallmatrix} 1&2&3&4&5&6\\4&3&2&1&6&5 \end{smallmatrix} \right)$, and we find the non-trivial result that the $3{+}5$ and $5{+}3$ channels both have the same integrand, only with different residues:
\begin{equation}
\mathcal{A}^{(0)}_{6,3\;(3{+}5)} = \Res{(234)=0} \int d^{3\times6}\Omega_8 \frac{\langle 16\rangle [45] \langle 23\rangle[32]}{(123)(561)(146)(236)}
\label{a53}
\end{equation}

Finally, we compute the $4{+}4$ channel diagram in Figure \ref{fig:N=86ptnmhvbcfw} using the orientation and labelling in figure~\ref{fig:N=46ptnmhv4+4labelled}. Using the algorithm in Appendix \ref{algorithm}, we obtain the following expression for the amplitude:
\begin{equation}
\mathcal{A}^{(0)}_{6,3\;(4{+}4)}= \Res{(612)=0} \int d^{3\times6}\Omega_8 \frac{\alpha_8}{\alpha_4} \frac{\langle\alpha_1\alpha_2\rangle \langle\alpha_5\alpha_7\rangle [\alpha_2\alpha_3] [\alpha_5\alpha_6] }{\alpha_6\prod_{i = 1}^8\alpha_i}
\label{a2}
\end{equation}
Writing the internal spinor brackets in terms of external ones then gives
\begin{align*}
 \langle\alpha_1\alpha_2\rangle &= \frac{\alpha_4}{\alpha_2}\langle 16\rangle \\
  \langle\alpha_5\alpha_7\rangle &= \frac{1}{\alpha_5\alpha_6\alpha_7}\left(\alpha_3\alpha_6\alpha_7\langle 32\rangle +\alpha_4\alpha_6\alpha_7\langle 36\rangle \right)\\
  [\alpha_2\alpha_3] &= \frac{\alpha_2}{\alpha_3} [12] \\
    [4\alpha_5] &= \alpha_5 [43].
\end{align*}
Plugging this into \ref{a2} and covariantizing then gives
\begin{equation}
\mathcal{A}^{(0)}_{6,3\;(4{+}4)} = \Res{(456)=0} \int d^{3\times6}\Omega_8 \frac{\langle 16\rangle [34](623) [12] \left((346)\langle 32\rangle+(432)\langle 36\rangle\right)}{(123)(561)(346)^2(256)}
\end{equation}
where the minors are computed using \eqref{eqn:4+4Cmatrix}. This expression for the integrand can then be simplified further, using the relations between spinor brackets and minors derived in Appendix~\ref{identities}. The relevant identities are
\begin{align*}
 \langle 32\rangle(346)+\langle 34\rangle(623) + \langle 36\rangle(432) &= 0 \\
 [43](145)^\perp - [41](453)^\perp - [45](314)^\perp &= \nonumber\\
 [43](623) + [41](612) + [45](256) &= 0. 
\end{align*}
On the support of residue at $(612) = 0$, the $[41]$ terms can be dropped, and we obtain the following simplified expression for the $4{+}4$ channel:
\begin{equation}
\mathcal{A}^{(0)}_{6,3\;(4{+}4)} = \Res{(612)=0} \int d^{3\times6}\Omega_8 \frac{\langle 16\rangle [45] \langle 34\rangle[12]}{(123)(561)(346)^2}.
\label{eqn:6ptN=84+4}
\end{equation}

Adding up the three contributions in \eqref{a35} \eqref{a53}, \eqref{eqn:6ptN=84+4}, we find that the sum of decorated planar on-shell diagrams in Figure \ref{fig:N=86ptnmhvbcfw} correspond to the following Grassmannian integral formula:
\begin{align}
\mathcal{A}^{(0)}_{6,3}= \left(\Res{(234)=0} + \Res{(456)=0}\right) \int d^{3\times6}\Omega_8 &\frac{\langle 16\rangle [45] \langle 23\rangle[32]}{(123)(561)(146)(236)} \nonumber\\ 
&+\Res{(612)=0} \int d^{3\times6}\Omega_8 \frac{\langle 16\rangle [45] \langle 34\rangle[12]}{(123)(561)(346)^2}.
\label{n8}
\end{align}
The Grassmannian integral formula for the full 6-point NMHV amplitude in $\mathcal{N}=8$ SUGRA is then given by summing \eqref{n8} over permutations of legs $1$ to $4$. Note that it is not possible to write \eqref{n8} as the sum of three residues of a single top form. To see this, first add and subtract the (612) residue of the first integral
\begin{align}
\mathcal{A}^{(0)}_{6,3} &= \left(\Res{(234)=0} + \Res{(456)=0}+\Res{(612)=0} \right) \int d^{3\times6}\Omega_8 \frac{\langle 16\rangle [45] \langle 23\rangle[32]}{(123)(561)(146)(236)} \nonumber \\ 
&+\Res{(612)=0} \int d^{3\times6}\Omega_8 \frac{\langle 16\rangle [45]}{(123)(561)(346)^2(146)(236)}\left( \langle 34\rangle[12](146)(236)  -  \langle 23\rangle[32](346)^2 \right)
\end{align}
and note that the second line does not vanish on the support of the residue (612) = 0. Indeed, in a certain gauge the solution to the delta functions and residue constraints is
\begin{equation}
C = \left( \begin{matrix}
  \lambda_1 & \lambda_2 & \lambda_3 & \lambda_4 & \lambda_5 & \lambda_6 \\
  0 & [45] & [53] & [34] &0 &0
 \end{matrix} \right).
\end{equation}
Evaluating the second term on this solution shows that it is not zero for generic momenta.

Hence, unlike in $\mathcal{N}=4$ SYM, the decorated planar on-shell diagrams from which the 6-point NMHV amplitude of $\mathcal{N}=8$ SUGRA can be deduced do not correspond to residues of a single top-form. This can also be understood diagramtically as follows. Whereas the three planar on-shell diagrams contributing to the 6-point NMHV amplitude in $\mathcal{N}=4$ SYM can be embedded in a single Postnikov diagram in Figure~\ref{postnikov33}, it is not possible to decorate this diagram in such a way that it encodes the three decorated on-shell diagrams in Figure~\ref{fig:N=86ptnmhvbcfw}. This is because the merger equivalence relation for $\mathcal{N}=8$ SUGRA is less flexible than the one in $\mathcal{N}=4$ SYM since it requires opposite edges to be decorated, as depicted in Figure \ref{merge2}. It would be interesting to see if a unique top-form can be deduced by solving the on-shell diagram recursion relations in a non-planar sector or incorporating the bonus relations. Note that one can obtain such a formula by covariantizing the formulae derived in \cite{Cachazo:2012pz,He:2012er}, however it is unclear how to relate this to a contour integral arising from on-shell diagrams.

\section{One-Loop} \label{1loopsection}

In this section, we derive worldsheet formulae for 1-loop four-point amplitudes $\mathcal{N}=4$ SYM and $\mathcal{N}=8$ SUGRA using on-shell diagrams. These worldsheet formulae are manifestly supersymmetric and supported on 1-loop scattering equations refined by MHV degree. The 1-loop formula in $\mathcal{N}=4$ SYM can be generalized to more complicated amplitudes using loop-level BCFW recursion.

\subsection{$\mathcal{N}=4$}

\begin{figure}
\centering
       \includegraphics[scale=0.2]{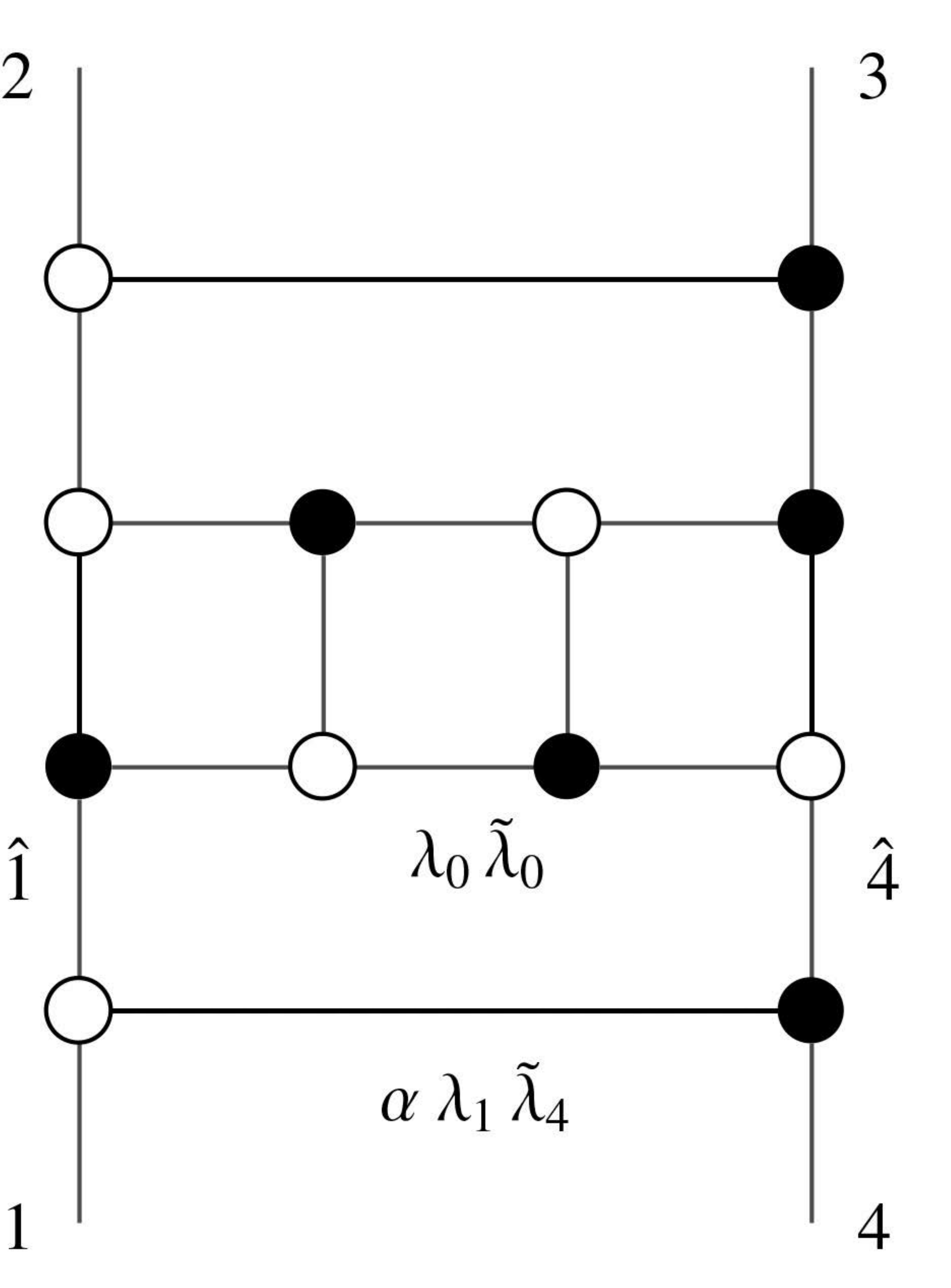}
    \caption{On-shell diagram for 1-loop four-point amplitude in $
\mathcal{N}$=4 SYM} 
    \label{1lp4ptn4}
\end{figure}

\begin{figure}
\centering
       \includegraphics[scale=0.2]{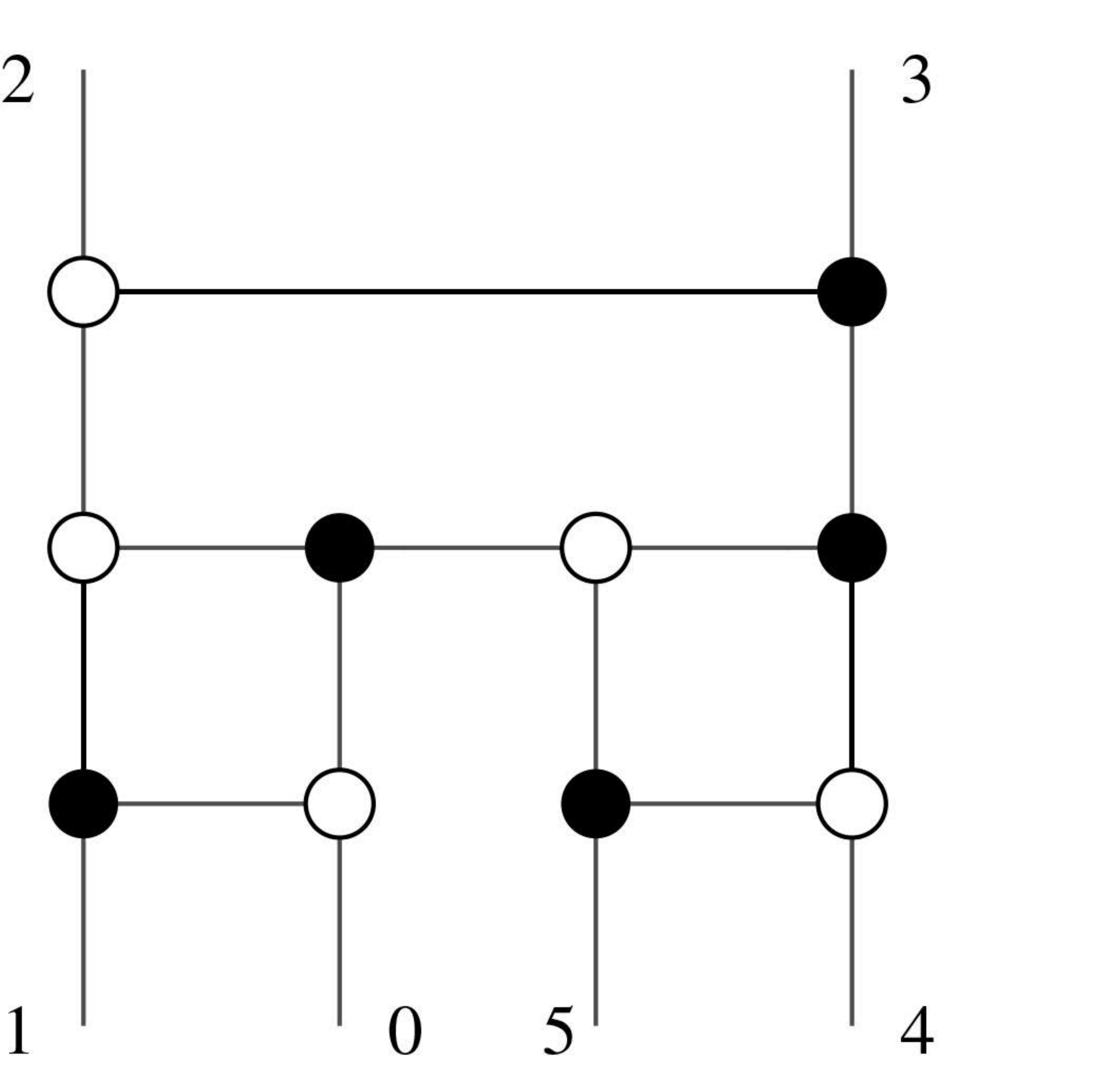}
    \caption{On-shell diagram from which Figure \ref{1lp4ptn4} can be obtained by taking a forward limit and adding a decorated BCFW bridge.} 
    \label{4p4n4}
\end{figure}

Using the on-shell diagram recursion in Figure \ref{n4recursion}, one finds that the 1-loop 4-point amplitude can be obtained by applying a forward limit and BCFW bridge to the tree-level 6-point NMHV amplitude, which is described by the three on-shell diagrams in Figure \ref{fig:N=46ptnmhvbcfw}. After doing so, only the $4+4$ channel diagram survives and using square moves and mergers the 1-loop 4-point amplitude can be described by the on-shell diagram in Figure \ref{1lp4ptn4} (for more details, see \cite{ArkaniHamed:2012nw}).  Note that this on-shell diagram can be obtained from Figure \ref{4p4n4} by taking the forward limit on legs $0$ and $5$, attaching a BCFW bridge to legs $1$ and $4$. Our strategy will therefore be to derive a Grassmannian integral formula for Figure \ref{4p4n4}, convert it to a worldsheet formula, and apply a forward limit and BCFW bridge to obtain a worldsheet formula for the 1-loop 4-point amplitude. We will subsequently define the loop
momentum to be the sum of the momenta in these two edges: 
\begin{equation}
l=\lambda_{0}\tilde{\lambda}_{0}+\alpha\lambda_{1}\tilde{\lambda}_{4}.
\label{loop1}
\end{equation} 

Note that Figure \ref{4p4n4} can be obtained from Figure~\ref{fig:N=46ptnmhv4+4labelled} by relabelling the external legs according to $P = \left( \begin{smallmatrix} 1&2&3&4&5&6\\1&0&5&4&3&2 \end{smallmatrix} \right)$. Applying this relabeling to \eqref{44diag} then gives the following Grassmannian integral formula for Figure \ref{4p4n4}:
\begin{equation}
\mathcal{A}_{6,3\;(4{+}4)'}^{(0)}=\Res{(012)=0}\int\frac{d^{3\times6}C}{GL(3)}\prod_{i=0}^{5}\frac{1}{(i\,i{+}1\,i{+}2)}\delta^{3\times(2|4)}\left(C\cdot\tilde{\lambda}\right)\delta^{3\times2}\left(\lambda\cdot C^{\perp}\right).\label{eq:4+4}
\end{equation}
To convert this into a worldsheet formula, we write it in terms of link
variables which can then be mapped into propagators on the 2-sphere. This can be accomplished by choosing coordinates on the Grassmannian
such that

\begin{equation}
C=\left(\begin{array}{cccccc}
c_{10} & 1 & 0 & c_{13} & c_{14} & 0\\
c_{20} & 0 & 1 & c_{23} & c_{24} & 0\\
0 & 0 & 0 & c_{53} & c_{54} & 1
\end{array}\right)\label{eq:link1loop}
\end{equation}
where the rows correspond to legs $1,2,5$. Note that the residue in \eqref{eq:4+4} sets $c_{50}=0$. Hence there are
eight link variables which are fixed by eight bosonic
delta functions in \eqref{eq:4+4}  (recall that the other four delta functions simply
enforce momentum conservation). We may therefore write \eqref{eq:4+4}
as 
\[
\mathcal{A}_{6,3\;(4{+}4)'}^{(0)}=\int d^{8}C\prod_{i=1}^{5}\frac{1}{(i\,i{+}1\,i{+}2)}\delta^{2\times(2|4)}\left(C\cdot\tilde{\lambda}\right)\delta^{3\times2}\left(\lambda\cdot C^{\perp}\right)
\]
where $d^{8}C$ is an integral over the eight link variables in \eqref{eq:link1loop}. 

The next step is to convert this to a worldsheet integral. Let's introduce
six punctures on the 2-sphere with homogeneous coordinates $\sigma_{i}^{\alpha}$,
$i\in\left\{ 0,...,5\right\}$, and set $\sigma_{1}=(0,1)$ and $\sigma_{2}=(1,0)$. The
coordinates of the remaining four punctures then provide eight integration
variables which precisely matches the number of link variables. To
map the link variables into worldsheet coordinates, simply insert
a factor of ``1'' into~(\ref{eq:4+4}) in the form 
\begin{equation}
1=\int\prod_{i\neq1,2}d^{2}\sigma_{i}\left(\prod_{r=0,3,4}\delta\left(\sigma_{r}^{1}+\frac{1}{c_{1r}}\right)\delta\left(\sigma_{r}^{2}-\frac{1}{c_{2r}}\right)\right)\delta^{2}\left(\sigma_{5}-\sigma_{5}^{*}\right)
\label{one}
\end{equation}
where

\[
\sigma_{5}^{*}=\frac{1}{c_{53}c_{54}\left(c_{14}c_{23}-c_{13}c_{24}\right)}\left(\begin{array}{c}
\left(c_{14}c_{53}-c_{13}c_{54}\right)c_{23}c_{24}\\
\left(c_{23}c_{54}-c_{24}c_{53}\right)c_{13}c_{14}
\end{array}\right).
\]
Noting that

\begin{equation}
\prod_{r=0,3,4}\delta\left(\sigma_{r}^{1}+\frac{1}{c_{1r}}\right)\delta\left(\sigma_{r}^{2}-\frac{1}{c_{2r}}\right)=\prod_{r=0,3,4}\frac{1}{(1r)^{2}(2r)^{2}}\delta\left(c_{1r}-\frac{1}{(1r)}\right)\delta\left(c_{2r}-\frac{1}{(2r)}\right)
\label{two}
\end{equation}
\begin{equation}
\delta^{2}\left(\sigma_{5}-\sigma_{5}^{*}\right)=(12)(34)\prod_{r=3,4}\frac{1}{(r5)^{2}}\delta\left(c_{5r}-\frac{1}{(5r)}\right)
\label{three}
\end{equation}
it is now straightforward to integrate out the link variables against
these delta functions, leaving an integral over worldsheet coordinates.
Covariantizing the resulting worldsheet integral, we obtain 
\begin{align*}
\mathcal{A}_{6,3\;(4{+}4)'}^{(0)}&=\int\frac{1}{GL(2)}\prod_{i=0}^{5}\frac{d^{2}\sigma_{i}}{(i\,i{+}1)}\frac{(14)(05)}{(15)(04)} \\
\times\delta^{2|4}&\left(\tilde{\lambda}_{5}-\sum_{r=3,4}\frac{\tilde{\lambda}_{r}}{(5r)}\right)\delta^{2}\left(\lambda_{0}-\sum_{l=1,2}\frac{\lambda_{l}}{(0l)}\right)\prod_{l=1,2}\delta^{2|4}\left(\tilde{\lambda}_{l}-\sum_{r=0,3,4}\frac{\tilde{\lambda}_{r}}{(lr)}\right)\prod_{r=3,4}\delta^{2}\left(\lambda_{r}-\sum_{l=1,2,5}\frac{\lambda_{l}}{(rl)}\right).
\end{align*}

To obtain a worldsheet formula for the 1-loop amplitude, we set $\left(\lambda_{5},\tilde{\lambda}_{5},\tilde{\eta}_{5}\right)=\left(-\lambda_{0},\tilde{\lambda}_{0},\tilde{\eta}_{0}\right)$
and BCFW shift legs 1 and 4, integrating over $\lambda_0 \tilde{\lambda}_0$
and the BCFW shift. Exchanging the definition of $\sigma_0$ and $\sigma_5$, we finally obtain
\begin{equation}
\mathcal{A}_{4,2}^{(1)}=\int\frac{d^{4}l}{l^{2}}\frac{1}{GL(2)}\prod_{i=0}^{5}\frac{d^{2}\sigma_{i}}{(i\, i{+}1)}\frac{(14)(05)}{(15)(04)}\delta^{2}(\tilde{S}_{0})\delta^{2}\left(S_{0}\right)\prod_{l}\delta^{2|4}\left(S_{l}\right)\prod_{r}\delta^{2}\left(S_{r}\right)
\label{4pt1pws}
\end{equation}
where the arguments of the delta functions are 1-loop scattering equations refined by MHV degree;
\[
\tilde{S}_{0}=\tilde{\lambda}_{0}-\sum_{r}\frac{\tilde{\lambda}_{r}}{(0r)},\,\,\, S_{0}=\lambda_{0}-\sum_{l}\frac{\lambda_{l}}{(5 l)},
\]
\begin{equation}
S_{l}=\hat{\tilde{\lambda}}_{l}-\sum_{r}\tilde{\lambda}_{r}\left(\frac{1}{(lr)}+\frac{1}{(l 5)(0r)}\right),\,\,\, S_{r}=\hat{\lambda}_{r}-\sum_{l}\lambda_{l}\left(\frac{1}{(rl)}-\frac{1}{(r0)(5 l)}\right),
\label{scatteringeq}
\end{equation}
with $l\in\left\{ 1,2\right\}$ and $r\in\{3,4\}$, $\hat{\lambda}_{4}=\lambda_{4}-\alpha\lambda_{1}$,
and $\left(\hat{\tilde{\lambda}}_{1},\hat{\tilde{\eta}}_{1}\right)=\left(\tilde{\lambda}_{1}+\alpha\tilde{\lambda}_{4},\tilde{\eta}_{1}+\alpha\tilde{\eta}_{4}\right)$ (the hats act trivially on the other spinors). From~(\ref{loop1}) the measure for the integral over loop momentum is
\[
\frac{d^{4}l}{l^{2}}=\frac{d^{2}\lambda_{0}d^{2}\tilde{\lambda}_{0}}{GL(1)}\frac{d\alpha}{\alpha}.
\]
The ratio of brackets multiplying the Parke-Taylor factor corresponds to summing over the exchange of $\sigma_0$ and $\sigma_5$: 
\begin{equation}
\prod_{i=0}^{5}\frac{d^{2}\sigma_{i}}{(i\, i{+}1)}\frac{(1 4)(0\, 5)}{(1\, 5)(0 4)}=\prod_{i=0}^{5}\frac{d^{2}\sigma_{i}}{(i\, i{+}1)}+\left(0\leftrightarrow 5\right).
\label{exchange}
\end{equation}

\begin{figure}
\centering
       \includegraphics[scale=0.65]{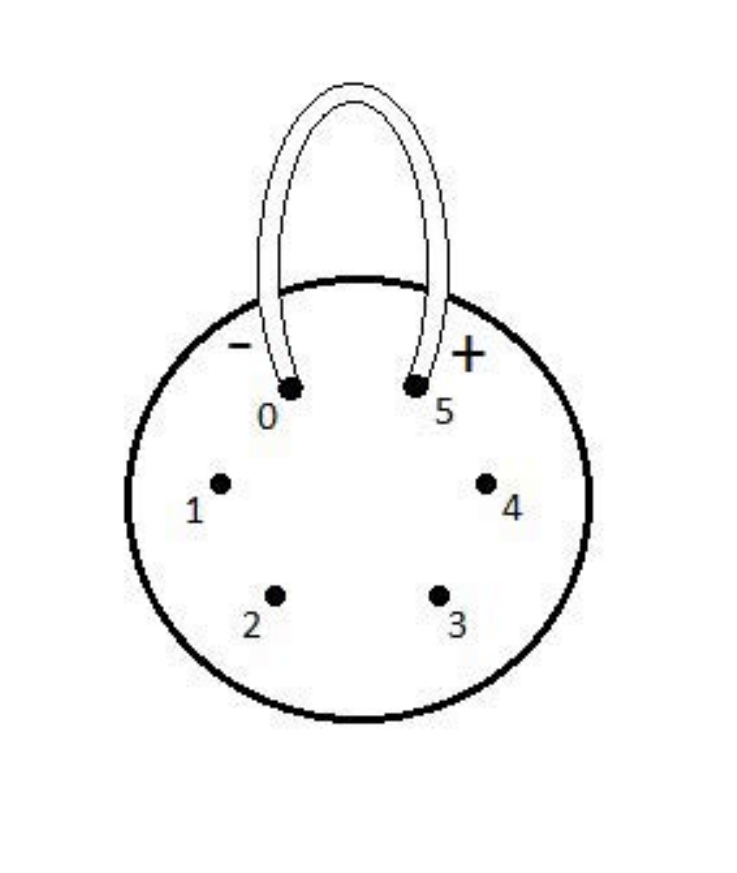}
    \caption{The worldsheet configuration describing a 1-loop 4-point amplitude in 4d ambitwistor string theory.} 
    \label{pinch}
\end{figure}

Let us point out some important features of the worldsheet formula for the 1-loop 4-point amplitude in \eqref{4pt1pws}. First note that it contains an integral over the locations of six punctures on a genus-0 worldsheet. Whereas the punctures $1,..,4$ are associated with the four external particles being scattered, punctures $0$ and $5$ are associated with the two internal particles participating in the forward limit. The worldsheet can therefore be visualized as Figure \ref{pinch}, which corresponds to a non-separating degeneration of a genus-1 worldsheet (similar to the 1-loop amplitudes of 10d ambitwistor string theory \cite{Geyer:2015jch}). The integral over loop momentum is implemented by decomposing it according to \eqref{loop1} and integrating over the forward limit momentum $\lambda_0 \tilde{\lambda}_0$ and BCFW shift parameter $\alpha$ which appear in the 1-loop scattering equations in \eqref{scatteringeq}. 
Note that \eqref{4pt1pws} is manifestly supersymmetric and does not contain Pfaffians, so is simpler than previous worldsheet formulae for 1-loop amplitudes. On the other hand, it must be regulated when integrating over loop momentum since it is intrinsically four-dimensional. In Appendix \ref{1loopb}, we show that \eqref{4pt1pws} is equivalent to the standard formula for the 1-loop 4-point amplitude in term of a scalar box integral. Although we have focused on 1-loop 4-point amplitude to make the discussion as simple and concrete as possible, there is no obstruction to generalizing this formula to more complicated amplitudes using BCFW recursion, which we leave for future work. In Appendix \ref{counting}, we consider a generalization of the 1-loop scattering equations refined by MHV degree to any number of legs, and analyze various properties of their solutions.  

\subsection{$\mathcal{N}=8$} \label{1lpn8}

\begin{figure}
\centering
       \includegraphics[scale=0.7]{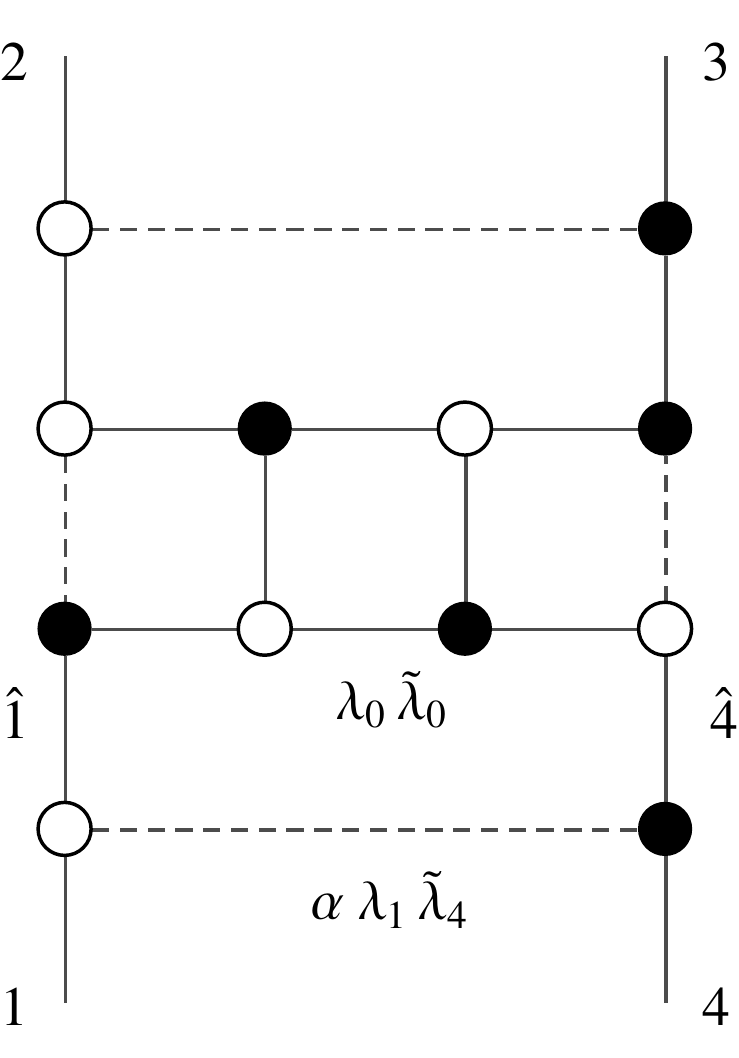}
    \caption{On-shell diagram for 1-loop four-point amplitude in $
\mathcal{N}$=8 SUGRA} 
    \label{1lp4pt}
\end{figure}

\begin{figure}
\centering
       \includegraphics[scale=0.65]{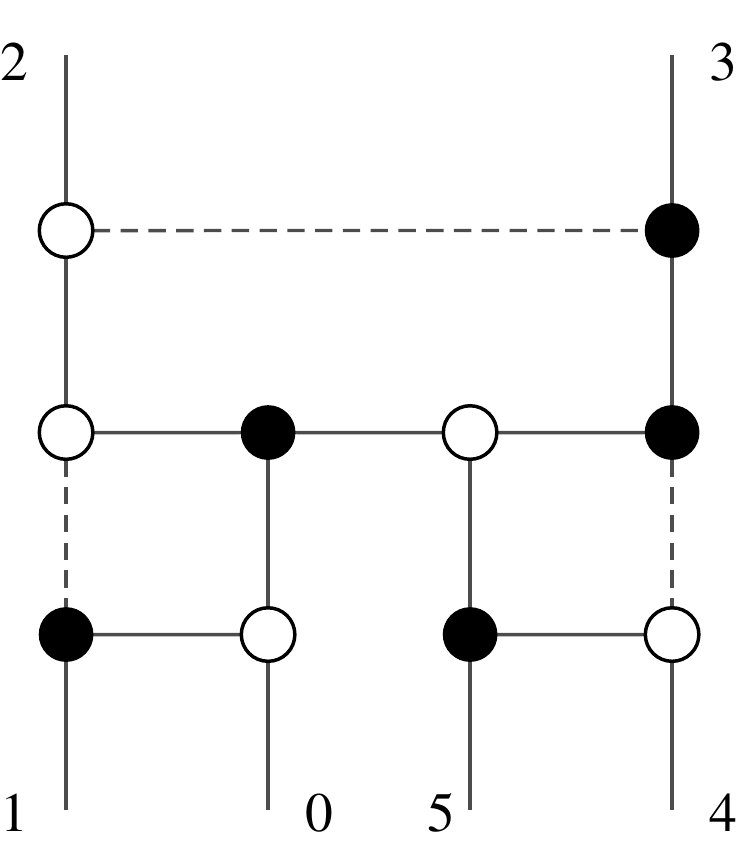}
    \caption{On-shell diagram from which Figure \ref{1lp4pt} can be obtained by taking a forward limit and adding a decorated BCFW bridge.} 
    \label{fig:6pt4+4N=8altered}
\end{figure}

In this section, we will deduce a worldsheet formula for the 1-loop
4-point amplitude of $\mathcal{N}=8$ SUGRA. Unlike in planar $\mathcal{N}=4$
SYM, a loop-level BCFW recursion relation is not known for $\mathcal{N}=8$
SUGRA. On the other hand, \cite{Heslop:2016plj} showed that this
amplitude corresponds to the on-shell diagram in Figure \ref{1lp4pt} after summing
over permutations of the external legs.
Equivalently, one can obtain the diagram in Figure \ref{1lp4pt} from the diagram
in Figure~\ref{fig:6pt4+4N=8altered} by taking the forward limit of legs $0$ and $5$ and attaching
a decorated BCFW bridge to legs $1$ and $4$. As we did in the previous section, we will define the loop
momentum to be the sum of the momenta in these two edges given by \eqref{loop1}. Moreover we will derive a Grassmannian integral formula
for Figure~\ref{fig:6pt4+4N=8altered}, convert it to a worldsheet formula, apply a forward limit and decorated BCFW bridge,
and sum over permutations of the external legs to obtain a worldsheet formula for the 1-loop 4-point amplitude.  

Note that the diagram in Figure \ref{fig:6pt4+4N=8altered} is the same as the $4+4$ diagram in Figure \ref{fig:N=86ptnmhvbcfw} up to the location of bridge decorations. Hence, we can compute it simply by multiplying the integrand in \eqref{a2} by the following ratio of bridge decorations and spinor brackets associated with the vertices: 
\[
 \frac{\mbox{new decorations}}{\mbox{old decorations}}\frac{\mbox{new brackets}}{\mbox{old brackets}} = \alpha_6^{-2}\alpha_1\alpha_{2}^{-1}\frac{\langle\beta \alpha_3\rangle [\gamma \alpha_7]}{\langle\alpha_1 \alpha_2 \rangle [4 \alpha_5]}.
\]
We then apply the relabeling $P = \left( \begin{smallmatrix} 1&2&3&4&5&6\\1&0&5&4&3&2 \end{smallmatrix} \right)$ to match the labelling between Figures \ref{fig:6pt4+4N=8altered} and \ref{fig:N=86ptnmhvbcfw}, which ultimately gives the following Grassmannian integral formula for the diagram in Figure \ref{fig:6pt4+4N=8altered}: 
\begin{equation}
\mathcal{A}_{6,3\;(4{+}4)'}^{(0)}= \Res{(012)=0}\int\frac{d^{3\times6}C}{GL(3)}\prod_{i=0}^{5}\frac{1}{(i\,i{+}1\,i{+}2)}\frac{\left\langle 12\right\rangle \left\langle 45\right\rangle \left[01\right]\left[34\right]}{(234)(450)(512)^{2}}\delta^{3\times(2|8)}\left(C\cdot\tilde{\lambda}\right)\delta^{3\times2}\left(\lambda\cdot C^{\perp}\right)
\label{eq:4+4altered}
\end{equation}
To map this into a worldsheet formula, we will first write it in terms
of link variables as we did in the previous section. Choosing coordinates on the Grassmannian according to \eqref{eq:link1loop}, we find that \eqref{eq:4+4altered} can be written as
as 
\begin{equation}
\mathcal{A}_{6,3\;(4{+}4)'}^{(0)}=\int d^{8}C\prod_{i=1}^{5}\frac{1}{(i\,i{+}1\,i{+}2)}\frac{\left\langle 12\right\rangle \left\langle 45\right\rangle \left[01\right]\left[34\right]}{(234)(450)(512)^{2}}\delta^{3\times(2|8)}\left(C\cdot\tilde{\lambda}\right)\delta^{2\times3}\left(\lambda\cdot C^{\perp}\right)
\label{w4}
\end{equation}
where $d^{8}C$ is the measure over the eight non-zero link variables
in \eqref{eq:link1loop}. 

The next step is to convert this to a worldsheet integral. Let us introduce
six punctures on the 2-sphere with homogeneous coordinates $\sigma_{i}^{\alpha}$,
$i=0,...,5$, and set $\sigma_{1}=(0,1)$ and $\sigma_{2}=(1,0)$. The
coordinates of the remaining four punctures then provide eight integration
variables which precisely matches the number of link variables. To
map the link variables into worldsheet coordinates, simply insert
a factor of ``1'' into \eqref{w4} in the form given by \eqref{one}.
Using equations \eqref{two} and \eqref{three} it is then straightforward to integrate out the link variables against
these delta functions, leaving an integral over worldsheet coordinates.
Covariantizing the resulting worldsheet integral, we obtain 

\begin{align*}
&\mathcal{A}_{6,3\;(4{+}4)'}^{(0)}=\left\langle 12\right\rangle \left\langle 45\right\rangle \left[01\right]\left[34\right]\int\frac{1}{GL(2)}\prod_{i=0}^{5}d^{2}\sigma_{i}\frac{(02)(13)(14)^{3}(24)(35)}{(04)^{2}(12)^{2}(15)^{2}(23)(34)^{2}}\times\\
&\delta^{2|8}\left(\tilde{\lambda}_{5}-\sum_{r=3,4}\frac{\tilde{\lambda}_{r}}{(5r)}\right)\delta^{2}\left(\lambda_{0}-\sum_{l=1,2}\frac{\lambda_{l}}{(0l)}\right)\prod_{l=1,2}\delta^{2|8}\left(\tilde{\lambda}_{l}-\sum_{r=0,3,4}\frac{\tilde{\lambda}_{r}}{(lr)}\right)\prod_{r=3,4}\delta^{2}\left(\lambda_{r}-\sum_{l=1,2,5}\frac{\lambda_{l}}{(rl)}\right)
\end{align*}

To obtain a worldsheet formula for the 1-loop amplitude, we set $\left(\lambda_{5},\tilde{\lambda}_{5},\tilde{\eta}_{5}\right)=\left(-\lambda_{0},\tilde{\lambda}_{0},\tilde{\eta}_{0}\right)$
and BCFW shift legs 1 and 4 (integrating over the forward limit momentum
and BCFW shift), and sum over permutations. After exchanging $\sigma_0$ with $\sigma_5$ and simplifying the integrand on the support of the scattering equations we obtain

\begin{align*}
\mathcal{M}_{4,2}^{(1)}=\sum_{\textrm{perms}\{1,2,3,4\}}\left\langle 12\right\rangle ^{2}\left[34\right]^{2}\int\frac{d^4 l}{l^{2}}\frac{1}{GL(2)}&\prod_{i=0}^{5}\frac{d^{2}\sigma_{i}}{(i\,i{+}1)}\frac{(14)(05)}{(15)(04)}\\
&\times\delta^{2}(\tilde{S}_{0})\delta^{2}\left(S_{0}\right)\prod_{l}\delta^{2|8}\left(S_{l}\right)\prod_{r}\delta^{2}\left(S_{r}\right)\label{eq:1-loop}
\end{align*}
where the arguments of the delta functions are the 1-loop scattering equations in \eqref{scatteringeq} with $l\in\{1,2\}$, $r\in\{3,4\}$, $\hat{\lambda}_{4}=\lambda_{4}-\alpha\lambda_{1}$,
and $\left(\hat{\tilde{\lambda}}_{1},\hat{\tilde{\eta}}_{1}\right)=\left(\tilde{\lambda}_{1}+\alpha\tilde{\lambda}_{n},\tilde{\eta}_{1}+\alpha\tilde{\eta}_{n}\right)$ (the hats act trivially on the other spinors). Note that the integrand can be written as a Parke-Taylor factor summed over the exchange of $\sigma_0$ and $\sigma_5$, as described in \eqref{exchange}. Furthermore, on the support of the scattering equations, we may write

\[
\left\langle 12\right\rangle ^{2}\left[34\right]^{2}=\frac{\prod_{i=1}^{4}(0i)(5i)}{1-(05)^{2}}\det H\det\tilde{H}
\]
where we have taken six point NMHV Hodges matrices defined in section \ref{ambrev} and removed the rows and columns associated with particles 0 and 5 to give
\[
H=\left(\begin{array}{cc}
-\frac{\left\langle 10\right\rangle }{\left(10\right)}-\frac{\left\langle 12\right\rangle }{(12)} & \frac{\left\langle 12\right\rangle }{(12)}\\
\frac{\left\langle 12\right\rangle }{(12)} & -\frac{\left\langle 20\right\rangle }{\left(20\right)}-\frac{\left\langle 21\right\rangle }{(21)}
\end{array}\right),\,\,\,\tilde{H}=\left(\begin{array}{cc}
-\frac{\left[30\right]}{(35)}-\frac{\left[34\right]}{(34)} & \frac{\left[34\right]}{(34)}\\
\frac{\left[34\right]}{(34)} & -\frac{[40]}{(45)}-\frac{\left[43\right]}{(43)}
\end{array}\right).
\]
Hence, we finally obtain the following worldsheet formula for the 1-loop 4-point amplitude of $\mathcal{N}=8$ SUGRA:
\[
\mathcal{M}_{4,2}^{(1)}=\sum_{\textrm{perms}\{1,2,3,4\}}\int\frac{d^{4}l}{l^{2}}\frac{1}{GL(2)}\prod_{i=0}^{5}\frac{d^{2}\sigma_{i}}{(i\, i{+}1)}\frac{(14)(05)}{(15)(04)}\frac{\prod_{i=1}^{4}(0i)(5i)}{1-(05)^{2}}\det H\det\tilde{H}
\]
\begin{equation}
\times \delta^{2}(\tilde{S}_{0})\delta^{2}\left(S_{0}\right)\prod_{l}\delta^{2|8}\left(S_{l}\right)\prod_{r}\delta^{2}\left(S_{r}\right).
\label{1ln8}
\end{equation}
The determinants can be thought of as arising from the forward limit of a tree-level 6-point
NMHV amplitude, but the remaining terms in the integrand are difficult
to interpret at present. Following the discussion in the end of section \ref{ambrev}, we see that \eqref{1ln8} has the expected scaling properties under the little group transformations. In particular, the term $1-(05)^{2}$ is invariant because $\sigma_0$ and $\sigma_5$ scale with opposite weight. 

\section{Conclusion}

In this paper we explore the relation between two approaches for computing scattering amplitudes known as 4d ambitwistor string theory and on-shell diagrams, focusing on the examples of $\mathcal{N}=4$ SYM and $\mathcal{N}=8$ SUGRA, which are believed to be the simplest quantum field theories in four dimensions. In the process, we obtain a number of new results at tree-level and 1-loop. For example, we obtain new Grassmannian integral formulae for tree-level amplitudes in $\mathcal{N}=8$ SUGRA by mapping ambitwistor string formulae into link variables as well as by solving the on-shell diagram recursion relations. For non-MHV amplitudes, we find that the decorated planar on-shell diagrams of $\mathcal{N}=8$ SUGRA do not arise from the residues of a single top form in the Grassmannian, in contrast to the planar on-shell diagrams of $\mathcal{N}=4$ SYM. We also derive new worldsheet formulae for 1-loop amplitudes in $\mathcal{N}=4$ SYM and $\mathcal{N}=8$ SUGRA which are manifestly supersymmetric and supported on scattering equations refined by MHV degree.

Based on these findings, there are a number of interesting directions for future research: 
\begin{itemize}
\item In Appendix \ref{bonusappendix}, we use the bonus relations to solve the on-shell diagram recursion relations in the planar sector for MHV amplitudes in $\mathcal{N}=8$ and obtain a compact expression for any number of legs. It would be interesting to see if this expression has a geometric interpretation, analogous to the Amplituhedron for planar $\mathcal{N}=4$ SYM. Beyond MHV, we find that the decorated planar on-shell diagrams from which the full amplitudes can be deduced do not correspond to residues of a single top-form, so it would be interesting to see if a unique top-form can be deduced by solving the recursion relations in a non-planar sector or incorporating the bonus relations. 

\item It would be interesting to generalize our worldsheet formulae for 1-loop amplitudes in $\mathcal{N}=4$ SYM and $\mathcal{N}=8$ SUGRA to higher loops and legs. Although there is no obstruction to doing this for planar $\mathcal{N}=4$ SYM using the loop-level BCFW recursion relations, there is not yet a systematic way to do this for $\mathcal{N}=8$ SUGRA. Furthermore, since the resulting worldsheet formulae are intrinsically four-dimensional, they will give rise to IR divergences when one integrates over the loop momentum, so it would be useful to find a simple prescription for regulating such divergences.

\item Since integrability is usually restricted to two-dimensional models, one would expect that reformulating perturbative scattering amplitudes as worldsheet integrals should provide new insight into the origin of such properties in a 4d theory like $\mathcal{N}=4$ SYM. It would therefore be interesting to investigate how Yangian symmetry is realized for the worldsheet formulae of $\mathcal{N}=4$ SYM. Moreover, if it is possible to generalize our worldsheet formulae for $\mathcal{N}=8$ SUGRA to higher loops, it would be interesting to investigate if they provide hints into the origin of unexpected UV cancellations.

\item There has recently been a great deal of progress in computing tree-level form factors in $\mathcal{N}=4$ SYM using on-shell diagrams \cite{Frassek:2015rka} and 4d ambitwistor string theory \cite{He:2016dol,He:2016jdg,Brandhuber:2016xue,Bork:2017qyh}, so it would interesting to see if our one-loop worldsheet formula for $\mathcal{N}=4$ SYM can be generalized to form-factors. 

\item Ultimately, one would like to derive perturbative loop amplitudes directly from the worldsheet theories for $\mathcal{N}=4$ SYM and $\mathcal{N}=8$ SUGRA. This may be challenging using the worldsheet models developed so far since the worldsheet theory for $\mathcal{N}=4$ SYM contains conformal supergravity in its spectrum \cite{Berkovits:2004jj} and the worldsheet theory for $\mathcal{N}=8$ SUGRA is not critical if one gauges the Virasoro symmetry \cite{Skinner:2013xp}. Nevertheless, the worldsheet formulae deduced in this paper may provide useful hints. 
\end{itemize}
In summary, we find that on-shell diagrams are intimately related to 4d ambitwistor string theory and that studying the interplay of these two approaches is rather fruitful.  

\begin{center}
\textbf{Acknowledgments}
\end{center}
We thank Daniele Dorigoni, Lionel Mason, and especially Paul Heslop for useful discussions. AL is supported by the Royal Society as a Royal Society University Research Fellowship holder, and JF is funded by EPSRC PhD scholarship EP/L504762/1. 
\appendix

\section{Algorithm for Computing On-Shell Diagrams}  \label{algorithm}
\begin{figure}[h]
\centering
\begin{tabular}{c c}
\includegraphics[scale = 0.7]{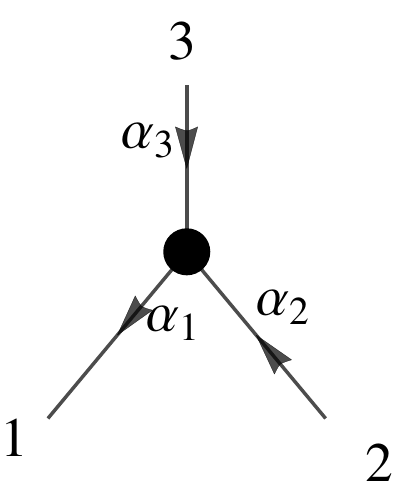} & \includegraphics[scale = 0.7]{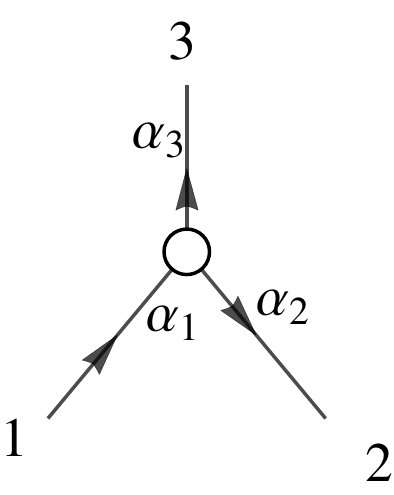}\\
  $\langle23\rangle = \frac{\langle12\rangle}{\alpha_1\alpha_3} = \frac{\langle31\rangle}{\alpha_1 \alpha_2}$
 & $[23] = \frac{[12]}{\alpha_1\alpha_3} = \frac{[31]}{\alpha_1 \alpha_2}$
\end{tabular}
\caption{Relations between spinor bracket factors at each vertex of an on-shell diagram in $\mathcal{N}=8$ SUGRA}
\label{fig:3ptvertex}
\end{figure}

In this section we provide a streamlined version of the algorithm described in \cite{Heslop:2016plj} for calculating on-shell diagrams in $\mathcal{N} = 8$ SUGRA in terms of Grassmannian integral formulae. In particular, given a decorated on-shell diagram computed from the recursion relations described in section \ref{osdrev}:

\begin{enumerate}
\item Choose a perfect orientation for the diagram by drawing arrows on each edge such that there are two arrows entering and one arrow leaving every black node, and two arrows leaving and one arrow entering every white node.

\item Label every half-edge with an edge variable $\alpha$ so that there are two variables for each internal edge (one associated with each of the two vertices attached to the edge). Then set one of the two edge variables on each internal edge to unity, and set one of the remaining variables associated with each vertex to unity. There will be $2n-4$ edge variables remaining after this step.

\item To construct the integrand, include a factor of $\rm{d} \alpha /\alpha^2$ for each edge variable leaving a white vertex or entering a black vertex and $\rm{d} \alpha /\alpha^3$ for each edge variable entering a white vertex or leaving a black vertex.

\item Now include decorations associated with the BCFW bridges and spinor bracket factors associated with the vertices. The spinor brackets at the bridges cancel with the bridge decoration to leave only edge variables. This step can be summarised as:

\begin{enumerate}
\item For each BCFW bridge, look at the sub-diagram formed only by this bridge, its two vertices, and the four legs attached to it. 
\begin{itemize}
\item If there is only one path through the sub diagram which includes the bridge, assign a factor of the edge variable on the bridge, divided by the two edge variables on the legs which are not on that path.

\item If there are four possible paths through the sub diagram, divide through by a factor of each of the edge variables on the external legs, and the edge variable on the bridge squared.
\end{itemize}
If there is no edge variable in any of the locations described above, then this edge variable was set to unity in step 2.

\item For each remaining black vertex not attached to a bridge, add a factor of $\langle ij\rangle$ where $i,j$ are the two edges with ingoing arrows. For each remaining white vertex not associated to a bridge, add a factor of $[ij]$ where $i,j$ are the two edges with outgoing arrows.
\end{enumerate}

\item Now it is necessary to relate all internal spinors to external spinors. This can be done algorithmically by noting that all spinors are related to each other via equations
\begin{align}
\lambda_i &= \sum_{\mathrm{paths\;j\rightarrow i}}\left(\prod_{\mathrm{edges\;in\;path\;e}}\alpha_e\right) \lambda_j \nonumber\\ 
\tilde{\lambda}_i &= \sum_{\mathrm{paths\;i\rightarrow j}}\left(\prod_{\mathrm{edges\;in\;path\;e}}\alpha_e\right) \tilde{\lambda}_j.
\label{paths}
\end{align}
In practice, one can often obtain simpler expressions using the relations between square and angle brackets at each vertex given in figure~\ref{fig:3ptvertex}. 

\item Calculate the $C$-matrix in terms of the coordinates assigned to the diagram, associating each column with an external leg and each row with an ingoing external leg. The element $C_{ij}$ can then be computed by summing over all paths from leg $i$ to leg $j$ taking the product of all the edge variables encountered along the path as in the first line of \eqref{paths}. Similarly, the $C^{\perp}$ matrix can be computed by summing over the reverse paths as in the second line of \eqref{paths}.    After doing so, include the following delta functions in the integrand
\[
\delta^{k\times (2|8)}(C \cdot\lambdat|C \cdot \eta)\delta^{(n-k)\times2}(C^{\perp}\cdot\lambda).
\]

\item If the diagram contains closed loops, include a factor of $\mathcal{J}^{\mathcal{N}-4}$, were $\mathcal{J}$ is a sum over products of disjoint closed loops \cite{Herrmann:2016qea}:
\[
\mathcal{J}=1+\sum_{i}f_{i}+\sum_{{\rm disjoint}\, i,j}f_{i}f_{j}+\sum_{{\rm disjoint}\, i,j,k}f_{i}f_{j}f_{k}+...
\]
and $f_i$ is minus the product of edge variables around the $i$'th closed loop. 

\item The above procedure gives an expression for the on-shell diagram as a Grassmannian integral in terms of specific coordinates. This can be uplifted to a covariant expression by expressing the rest of the integrand in terms of minors. This results in an SL(k) invariant expression, but the overall GL(1) scaling of the GL(k) gauge freedom will not be correct in general. There will always be one minor which is gauged fixed to be equal to unity, and the correct number of factors of this minor should be included in the integrand to give an overall GL(1) weight of zero to the integrand. Note that $d^{k \times n} \Omega_\mathcal{N}$ in \eqref{meas} has GL(1) weight $\mathcal{N}-4$. For on-shell diagrams contributing to non-MHV amplitudes, this lift will specify a nontrivial contour in the Grassmannian. Details of this process for 6 point NMHV amplitude are explained in section~\ref{nmhv}.

\end{enumerate}

\section{Bonus Relations} \label{bonusappendix} 

\begin{figure}
\centering
\begin{tabular}{m{3cm} m{1cm} m{.5cm} m{1cm} m{10cm}}
      &$\cM_{n,2}^{(0)}$ & = & \[\mathlarger{\sum}_{i=2}^{n-1}\] & \includegraphics[scale=0.6]{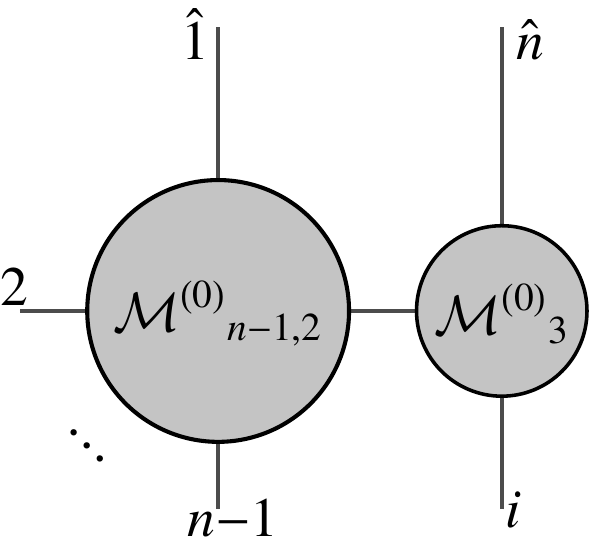}
\end{tabular}
    \caption{BCFW recursion for tree-level MHV amplitudes in $\mathcal{N}=8$ SUGRA} 
    \label{mhvre}
\end{figure}

In this appendix, we will explain how to incorporate the bonus relations into on-shell diagram recursion for $\mathcal{N}=8$ SUGRA, focusing on the example of MHV amplitudes for simplicity. Consider shifting legs $1$ and $n$ of an $n$-point amplitude as follows:
\[
\hat{\tilde{\lambda}}_{1}=\tilde{\lambda}_{1}+z\tilde{\lambda}_{n},\,\,\,\hat{\lambda}_{n}=\lambda_{n}-z\lambda_{1}
\]
The momenta are then shifted as $\hat{p}_{1}=p_{1}+zq$ and $\hat{p}_n=p_{n}-zq$,
where $q=\lambda_{1}\tilde{\lambda}_{n}$. For
an MHV amplitude, each factorization channel will consist of a 3-point
amplitude containing leg $n$ times an $(n-1)$-point amplitude containing
leg 1 and can be labelled by the unshifted external leg appearing
on the 3-point amplitude, as depicted in Figure \ref{mhvre}. The value of $z$
corresponding to the $i$th factorization channel is determined by
solving the equation $\left(\hat{p}_{n}+p_{i}\right)^{2}=0$ and is
given by 
\[
z_{i}=\frac{p_{n}\cdot p_{i}}{q\cdot p_{i}}=\frac{\left\langle ni\right\rangle }{\left\langle 1i\right\rangle }.
\]

\begin{figure}
\centering
\begin{tabular}{m{.8cm} m{3.5cm} m{.8cm} m{.3cm} m{3.5cm} m{2cm} m{3.5cm}}
   \[\mathlarger{\sum}_{i=2}^{n-1} \:z_i\] & \includegraphics[scale=0.6]{Figures/gravityBCFWbonus1.pdf}  & $= 0$  
 &$\Rightarrow$ & \includegraphics[scale=0.6]{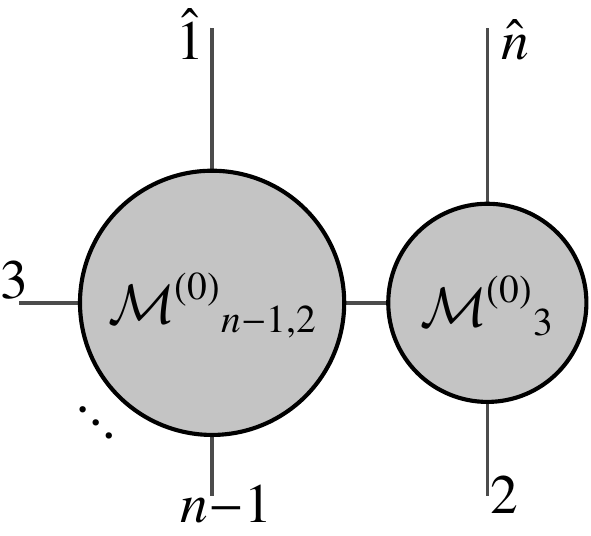} & \[=\;\;-\mathlarger{\sum}_{i=3}^{n-1} \:\frac{z_i}{z_2}\] &  \includegraphics[scale=0.6]{Figures/gravityBCFWbonus1.pdf}
\end{tabular}

\begin{tabular}{m{3cm}m{1cm} m{1cm} m{.5cm} m{2.3cm} m{10cm}}
      &$\Rightarrow$ &$\cM_{n,2}^{(0)}$ & = & \[\mathlarger{\sum}_{i=3}^{n-1}\left(1-\frac{z_i}{z_2}\right)\] & \includegraphics[scale=0.6]{Figures/gravityBCFWbonus1.pdf}
\end{tabular}
    \caption{The MHV bonus relations in $\mathcal{N}=8$ SUGRA can be used to eliminate one channel from the recursion in Figure~\ref{mhvre}, reducing the number of channels from $(n - 2)$ to $(n - 3)$} 
    \label{bonus}
\end{figure}

For $\mathcal{N}=8$ SUGRA, the superamplitudes vanish like $\mathcal{O}\left(z^{-2}\right)$
as $z\rightarrow\infty$, which implies that the sum over factorization
channels weighted by the value $z$ for each channel should vanish.
These constraints, known as bonus relations, allow one to express
one factorization channel as a sum over the others \cite{ArkaniHamed:2008gz,Spradlin:2008bu}. In particular
for an $n$-point MHV amplitude BCFW shifted as a described above,
we can express the $i=2$ channel as a sum over the other $n-3$ channels
as depicted in Figure \ref{bonus}. Plugging this into the BCFW recursion relation,
one subsequently finds that the amplitude can be expressed as a sum
over the channels $i\in\left\{ 3,...,n\right\} $ each weighted by the
factor
\[
\beta_{12n;i}=\frac{\left\langle 1n\right\rangle \left\langle i2\right\rangle }{\left\langle 1i\right\rangle \left\langle n2\right\rangle }.
\label{eq:bonussimpfac}
\]
For non-MHV amplitudes, the corresponding factor will be more complicated \cite{He:2010ab}.

Let us now apply this observation to on-shell diagrams. Recall that
an $n$-point MHV amplitude can be obtained by attaching an $(n-1)$-point amplitude
to a 3-point vertex and adding a decorated BCFW bridge as depicted
in Figure \ref{n8f} and summing over $i \in \left\{ 2,...,n-1\right\} $.
On the other hand, if we multiply this diagram by the factor above, the amplitude can be obtained by summing over $i \in \left\{ 3,...,n-1\right\} $. Hence,
one can incorporate the bonus relations into on-shell diagram recursion by using the modified bridge
decoration 
\[
B_{12n;i}=\frac{\left\langle i2\right\rangle }{\left[1n\right]\left\langle 1i\right\rangle \left\langle 2n\right\rangle }
\]
where the subscript $12n$ indicates that one must hold legs $1,2,n$ fixed
when summing over permutations of the external legs to obtain the full amplitude. 

We now use the modified bridge decoration proposed above to find a solution to the planar on-shell diagram recursion in the MHV sector. Using the bonus relations, this will yield a decorated planar on-shell diagram from which the full amplitude can be obtained by summing over permutations of the external legs holding three legs fixed. This is reminiscent of the CHY formulation for scattering, where 3 punctures are fixed and there are $(n - 3)!$ solutions to the scattering equations at $n$ points. We define the bonus-simplified planar on-shell diagram with $n$ external legs to be $\mathcal{A}^{(0)*}_{n,2}$.

\begin{figure}[h]
\centering
\includegraphics[scale = 0.5]{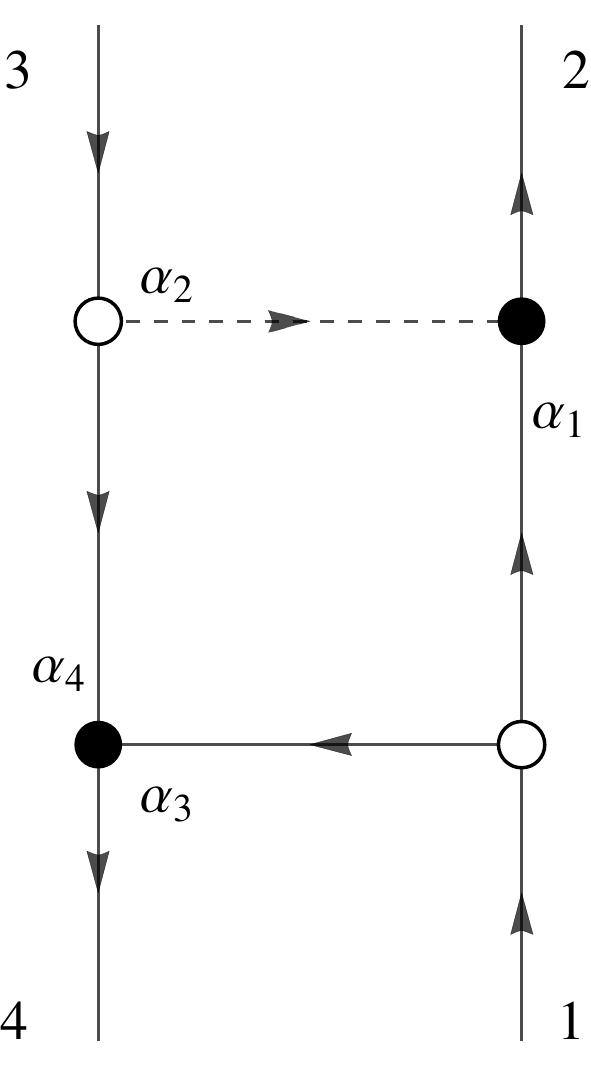} 
\caption{On-shell diagram for 4-point seed amplitude}
\label{fig:4ptbase}
\end{figure}

At $n$ points in the planar MHV sector there is only one BCFW diagram to consider, which is generated from the $(n-1)$ diagram by adding an inverse soft factor, as depicted in Figure \ref{n8f}. This gives a simple recursive way to generate all  $\mathcal{A}^{(0)*}_{n,2}$. To be able to calculate the diagrams, we would also like to choose an orientation and labelling which can be extended to higher points in a similar recursive fashion. We start with $\mathcal{A}^{(0)*}_{4,2}$ as the seed, with the orientation and labelling in figure~\ref{fig:4ptbase}.

\begin{figure}[h!]
\centering
\includegraphics[scale = 0.65]{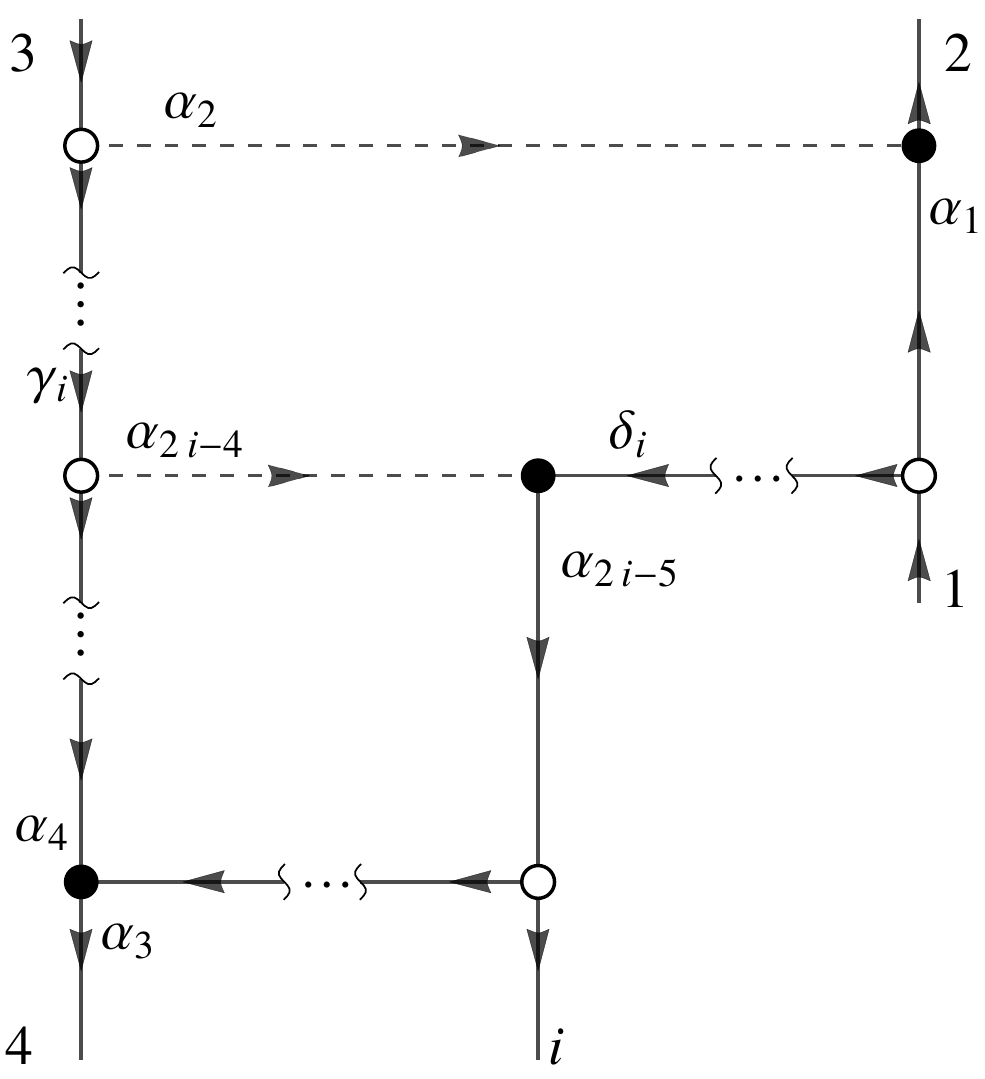} 
\caption{On-shell diagram showing how 4-point seed amplitude orientation is cut, and an inverse soft factor is inserted for $i \in \{5, ... n\}$}
\label{fig:nptgenerated}
\end{figure}

To get the $n$-point diagram, as explained above, we add the appropriate inverse soft factor. The planar BCFW recursion adds this factor onto legs one and two, but in terms of the orientation, to provide an easily extendible diagram where the paths remain mostly unchanged, we can think of "cutting" the diagram to insert the new factor, as shown in Figure~\ref{fig:nptgenerated}. The labelling of the legs then always remains the same such that the top inverse soft factor is always labelled with legs 1, 2 and 3, and the bottom left black vertex is always labelled as leg 4. This allows all calculations of these parts of the diagram to remain almost independent of the newly added leg, which always comes in to the left of leg one. Note that in this process, BCFW recursion is always carried out in the standard way and the diagram itself is never really cut, it is only the unphysical orientation of the diagram which is cut. 

Constructing the diagrams in this way allows for a simple recursive calculation of the $C$ matrix, which we will denote as $C^n$ at $n$-points. The 4-point $C$ matrix can be read off as

\begin{equation}
C^{4} = \left(
\begin{array}{cccc}
 1 & \alpha _1 & 0 & \alpha _3 \\
 0 & \alpha _2 & 1 & \alpha _4
\end{array}
\right).
\end{equation}
As the orientation and labelling remain the same on the top inverse soft factor, the first three rows of the $C$ matrix are the same for all $n$. Now consider $C^{n}$. Each path from 1 to $i$, for $i \in \{4, ... , n\}$, remains the same except for an extra factor of $\alpha_{2n-5}$ from the new inverse soft factor.  There is now a new path from 3 to $i$, for $i \in \{4, ... , n\}$ additional to all of the previous paths. The new path is the same as the longest path from 1 to $i$ in the $n$ point diagram, with an extra factor of $\alpha_{2n-5}\alpha_{2n-4}$ from the new inverse soft factor. Finally, the path from 1 to $n$ is  $\alpha_{2n-5}$ and from 3 to $n$ is $\alpha_{2n-5}\alpha_{2n-4}$. From these rules we can see that $C^n$ can be defined recursively from $C^{n-1}$ as

\begin{align}
(C^n)_{a i} &= (C^4)_{a i}, i \in \{1, 2, 3\}, a \in \{1, 2\} \nonumber\\
(C^n)_{1 i} &= \alpha_{2n-5}(C^{n-1})_{1 i}, i \in \{4, ..., n-1\} \nonumber\\
(C^n)_{2 i} &= (C^{n-1})_{2 i}+\alpha_{2n-5}\alpha_{2n-4}(C^{n-1})_{1 i}, i \in \{4, ..., n-1\}\nonumber \\
(C^n)_{1 n} &= \alpha_{2n-5} \nonumber\\
(C^n)_{2 n} &= \alpha_{2n-5}\alpha_{2n-4}. 
\end{align}

Let us calculate some useful minors of $C^n$, which will be written $(ab)_n$. Note that for $n = 4$, we define leg 5 to be leg $1$ so for example $(45)_4 := (41)_4 = \alpha _4$. We can then calculate $(45)_n$ using the recursive definition for $C^n$ as follows. 
\begin{equation}
(45)_n = (C^n)_{41} (C^n)_{52} - (C^n)_{51} (C^n)_{42} = \alpha_{2n-5} (45)_{n-1}
\end{equation}
and using the fact that $(45)_4 = \alpha_4$, we solve for $(45)_n$ by induction. The remaining minors we will need can be read off from the $n$ point $C$ matrix as
\begin{align}
\label{eqn:mhvminors}
(13)_n &= 1 \nonumber\\
(23)_n &= \alpha_1 \nonumber\\
\frac{(3j)_n}{(3i)_n} &= \prod_{k = j}^{i-1}\alpha_{2k-5},  j < i \in \{4, ..., n\}.
\end{align}

Now let us calculate the seed of the recursion $\mathcal{A}^{(0)*}_{4,2}$. We have a factor $\ang{\alpha_2 \alpha_1}[\alpha_2 \alpha_4]B_{234;1} = \frac{\alpha_2}{\alpha_1}\frac{\ang{23}\ang{14}}{\ang{31}\ang{24}}$ from the bridge, a factor of one over each edge variable, and spinor bracket factors of $\langle 13 \rangle$ and $\alpha_1[12]$ which can be uplifted to a covariant expression using~(\ref{eqn:mhvminors}) and the relations between spinor brackets and minors to give

\begin{equation}
\label{eq:4ptbasecase}
\mathcal{A}^{(0)*}_4 = \int d^{2 \times 4} \Omega_8 \frac{\langle 13 \rangle [12]}{\alpha_1\alpha_3\alpha_4}\frac{\ang{23}\ang{14}}{\ang{31}\ang{24}} = \int d^{2 \times 4} \Omega_8 \frac{\langle 13 \rangle [12]}{(13)^2(34)(24)}
\end{equation}
where $d^{2\times4}\Omega_4$ is defined in~(\ref{meas}). Since we have computed this using the bonus relations, this is equal to the full 4 point amplitude in $\mathcal{N} = 8$ SUGRA without summing over any permutations, in terms of any two distinct $a$ and $b$ and distinct $c$ and $d$:
\[
\mathcal{A}^{(0)*}_{4,2} = \mathcal{M}^{(0)}_{4,2} = \int \frac{d^{2 \times 4} C}{\mbox{Vol(GL}(2))} 
\frac{\delta^{2\times (2|8)}(C \cdot\lambdat
)\delta^{2\times2}(\lambda\cdot C^{\perp})}
{\prod_{i<j} (ij)} \frac{\ang{ab}}{(ab)}\frac{[cd]}{(cd)^\perp}.
\]
Equation~(\ref{eq:4ptbasecase}) will be used as the base case for the recursion, and can be rearranged using the relations between angle brackets and minors, and identifying 5 with 1  to give the following expression: 

\begin{equation}
\mathcal{A}^{(0)*}_{4,2} = \int d^{2 \times 4} \Omega_8 \frac{\langle ab\rangle}{(ab)} \frac{ [12] (35)(14)}{(13)^2(34)(45)(24)}.
\end{equation}

To calculate $\mathcal{A}^{(0)*}_{n,2}$ from $\mathcal{A}^{(0)*}_{n{-}1,2}$, first we must simplify the new bridge decoration times the new spinor bracket factors. The previous bridge decoration without the bonus relation simplification was local, in that it only involved the spinors of the two vertices attached to the bridge. This allowed the decoration to be expressed directly in terms of edge variables only. When incorporating the bonus relations we must now relate the bridge to a further leg of the diagram; the extra leg that will remain fixed when we now sum over only $(n-3)!$ instead of $(n-2)!$ permutations of the external legs. This results in the bridge factor becoming non-local, and we must calculate the relevant spinor bracket factors associated with the bridge. 
We use the local rules from the algorithm to simplify some bracket factors, and we find that

\[
\ang{\alpha_{2n-4} \delta_n}[\alpha_{2n-4} \gamma_{n-1}] B_{\gamma_n \delta_n 4;n} = \alpha_{2n-4}\frac{\ang{\gamma_n \delta_n}\ang{n4}}{\ang{\delta_n 4}\ang{\gamma_n n}}
\]
which we calculate in terms of the spinors $|\gamma_i\rangle$ and $|\delta_i\rangle$ denoted in Figure~\ref{fig:nptgenerated}. As constructed from the diagram, $|\gamma_i\rangle = |3\rangle$ for all $i \in \{5,...,n\}$. Following through the paths for $|\delta_i\rangle$, uplifting to a covariant expression in terms of minors, and simplify using the following relation derived in Appendix~\ref{identities}
\[
|\delta_i\rangle = |1\rangle(i{+}1\:3) + |3\rangle(1\:i{+}1) = |i{+}1\rangle(13)
\]
we obtain the following expression for the bridge factor encoding the bonus relations:

\[
\ang{\alpha_{2n-4} \delta_n}[\alpha_{2n-4} \gamma_{n-1}] B_{\gamma_n \delta_n 4;n} = \alpha_{2n-4}\frac{\ang{3\:i{+}1}\ang{4i}}{\ang{3i}\ang{4\:i{+}1}}
\]
In addition to the bridge factor we gain an extra factor of $\alpha_{2n-4}^{-1}\alpha_{2n-5}^{-2}$ coming from the new edge variables, a factor $\alpha_{2n-5}$ from the calculation of the spinor bracket $\langle 13 \rangle$, and a factor $\sum_{j=4}^{n-1}\left(\prod_{k=j}^{n-1}\alpha_{2n-5}\right)[ji]$ from the new vertex, and we obtain the following recursion relation:

\begin{equation}
\mathcal{A}_{n,2}^{(0)*}  = \mathcal{A}_{n{-}1,2}^{(0)*}  \frac{1}{\alpha_{2n-5}}\frac{\ang{3\:n{+}1}\ang{4n}}{\ang{3n}\ang{4\:n{+}1}}\sum_{j=4}^{n-1}\left(\prod_{k=j}^{n-1}\alpha_{2n-5}\right)[jn]
\end{equation}

Solving this simple recursion relation and uplifting to a covariant expression in terms of minors using~(\ref{eqn:mhvminors}) then gives the final result: 

\begin{align}
\mathcal{A}_{n,2}^{(0)*} &= \int d^{2 \times n} \Omega_8 \frac{\langle ab\rangle}{(ab)} \frac{ [12] (35)(14)}{(13)^2(24)(34)(45)}\prod_{i=5}^n\frac{(3\:i{+}1)(4i)}{(3i)(4\:i{+}1)}\frac{\sum_{j=4}^{i-1}(3j)[ji]}{(3i)}   \nonumber\\ 
&= \int d^{2 \times n} \Omega_8 \frac{\langle ab\rangle}{(ab)} \frac{[12]}{(13)(24)(34)}\prod_{i=5}^n\frac{\sum_{j=4}^{i-1}(3j)[ji]}{(3i)}   \nonumber\\
&= 
\int d^{2 \times n} \Omega_8 \frac{\langle ab\rangle}{(ab)} \frac{1}{(23)(34)(42)}\prod_{i=4}^n\frac{\sum_{j=4}^{i}(3j)[j\,i{+1}]}{(3\,i{+}1)}  \nonumber \\
&= \frac{\delta^{4|16}(P)}{\prod_{i}\ang{i\,i{+}1}}\frac{1}{\ang{23}\ang{34}\ang{42}}\prod_{i=4}^n\frac{\langle3|P_{4...i}|i+1]}{\langle3\,i{+}1\rangle}
\label{eq:MHVplanarsol}
\end{align}
Note that the factor of $\alpha_{2n-5}^{-1}$ was absorbed in changing $(45)_{n{-}1}$ to $(45)_n$. The full $n$-point MHV amplitude in $\mathcal{N}=8$ SUGRA can be obtained from the above formula by summing over permutations of the legs $1, 5, ..., n$, which we have verified numerically up to 10 points. Equation \eqref{eq:MHVplanarsol} can be easily related to the BGK formula for MHV graviton scattering \cite{Berends:1988zp}. Note that as written with the Parke Taylor factor in the denominator it is clear that this formula comes from a planar object, a property which is not obvious from BGK's original form. Our formula is also valid for $n = 3,4$ whereas the original BGK formula only holds for $n\ge5$. A similar form was obtained in \cite{Mason:2008jy}.

\section{Relations between spinors and minors} \label{identities}
The Grassmannian approach to scattering amplitudes provides a geometrical way to view the $\lambda$ and $\lambdat$ spinors. For an $n$-point N$^{k-2}$MHV amplitude, the $\lambda$ spinors can be seen to lie inside $k$-planes in $n$ dimensions and the $\lambdat$ spinors lie in the orthogonal $(n-k)$-planes. Representing the $k$-planes by an $k \times n$ matrix $C$ and the $(n-k)$-planes by an $n \times (n-k)$ matrix $C^{\perp}$, this is implied by delta functions in the Grassmannian integral formulae for scattering amplitudes enforcing $C\cdot\lambdat = 0$ $C^\perp\cdot\lambda = 0$. We will now show that this gives rise to nontrivial relations between spinor brackets and minors of $C$ and $C^{\perp}$. 

Let the rows of the $k\times n$ matrix $C$ be denoted $C_{\alpha i}$, $i \in \{1, ..., n\}, \alpha\in\{1,...,k\}$. Then Cramer's rule for the linear dependence of the distinct set of rows from 1 to $k+1$ can be written succinctly as follows: 
\[
\sum_{\sigma \in \mathbb{Z}_{k+1}} (-1)^{1_\sigma k} C_{\alpha 1_\sigma} (2_\sigma ... (k+1)_\sigma) = 0.
\]
Analogous formulae exist for any distinct set of $k+1$ rows. Taking the product of this vector relation with a $(k-1)$ blade formed of rows of the $C$ matrix generates all possible Pl\"{u}cker identities for $Gr(k,n)$. As an example, consider $Gr(3,n)$. Choose four distinct rows $a, b, c$ and $d$ of $C$. Then Cramer's rule can be written
\[
 C_{\alpha a} (bcd) - C_{\alpha b} (cda) + C_{\alpha c} (dab) - C_{\alpha d} (abc)  = 0
\]
and taking the product with the $(k-1)$ blade $\epsilon^{\alpha\beta\gamma}C_{\beta d}C_{\gamma e}$ gives the Pl\"{u}cker relations
\begin{equation*} 
(dea) (bcd) - (deb) (cda) + (dec) (dab)  = 0
\end{equation*}

The constraints on $C$ and the $\lambda$ spinors imply that we can use the $GL(k)$ symmetry to set the first two rows of $C$ equal to $\lambda$, i.e. $C_{\beta i} = \lambda_{\beta i}, \beta \in \{1, 2\}$ and $C_{\alpha i}, a \in \{3, ..., k\}$ are unspecified. Doing so then gives the mixed Cramer's rule:
\begin{equation} 
\label{eqn:mixedcr1}
\sum_{\sigma \in \mathbb{Z}_{k+1}} (-1)^{1_\sigma k} \lambda_{1_\sigma} (2_\sigma ... (k+1)_\sigma) = 0
\end{equation}
For $k=2$, this is equivalent to the statement that $\left\langle ij\right\rangle / \left(ij\right)$ is invariant for all $i \neq j$ \cite{Heslop:2016plj}. 
Consider again the same example of $Gr(3,n)$. Choose four distinct rows $a, b, c$ and $d$ of $C$. Then the mixed Cramer's rule can be written
\begin{equation*} 
 \lambda_{a} (bcd) - \lambda_{b} (cda) + \lambda_{c} (dab) - \lambda_{d} (abc)  = 0
\end{equation*}
and taking the product with spinor $\lambda_d$ gives the mixed Pl\"{u}cker relations
\begin{equation*} 
 \langle da\rangle (bcd) - \langle db\rangle (cda) + \langle dc\rangle (dab)  = 0
\end{equation*}

We have so far used only the relations $C\cdot\lambdat = 0$. On the support of $C^\perp\cdot\lambda = 0$, we can derive another mixed Cramer's rule which relates $\tilde{\lambda}$ spinors and the minors of $C^\perp$:
\begin{equation} 
\label{eqn:mixedcr2}
\sum_{\sigma \in \mathbb{Z}_{n-k+1}} (-1)^{1_\sigma (n-k)} \tilde{\lambda}_{1_\sigma} (2_\sigma ... (n-k+1)_\sigma)^\perp = 0.
\end{equation}

\section{One-Loop 4-point Amplitude} \label{1loopb}

\begin{figure}
\centering
       \includegraphics[scale=0.7]{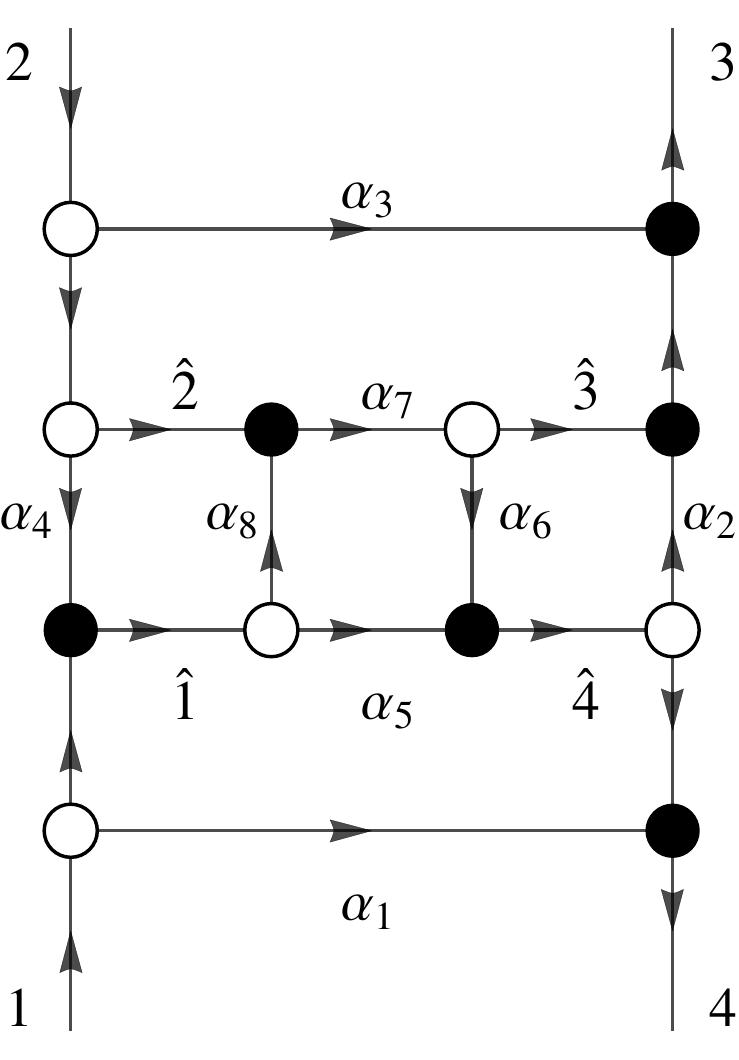}
    \caption{On-shell diagram for the 1-loop 4 point amplitude in $\mathcal{N}=4$ SYM} 
    \label{n41los1}
\end{figure}

\begin{figure}
\centering
       \includegraphics[scale = 0.6]{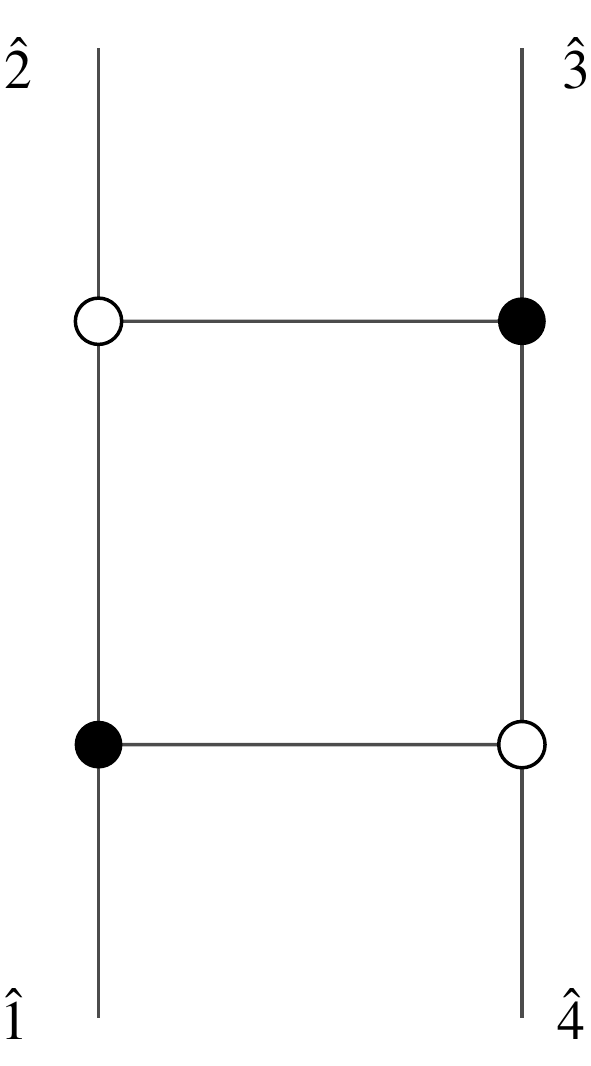}
    \caption{On-shell diagram for the tree-level 4 point amplitude in $\mathcal{N}=4$ SYM with BCFW shifted legs} 
    \label{4ptosd}
\end{figure}

\begin{figure}
\centering
       \includegraphics[scale=0.8]{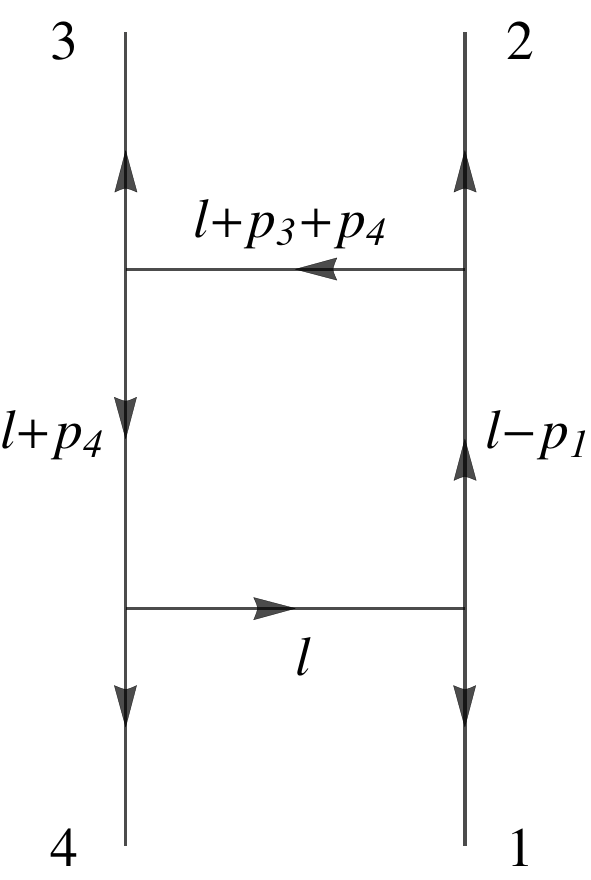}
    \caption{One-loop scalar box integral} 
    \label{1lpbox}
\end{figure}

In this Appendix, we will show that \eqref{4pt1pws} is equivalent to the standard expression for the 1-loop 4-point amplitude in $\mathcal{N}=4$ SYM in terms of a scalar box integral, with the loop momentum given by \eqref{loop1}. Let us assign arrows and edge variables to the on-shell diagram corresponding to the 1-loop 4-point amplitude as shown in Figure \ref{n41los1}. 
In terms of edge variables, this diagram is given by
\begin{equation}
\mathcal{A}_{4,2}^{(1)} =\int\prod_{i=1}^{4}\frac{d\alpha_{i}}{\alpha_{i}}\, \mathcal{A}_{4,2}^{(0)}\left(\hat{1},\hat{2},\hat{3},\hat{4}\right)\label{eq:m4}
\end{equation}
where $\mathcal{A}_{4}^{(0)}$ is the on-shell
diagram in Figure \ref{4ptosd}, which is simply a tree-level 4-point amplitude with BCFW shifted arguments:
\[
\mathcal{A}_{4,2}^{(0)}\left(\hat{1},\hat{2},\hat{3},\hat{4}\right)=\frac{\delta^{4|8}\left(P\right)}{\left\langle \hat{1}\hat{2}\right\rangle \left\langle \hat{2}\hat{3}\right\rangle \left\langle \hat{3}\hat{4}\right\rangle \left\langle \hat{4}\hat{1}\right\rangle }.
\]
Dividing equation~(\ref{eq:m4}) by the unshifted tree-level 4-point amplitude implies that
\begin{equation}
\frac{\mathcal{A}_{4,2}^{(1)}}{\mathcal{A}_{4,2}^{(0)}}=\int\prod_{i=1}^{4}\frac{d\alpha_{i}}{\alpha_{i}}\frac{\left\langle 12\right\rangle \left\langle 23\right\rangle \left\langle 34\right\rangle \left\langle 41\right\rangle }{\left\langle \hat{1}\hat{2}\right\rangle \left\langle \hat{2}\hat{3}\right\rangle \left\langle \hat{3}\hat{4}\right\rangle \left\langle \hat{4}\hat{1}\right\rangle }.
\label{n41}
\end{equation}
In order to show that \eqref{n41} is the standard formula for the 1-loop 4-point
amplitude, we must convert the integral over edge variables to an
integral over loop momentum. In fact, this integral is simply the
scalar box integral in Figure \ref{1lpbox}
\begin{equation}
\int\prod_{i=1}^{4}\frac{d\alpha_{i}}{\alpha_{i}}\frac{\left\langle 12\right\rangle \left\langle 23\right\rangle \left\langle 34\right\rangle \left\langle 41\right\rangle }{\left\langle \hat{1}\hat{2}\right\rangle \left\langle \hat{2}\hat{3}\right\rangle \left\langle \hat{3}\hat{4}\right\rangle \left\langle \hat{4}\hat{1}\right\rangle }=\int\frac{d^{4}l\left(p_{1}+p_{2}\right)^{2}\left(p_{1}+p_{4}\right)^{2}}{l^{2}\left(l+p_{4}\right)^{2}\left(l+p_{3}+p_{4}\right)^{2}\left(l-p_{1}\right)^{2}},\label{eq:1loopbox}
\end{equation}
where $l=\lambda_{\alpha_{5}}\tilde{\lambda}_{\alpha_{5}}+\alpha_{1}\lambda_{1}\tilde{\lambda}_{4}$. 
To prove this, note that $\lambda_{\alpha_{5}}\tilde{\lambda}_{\alpha_{5}}=\alpha_{5}\hat{\lambda}_{1}\hat{\tilde{\lambda}}_{4}$
and $\hat{\lambda}_{4}=\alpha_{5}\hat{\lambda}_{1}+\alpha_{6}\hat{\lambda}_3$,
from which it follows that $\alpha_{5}=\frac{\left\langle \hat{4}\hat{3}\right\rangle }{\left\langle \hat{1}\hat{3}\right\rangle }$
and
\begin{equation}
l=\frac{\left\langle \hat{4}\hat{3}\right\rangle }{\left\langle \hat{1}\hat{3}\right\rangle }\hat{\lambda}_{1}\hat{\tilde{\lambda}}_{4}+\alpha_{1}\lambda_{1}\tilde{\lambda}_{4}.\label{eq:loop}
\end{equation}
From this result, we then find that
\[
l-p_{1}=\frac{\left\langle \hat{4}\hat{3}\right\rangle }{\left\langle \hat{1}\hat{3}\right\rangle }\hat{\lambda}_{1}\hat{\tilde{\lambda}}_{4}+\alpha_{1}\lambda_{1}\tilde{\lambda}_{4}-\lambda_{1}\tilde{\lambda}_{1}
\]
Noting that $\tilde{\lambda}_{1}=\hat{\tilde{\lambda}}_{1}+\alpha_{1}\tilde{\lambda}_{4}$,
this simplifies to
\[
l-p_{1}=\frac{\left\langle \hat{4}\hat{3}\right\rangle }{\left\langle \hat{1}\hat{3}\right\rangle }\hat{\lambda}_{1}\hat{\tilde{\lambda}}_{4}-\lambda_{1}\hat{\tilde{\lambda}}_{1}.
\]
Plugging in the relation $\lambda_{1}=\alpha_{4}\hat{\lambda}_{2}-\hat{\lambda}_{1}$
and using the Schouten identity finally gives
\[
l-p_{1}=-\frac{\left\langle \hat{2}\hat{3}\right\rangle }{\left\langle \hat{1}\hat{3}\right\rangle }\hat{\lambda}_{1}\hat{\tilde{\lambda}}_{2}-\alpha_{4}\hat{\lambda}_{2}\hat{\tilde{\lambda}}_{1}.\label{eq:l-p1}
\]
Using similar manipulations, one finds that
\[
l+p_{4}=\frac{\left\langle \hat{4}\hat{1}\right\rangle }{\left\langle \hat{1}\hat{3}\right\rangle }\hat{\lambda}_{3}\hat{\tilde{\lambda}}_{4}+\alpha_{4}\hat{\lambda}_{4}\hat{\tilde{\lambda}}_{3}
\]
\[
l+p_{3}+p_{4}=-\frac{\left\langle \hat{2}\hat{1}\right\rangle }{\left\langle \hat{1}\hat{3}\right\rangle }\hat{\lambda}_{3}\hat{\tilde{\lambda}}_{2}+\alpha_{3}\hat{\lambda}_{2}\hat{\tilde{\lambda}}_{3}.
\]
Noting that the integral over loop momentum can be written as 

\[
\frac{d^{4}l}{l^{2}}=\frac{d^{2}\lambda_{\alpha_{5}}d^{2}\tilde{\lambda}_{\alpha_{5}}}{GL(1)}\frac{d\alpha_{1}}{\alpha_{1}}
\]
we then relate the loop integral to an integral over the variables $\alpha_i$ by taking the wedge product of the exterior derivative of equation~(\ref{eq:loop}), resulting in
\[
\int\frac{d^{4}l\left(p_{1}+p_{2}\right)^{2}\left(p_{1}+p_{4}\right)^{2}}{l^{2}\left(l+p_{4}\right)^{2}\left(l+p_{3}+p_{4}\right)^{2}\left(l-p_{1}\right)^{2}}=\int\prod_{i=1}^{4}\frac{d\alpha_{i}}{\alpha_{i}}\frac{\left\langle 43\right\rangle \left\langle 14\right\rangle }{\left\langle \hat{1}\hat{2}\right\rangle \left\langle \hat{2}\hat{3}\right\rangle \left\langle \hat{3}\hat{4}\right\rangle \left\langle \hat{4}\hat{1}\right\rangle }\frac{\left[43\right]\left\langle \hat{4}\hat{3}\right\rangle \left[14\right]\left\langle \hat{1}\hat{4}\right\rangle }{\left[\hat{3}\hat{2}\right]\left[\hat{1}\hat{2}\right]}.
\]
Equation \eqref{eq:1loopbox} then follows from using the identities
$\left[43\right]\left\langle \hat{4}\hat{3}\right\rangle =\left\langle 12\right\rangle \left[\hat{1}\hat{2}\right]$ and
$\left[14\right]\left\langle \hat{1}\hat{4}\right\rangle =\left\langle 32\right\rangle \left[\hat{3}\hat{2}\right]$. Combining equations \eqref{n41} and \eqref{eq:1loopbox} finally gives
\[
\frac{\mathcal{A}_{4,2}^{(1)}}{\mathcal{A}_{4,2}^{(0)}}=\int\frac{d^{4}l\left(p_{1}+p_{2}\right)^{2}\left(p_{1}+p_{4}\right)^{2}}{l^{2}\left(l+p_{4}\right)^{2}\left(l+p_{3}+p_{4}\right)^{2}\left(l-p_{1}\right)^{2}}
\]
which demonstrates that \eqref{4pt1pws} is equivalent to the standard expression for the 1-loop 4-point in terms of a scalar box integral.

\section{1-Loop Scattering Equations} \label{counting}

In section \ref{1loopsection}, we derived formulas for 1-loop 4-point amplitudes supported on scattering equations refined by MHV degree. In this appendix, we will consider a generalization of these scattering equations for any number of legs and analyze various properties of the solutions to these equations. The generalization can be obtained by applying a forward limit and BCFW shift to the tree-level scattering equations refined by MHV degree described in section \ref{ambrev}:
\[
\tilde{\lambda}_{l}-\sum_{r}\frac{\tilde{\lambda}_{r}}{(lr)}-\frac{\tilde{\lambda}_{n+1}}{(l\, n{+}1)}=0,\,\,\,\lambda_{r}-\sum_{l}\frac{\lambda_{l}}{(rl)}-\frac{\lambda_{0}}{(r0)}=0,
\]
where $l\in\left\{ 1,...,k\right\}$, $r\in\left\{ k+1,...,n\right\}$. In particular, setting $\left(\lambda_{n+1},\tilde{\lambda}_{n+1},\tilde{\eta}_{n+1}\right)=\left(-\lambda_{0},\tilde{\lambda}_{0},\tilde{\eta}_{0}\right)$ and BCFW shifting legs $1$ and $n$ according to  $\left(\hat{\tilde{\lambda}}_{1},\hat{\tilde{\eta}}_{1}\right)=\left(\tilde{\lambda}_{1}+\alpha\tilde{\lambda}_{n},\tilde{\eta}_{1}+\alpha\tilde{\eta}_{n}\right)$ and $\hat{\lambda}_{n}=\lambda_{n}-\alpha\lambda_{1}$, we obtain the following 1-loop scattering equations refined by MHV degree:
\[
\tilde{\lambda}_{0}-\frac{1}{\Phi}\sum_{r}\frac{\tilde{\lambda}_{r}}{(0r)}=0,\,\,\,\lambda_{0}+\frac{1}{\Phi}\sum_{l}\frac{\lambda_{l}}{(n{+}1\, l)}=0
\]
\begin{equation}
\hat{\tilde{\lambda}}_{l}-\sum_{r}\tilde{\lambda}_{r}\left(\frac{1}{(lr)}+\frac{1}{\Phi(l\, n{+}1)(0r)}\right)=0,\,\,\,\hat{\lambda}_{r}-\sum_{l}\lambda_{l}\left(\frac{1}{(rl)}-\frac{1}{\Phi(r0)(n{+}1\, l)}\right)=0
\label{1loopgeneralscatt}
\end{equation}
where $\Phi=1-(0 \,n{+}1)^{-1}$ and the hats act trivially on legs other than $1$ and $n$. These equations arise from computing a 1-loop $n$-point N$^{k-2}$MHV amplitude by plugging a tree-level $(n+2)$-point N$^{k-1}$MHV amplitude in \eqref{n=4ambi} into the first term of the loop-level BCFW recursion relation depicted in Figure \ref{n4recursion}. For $n=4$, the second term in the recursion relation does not contribute and the scattering equations above can be mapped into \eqref{scatteringeq} by taking $\left(\sigma_{0},\sigma_{5}\right)\rightarrow\left(\beta\sigma_{0},\gamma\sigma_{5}\right)$ and $\left(\lambda_{0},\tilde{\lambda}_{0}\right)\rightarrow\left(-\beta\lambda_{0},\gamma\tilde{\lambda}_{0}\right)$, where $\beta$ and $\gamma$ satisfy the constraint $\beta\gamma=1+(05)^{-1}$, with $(05)$ defined prior to the transformation. Although the second term in the recursion relation will contribute for $n>4$, we will neglect it for simplicity. 

For fixed loop momentum, we can treat all the spinors appearing in \eqref{1loopgeneralscatt} as external data and solve for the worldsheet coordinates $\sigma_i^\alpha$. Although there are $2n+4$ equations, four of them encode momentum conservation and can be discarded. One can also fix four of the worldsheet coordinates, leaving $2n$ equations for $2n$ unknowns. This can be achieved fixing $\sigma_i^\alpha$ for two particles of the same helicity, and discarding the scattering equations for two (possibly different) particles of the same helicity.

We will now derive a formula for the number of solutions to \eqref{1loopgeneralscatt}. First recall that the number of solutions to the tree level N$^{k-2}$MHV equations at $n$ points are given by the Eulerian numbers,~\cite{Spradlin:2009qr,Cachazo:2013iaa,Lipstein:2015rxa} 
\[E_{n,k}^{(0)} = \left\langle\begin{smallmatrix} n-3\\k-2 \end{smallmatrix}  \right\rangle.\]
Now consider adding two extra particles and taking a forward limit. We must assign helicities to the two new particles, which can be done in three distinct ways. If both particles are in the left set, then after taking the forward limit the scattering equations become degenerate and correspond to the tree level equations with $n+1$ particles in total and $k+1$ particles in the left set, giving $\left\langle\begin{smallmatrix} n-2\\k-1 \end{smallmatrix}  \right\rangle$ solutions which sum over $k$ to $(n-2)!$. All of these solutions are singular in the sense that $\sigma_0^\alpha$ and $\sigma_{n+1}^\alpha$ degenerate to the same point on the Riemann sphere. Following through the same analysis for both particles in the right set we find $\left\langle\begin{smallmatrix} n-2\\k-2 \end{smallmatrix}  \right\rangle$ solutions which again are all singular and sum over $k$ to $(n-2)!$. Hence, if we take the two particles to have opposite helicity, as we did to obtain the 1-loop scattering equations in \eqref{1loopgeneralscatt}, the number of solutions is given by
\[E_{n,k}^{(1)} = \left\langle\begin{smallmatrix} n-1\\k-1 \end{smallmatrix}  \right\rangle-\left\langle\begin{smallmatrix} n-2\\k-1 \end{smallmatrix}  \right\rangle-\left\langle\begin{smallmatrix} n-2\\k-2 \end{smallmatrix}  \right\rangle,\] 
which we have verified numerically up to 7 points and enumerate in figure~\ref{fig:oneloopsols}. As we have considered all of the $2(n-1)!$ singular solutions found in general dimensions~\cite{He:2015yua}, we conclude that \eqref{1loopgeneralscatt} must produce only regular solutions, which we have also verified numerically up to 7 points. Summing $E_{n,k}^{(1)}$ over $k$ then gives $(n-1)!-2(n-2)!$ solutions, in agreement with the number of regular solutions to the 1-loop scattering equations in general dimensions.  For supersymmetric theories, singular solutions do not contribute to the calculation of the amplitude~\cite{He:2015yua}, and we need consider only the equations in~\eqref{1loopgeneralscatt}.
\begin{figure}[h]
\centering
\begin{tabular}{cc}
$n$ & $E_{n,k}^{(1)}$ \\
4 & 2\\    
5 & 6 6	\\    
6 & 14 44 14\\ 
7 & 30 210 210 30\\   
8 & 62 832 1812 832 62 \\
9 & 126 2982 12012 12012 2982 126
\end{tabular}
\caption{Number of solutions to 1-loop scattering equations refined by MHV degree in \eqref{1loopgeneralscatt}.}
\label{fig:oneloopsols}
\end{figure} 
\bibliographystyle{ieeetr} 
\bibliography{ambi2osdreferences}

\begin{thebibliography}{10}

\bibitem{Drummond:2009fd}
J.~M. Drummond, J.~M. Henn, and J.~Plefka, ``{Yangian symmetry of scattering
  amplitudes in N=4 super Yang-Mills theory},'' {\em JHEP}, vol.~05, p.~046,
  2009.

\bibitem{Bern:2006kd}
Z.~Bern, L.~J. Dixon, and R.~Roiban, ``{Is N = 8 supergravity ultraviolet
  finite?},'' {\em Phys. Lett.}, vol.~B644, pp.~265--271, 2007.

\bibitem{Mason:2013sva}
L.~Mason and D.~Skinner, ``{Ambitwistor strings and the scattering
  equations},'' {\em JHEP}, vol.~07, p.~048, 2014.

\bibitem{Cachazo:2013hca}
F.~Cachazo, S.~He, and E.~Y. Yuan, ``{Scattering of Massless Particles in
  Arbitrary Dimensions},'' {\em Phys. Rev. Lett.}, vol.~113, no.~17, p.~171601,
  2014.

\bibitem{Fairlie:1972zz}
D.~B. Fairlie and D.~E. Roberts, ``{DUAL MODELS WITHOUT TACHYONS - A NEW
  APPROACH},'' 1972.

\bibitem{Gross:1987kza}
D.~J. Gross and P.~F. Mende, ``{The High-Energy Behavior of String Scattering
  Amplitudes},'' {\em Phys. Lett.}, vol.~B197, pp.~129--134, 1987.

\bibitem{Adamo:2013tsa}
T.~Adamo, E.~Casali, and D.~Skinner, ``{Ambitwistor strings and the scattering
  equations at one loop},'' {\em JHEP}, vol.~04, p.~104, 2014.

\bibitem{Geyer:2015bja}
Y.~Geyer, L.~Mason, R.~Monteiro, and P.~Tourkine, ``{Loop Integrands for
  Scattering Amplitudes from the Riemann Sphere},'' {\em Phys. Rev. Lett.},
  vol.~115, no.~12, p.~121603, 2015.

\bibitem{Geyer:2015jch}
Y.~Geyer, L.~Mason, R.~Monteiro, and P.~Tourkine, ``{One-loop amplitudes on the
  Riemann sphere},'' {\em JHEP}, vol.~03, p.~114, 2016.

\bibitem{Geyer:2014fka}
Y.~Geyer, A.~E. Lipstein, and L.~J. Mason, ``{Ambitwistor Strings in Four
  Dimensions},'' {\em Phys. Rev. Lett.}, vol.~113, no.~8, p.~081602, 2014.

\bibitem{Nair:1988bq}
V.~P. Nair, ``{A Current Algebra for Some Gauge Theory Amplitudes},'' {\em
  Phys. Lett.}, vol.~B214, pp.~215--218, 1988.

\bibitem{Witten:2003nn}
E.~Witten, ``{Perturbative gauge theory as a string theory in twistor space},''
  {\em Commun. Math. Phys.}, vol.~252, pp.~189--258, 2004.

\bibitem{Berkovits:2004hg}
N.~Berkovits, ``{An Alternative string theory in twistor space for N=4
  superYang-Mills},'' {\em Phys. Rev. Lett.}, vol.~93, p.~011601, 2004.

\bibitem{Roiban:2004yf}
R.~Roiban, M.~Spradlin, and A.~Volovich, ``{On the tree level S matrix of
  Yang-Mills theory},'' {\em Phys. Rev.}, vol.~D70, p.~026009, 2004.

\bibitem{Skinner:2013xp}
D.~Skinner, ``{Twistor Strings for N=8 Supergravity},'' 2013.

\bibitem{Spradlin:2009qr}
M.~Spradlin and A.~Volovich, ``{From Twistor String Theory To Recursion
  Relations},'' {\em Phys. Rev.}, vol.~D80, p.~085022, 2009.

\bibitem{ArkaniHamed:2012nw}
N.~Arkani-Hamed, J.~L. Bourjaily, F.~Cachazo, A.~B. Goncharov, A.~Postnikov,
  and J.~Trnka, {\em {Scattering Amplitudes and the Positive Grassmannian}}.
\newblock Cambridge University Press, 2016.

\bibitem{Britto:2004ap}
R.~Britto, F.~Cachazo, and B.~Feng, ``{New recursion relations for tree
  amplitudes of gluons},'' {\em Nucl. Phys.}, vol.~B715, pp.~499--522, 2005.

\bibitem{Britto:2005fq}
R.~Britto, F.~Cachazo, B.~Feng, and E.~Witten, ``{Direct proof of tree-level
  recursion relation in Yang-Mills theory},'' {\em Phys. Rev. Lett.}, vol.~94,
  p.~181602, 2005.

\bibitem{ArkaniHamed:2010kv}
N.~Arkani-Hamed, J.~L. Bourjaily, F.~Cachazo, S.~Caron-Huot, and J.~Trnka,
  ``{The All-Loop Integrand For Scattering Amplitudes in Planar N=4 SYM},''
  {\em JHEP}, vol.~01, p.~041, 2011.

\bibitem{ArkaniHamed:2009dn}
N.~Arkani-Hamed, F.~Cachazo, C.~Cheung, and J.~Kaplan, ``{A Duality For The S
  Matrix},'' {\em JHEP}, vol.~03, p.~020, 2010.

\bibitem{Arkani-Hamed:2013jha}
N.~Arkani-Hamed and J.~Trnka, ``{The Amplituhedron},'' {\em JHEP}, vol.~10,
  p.~030, 2014.

\bibitem{Heslop:2016plj}
P.~Heslop and A.~E. Lipstein, ``{On-shell diagrams for $ \mathcal{N} $ = 8
  supergravity amplitudes},'' {\em JHEP}, vol.~06, p.~069, 2016.

\bibitem{Herrmann:2016qea}
E.~Herrmann and J.~Trnka, ``{Gravity On-shell Diagrams},'' {\em JHEP}, vol.~11,
  p.~136, 2016.

\bibitem{Baadsgaard:2015twa}
C.~Baadsgaard, N.~E.~J. Bjerrum-Bohr, J.~L. Bourjaily, S.~Caron-Huot, P.~H.
  Damgaard, and B.~Feng, ``{New Representations of the Perturbative
  S-Matrix},'' {\em Phys. Rev. Lett.}, vol.~116, no.~6, p.~061601, 2016.

\bibitem{Dolan:2009wf}
L.~Dolan and P.~Goddard, ``{Gluon Tree Amplitudes in Open Twistor String
  Theory},'' {\em JHEP}, vol.~12, p.~032, 2009.

\bibitem{Nandan:2009cc}
D.~Nandan, A.~Volovich, and C.~Wen, ``{A Grassmannian Etude in NMHV Minors},''
  {\em JHEP}, vol.~07, p.~061, 2010.

\bibitem{ArkaniHamed:2009dg}
N.~Arkani-Hamed, J.~Bourjaily, F.~Cachazo, and J.~Trnka, ``{Unification of
  Residues and Grassmannian Dualities},'' {\em JHEP}, vol.~01, p.~049, 2011.

\bibitem{Berends:1988zp}
F.~A. Berends, W.~T. Giele, and H.~Kuijf, ``{On relations between multi - gluon
  and multigraviton scattering},'' {\em Phys. Lett.}, vol.~B211, pp.~91--94,
  1988.

\bibitem{Green:1982sw}
M.~B. Green, J.~H. Schwarz, and L.~Brink, ``{N=4 Yang-Mills and N=8
  Supergravity as Limits of String Theories},'' {\em Nucl. Phys.}, vol.~B198,
  pp.~474--492, 1982.

\bibitem{Bandos:2014lja}
I.~Bandos, ``{Twistor/ambitwistor strings and null-superstrings in spacetime of
  D=4, 10 and 11 dimensions},'' {\em JHEP}, vol.~09, p.~086, 2014.

\bibitem{Parke:1986gb}
S.~J. Parke and T.~R. Taylor, ``{An Amplitude for $n$ Gluon Scattering},'' {\em
  Phys. Rev. Lett.}, vol.~56, p.~2459, 1986.

\bibitem{Hodges:2012ym}
A.~Hodges, ``{A simple formula for gravitational MHV amplitudes},'' 2012.

\bibitem{Hodges:2011wm}
A.~Hodges, ``{New expressions for gravitational scattering amplitudes},'' {\em
  JHEP}, vol.~07, p.~075, 2013.

\bibitem{Bullimore:2009cb}
M.~Bullimore, L.~J. Mason, and D.~Skinner, ``{Twistor-Strings, Grassmannians
  and Leading Singularities},'' {\em JHEP}, vol.~03, p.~070, 2010.

\bibitem{Postnikov:2006kva}
A.~Postnikov, ``{Total positivity, Grassmannians, and networks},'' 2006.

\bibitem{Ferro:2013dga}
L.~Ferro, T.~Łukowski, C.~Meneghelli, J.~Plefka, and M.~Staudacher,
  ``{Spectral Parameters for Scattering Amplitudes in N=4 Super Yang-Mills
  Theory},'' {\em JHEP}, vol.~01, p.~094, 2014.

\bibitem{Cachazo:2012pz}
F.~Cachazo, L.~Mason, and D.~Skinner, ``{Gravity in Twistor Space and its
  Grassmannian Formulation},'' {\em SIGMA}, vol.~10, p.~051, 2014.

\bibitem{He:2012er}
S.~He, ``{A Link Representation for Gravity Amplitudes},'' {\em JHEP}, vol.~10,
  p.~139, 2013.

\bibitem{Frassek:2015rka}
R.~Frassek, D.~Meidinger, D.~Nandan, and M.~Wilhelm, ``{On-shell diagrams,
  Graßmannians and integrability for form factors},'' {\em JHEP}, vol.~01,
  p.~182, 2016.

\bibitem{He:2016dol}
S.~He and Y.~Zhang, ``{Connected formulas for amplitudes in standard model},''
  {\em JHEP}, vol.~03, p.~093, 2017.

\bibitem{He:2016jdg}
S.~He and Z.~Liu, ``{A note on connected formula for form factors},'' {\em
  JHEP}, vol.~12, p.~006, 2016.

\bibitem{Brandhuber:2016xue}
A.~Brandhuber, E.~Hughes, R.~Panerai, B.~Spence, and G.~Travaglini, ``{The
  connected prescription for form factors in twistor space},'' {\em JHEP},
  vol.~11, p.~143, 2016.

\bibitem{Bork:2017qyh}
L.~V. Bork and A.~I. Onishchenko, ``{Four dimensional ambitwistor strings and
  form factors of local and Wilson line operators},'' 2017.

\bibitem{Berkovits:2004jj}
N.~Berkovits and E.~Witten, ``{Conformal supergravity in twistor-string
  theory},'' {\em JHEP}, vol.~08, p.~009, 2004.

\bibitem{ArkaniHamed:2008gz}
N.~Arkani-Hamed, F.~Cachazo, and J.~Kaplan, ``{What is the Simplest Quantum
  Field Theory?},'' {\em JHEP}, vol.~09, p.~016, 2010.

\bibitem{Spradlin:2008bu}
M.~Spradlin, A.~Volovich, and C.~Wen, ``{Three Applications of a Bonus Relation
  for Gravity Amplitudes},'' {\em Phys. Lett.}, vol.~B674, pp.~69--72, 2009.

\bibitem{He:2010ab}
S.~He, D.~Nandan, and C.~Wen, ``{Note on Bonus Relations for N=8 Supergravity
  Tree Amplitudes},'' {\em JHEP}, vol.~02, p.~005, 2011.

\bibitem{Mason:2008jy}
L.~J. Mason and D.~Skinner, ``{Gravity, Twistors and the MHV Formalism},'' {\em
  Commun. Math. Phys.}, vol.~294, pp.~827--862, 2010.

\bibitem{Cachazo:2013iaa}
F.~Cachazo, S.~He, and E.~Y. Yuan, ``{Scattering in Three Dimensions from
  Rational Maps},'' {\em JHEP}, vol.~10, p.~141, 2013.

\bibitem{Lipstein:2015rxa}
A.~E. Lipstein, ``{Soft Theorems from Conformal Field Theory},'' {\em JHEP},
  vol.~06, p.~166, 2015.

\bibitem{He:2015yua}
S.~He and E.~Y. Yuan, ``{One-loop Scattering Equations and Amplitudes from
  Forward Limit},'' {\em Phys. Rev.}, vol.~D92, no.~10, p.~105004, 2015.

\end{thebibliography}
\end{document}